\documentclass[aps,prl,twocolumn,superscriptaddress,nofootinbib]{revtex4-2}
\usepackage{amsmath,amssymb}
\usepackage{graphicx}
\usepackage{hyperref}
\usepackage{booktabs}

\usepackage[T1]{fontenc}
\usepackage{multirow}
\usepackage{enumitem}
\usepackage{array}

\usepackage{bibunits}

\usepackage[pass]{geometry}

\makeatletter
\newcommand{\beginsupplement}{%
  \clearpage
  \onecolumngrid
  \newgeometry{margin=1in}
  \setcounter{page}{1}%
  \renewcommand{\thepage}{S\arabic{page}}%
  \setcounter{secnumdepth}{3}%
  \setcounter{section}{0}%
  \renewcommand{\thesection}{S\arabic{section}}%
  \renewcommand{\thesubsection}{\thesection.\arabic{subsection}}%
  \renewcommand{\thesubsubsection}{\thesubsection.\arabic{subsubsection}}%
  \renewcommand{\p@subsection}{}%
  \renewcommand{\p@subsubsection}{}%
  \setcounter{equation}{0}\renewcommand{\theequation}{S\arabic{equation}}%
  \setcounter{table}{0}\renewcommand{\thetable}{S\arabic{table}}%
  \setcounter{figure}{0}\renewcommand{\thefigure}{S\arabic{figure}}%
  \renewcommand{\bibnumfmt}[1]{[S##1]}%
  \renewcommand{\citenumfont}[1]{S##1}%
}
\makeatother

\begin{document}

\title{Absorbing Many-Body Correlations into Core-Optimized Orbitals}

\author{Hao Zhang}
\email{hao.zhang.quantum@gmail.com}
\affiliation{Department of Physics, University of Wisconsin--Madison, Madison, WI 53706, USA}
\author{Matthew Otten}
\email{mjotten@wisc.edu}
\affiliation{Department of Physics, University of Wisconsin--Madison, Madison, WI 53706, USA}
\affiliation{Department of Chemistry, University of Wisconsin--Madison, Madison, WI 53706, USA}

\date{\today}

\begin{abstract}
The cost of simulating quantum many-body systems---on
classical or quantum hardware---scales with the number of
variational parameters, so progress at fixed computational budget hinges on more
parameter-efficient ans\"atze. Configuration
Interaction (CI) is widely dismissed as parameter-heavy; we
show this verdict is an artifact of the orbital basis.
Co-optimizing the orbital basis with a sparse CI
wavefunction---a method we call \emph{Core-Optimized Orbitals}
(COO)---absorbs a large fraction of the dynamical correlation
directly into the single-particle basis, cutting the
determinant count by several orders of magnitude beyond the
already compact TrimCI ansatz on which it builds.
On \textrm{[Fe$_4$S$_4$]} (54e,\,36o), a billion-determinant
TrimCI~+~COO wavefunction reaches accuracy that would require
$3\!\times\!10^{14}$ determinants in a localized basis. At
matched accuracy, it is $8\times$ more compact than the
largest unrestricted-DMRG benchmark ($25\times$ with PT2).
Across the iron-sulfur series---from \textrm{[Fe$_2$S$_2$]}
(30e,\,20o) to the P-cluster (114e,\,73o)---TrimCI~+~COO is
$10$--$100\times$ more compact than SU(2)-adapted DMRG with
entanglement-minimized orbitals at matched accuracy.
A tunable Hubbard-on-graph model factorizes the advantage into an
orbital-basis gain and an ansatz gain, the latter capturing
multi-center entanglement that resists MPS localization. COO
therefore changes the picture of CI efficiency: sparse CI with optimized orbitals can outperform state-of-the-art tensor networks on strongly correlated multi-center systems.
\end{abstract}

\maketitle

\begin{bibunit}[apsrev4-2]


Over the past several decades, numerical many-body
calculations~\cite{KnowlesHandy1984,Dagotto1994,Ceperley1980,White1992,ZhangKrakauer2003,BartlettMusial2007,Schollwoeck2011,VerstraeteCirac2004,Verstraete2008,Booth2009,Carleo2017}
have become the main tool for strongly correlated quantum systems. Progress has largely been a story of compactness: how many adjustable parameters are needed to describe the state accurately.
At the frontier,
optimizable variational ansatzes now carry roughly $10^{9}$--$10^{12}$
parameters~\cite{Larsson2022,xiangDistributedMultiGPUInitio2024,Liu2025FPEPS,Menczer2024DMRG,Brower2026DMRG,Lee2026},
while CI methods based on product-space can manipulate
$10^{12}$--$10^{15}$ Slater
determinants~\cite{Gao2024,Xu2025SCI,Shayit2025}; yet Hilbert space still grows exponentially, far faster than practical
computing resources.  Every parameter therefore has
to count.  One path to a compact representation is to explore a better
many-body ansatz --- neural-network quantum
states~\cite{Carleo2017,Pfau2020,Hermann2020,guSolvingHubbardModel2025,liuEfficientOptimizationNeural2025,fuFermiSetsUniversal2026}, parameterized quantum
circuits~\cite{Peruzzo2014,McClean2016,grimsleyAdaptiveVariationalAlgorithm2019,zhangCyclicVariationalQuantum2025}, or
direct search over sparse determinant
expansions~\cite{TrimCI_paper}.
Here, we instead seek compactness by optimizing the single-particle basis in which the many-body ansatz is expressed.

The single-particle basis has been a variational degree of freedom from
the beginning: Hartree--Fock and Kohn--Sham
theory~\cite{Hartree1928,Fock1930,Roothaan1951,KohnSham1965} optimize
the orbitals at the mean-field level, and CASSCF and
DMRG-SCF~\cite{Roos1980,Werner1985,Zgid2008,Legeza2025OrbOpt} extend
the same idea to strongly correlated active spaces. Beyond this
common framework, many-body methods developed distinct
basis-handling strategies.
One line identifies the orbitals that matter. Active-space, downfolding, and embedding
constructions~\cite{Georges1996,Kotliar2006,KniziaChan2012,Sayfutyarova2017}
carve out which orbitals the expensive solver sees; natural
orbitals~\cite{Loewdin1955} diagonalize the one-body density matrix
to concentrate occupation on a few modes; localized molecular
orbitals and Wannier
functions~\cite{Boys1960,PipekMezey1989,MarzariVanderbilt1997}
rotate for spatial locality.
A second line is specific to DMRG, where the cost is set by entanglement
across an MPS chain; orbital ordering and entanglement-minimizing rotations
therefore reduce the bond dimension required by the matrix-product
representation~\cite{Legeza2003,Rissler2006,Krumnow2016,Li2025}.  A
third line appears in selected CI~\cite{Huron1973,Holmes2016,Smith2017,Sharma2017,Li2018}, where the bottleneck is the length of
the determinant expansion and orbital rotations can make the selected
space more compact, typically by up to one order of
magnitude~\cite{YaoUmrigar2021}.  A more recent motivation comes from quantum algorithms, where inefficient orbitals translate directly into deeper circuits, larger measurement
budgets, and a more expensive initial-state-preparation
stage~\cite{Kivlichan2018,Kanno2023,RobledoMoreno2025SQD,Danilov2025SQD,Wang2025LASSQD,Shajan2025Protein}.
Orbital-optimized VQE and UCC use a classical orbital rotation
before the quantum circuit is applied~\cite{Mizukami2020,Sokolov2020,Bierman2023}.
Single-determinant-overlap optimization targets phase estimation from a
different angle: since the cost depends on the initial overlap with the
eigenstate~\cite{Reiher2017}, it rotates the basis so that a cheap Slater
determinant becomes a better trial state~\cite{Ollitrault2024}.  The recent entanglement-minimized orbitals (EMO) of
Li~\cite{Li2025} are the state of the art on this front: by
rotating and reordering orbitals to minimize the entanglement an
MPS initial state must carry, EMO improves the dominant-determinant
overlap on iron-sulfur clusters by two to five orders of magnitude
over localized orbitals, significantly reducing the cost of QPE state preparation.

In this work, we optimize orbitals directly for many-body compactness. The usual
route --- CASSCF, DMRG-SCF, selected-CI orbital
optimization~\cite{Roos1980,Werner1985,Zgid2008,Greiner2024,Legeza2025OrbOpt,YaoUmrigar2021}
--- extracts the orbital gradient from a large variational
wavefunction, typically $10^{6}$--$10^{9}$ parameters, on the
implicit assumption that more parameters give a cleaner gradient.
But this large-wavefunction approach makes the orbital
optimization itself harder: the orbital parameters are coupled to
all of those variational parameters, and the sheer number of
coupled parameters tends to trap the orbital optimization in a
local basin, far from the globally optimal orbitals. We take the
opposite approach. Starting from random determinants, the recent
TrimCI~\cite{TrimCI_paper} algorithm finds a highly compact
$10^{2}$--$10^{4}$-determinant core that already captures the
structure of the ground state. We use that small but high-quality
core to set the orbital gradient and rotate the basis. The new basis
feeds back into TrimCI for a fresh core, and the loop iterates.
Typically 5--10 cycles converge to a high-quality basis. We call the
result Core Optimized Orbitals (COO).

The result reverses the expected compactness hierarchy on multi-center
strongly correlated molecules. On the iron-sulfur clusters Fe$_2$S$_2$, Fe$_4$S$_4$, and the
P-cluster~\cite{Sharma2014,Li2019Pcluster,xiangDistributedMultiGPUInitio2024}, TrimCI in COO orbitals shows
higher parameter efficiency than the best DMRG variants we are
aware of --- the unrestricted DMRG of Ref.~\cite{Lee2026} and the
SU(2)-DMRG with entanglement-minimized orbitals of
Ref.~\cite{Li2025} --- despite the substantial compression
machinery in those methods. Comparing COO directly to its starting
localized molecular orbital (LMO) basis on \textrm{[Fe$_4$S$_4$]},
a billion-determinant COO expansion reaches an energy that LMO would require
$\sim\!3\!\times\!10^{14}$ determinants for. On these multi-center strongly correlated
systems, sparse CI in COO is more compressed than the structured
MPS ansatz --- inverting the standard picture of tensor networks
as more compressed CI representations.


\begin{figure}[t]
\centering
\includegraphics[width=\columnwidth]{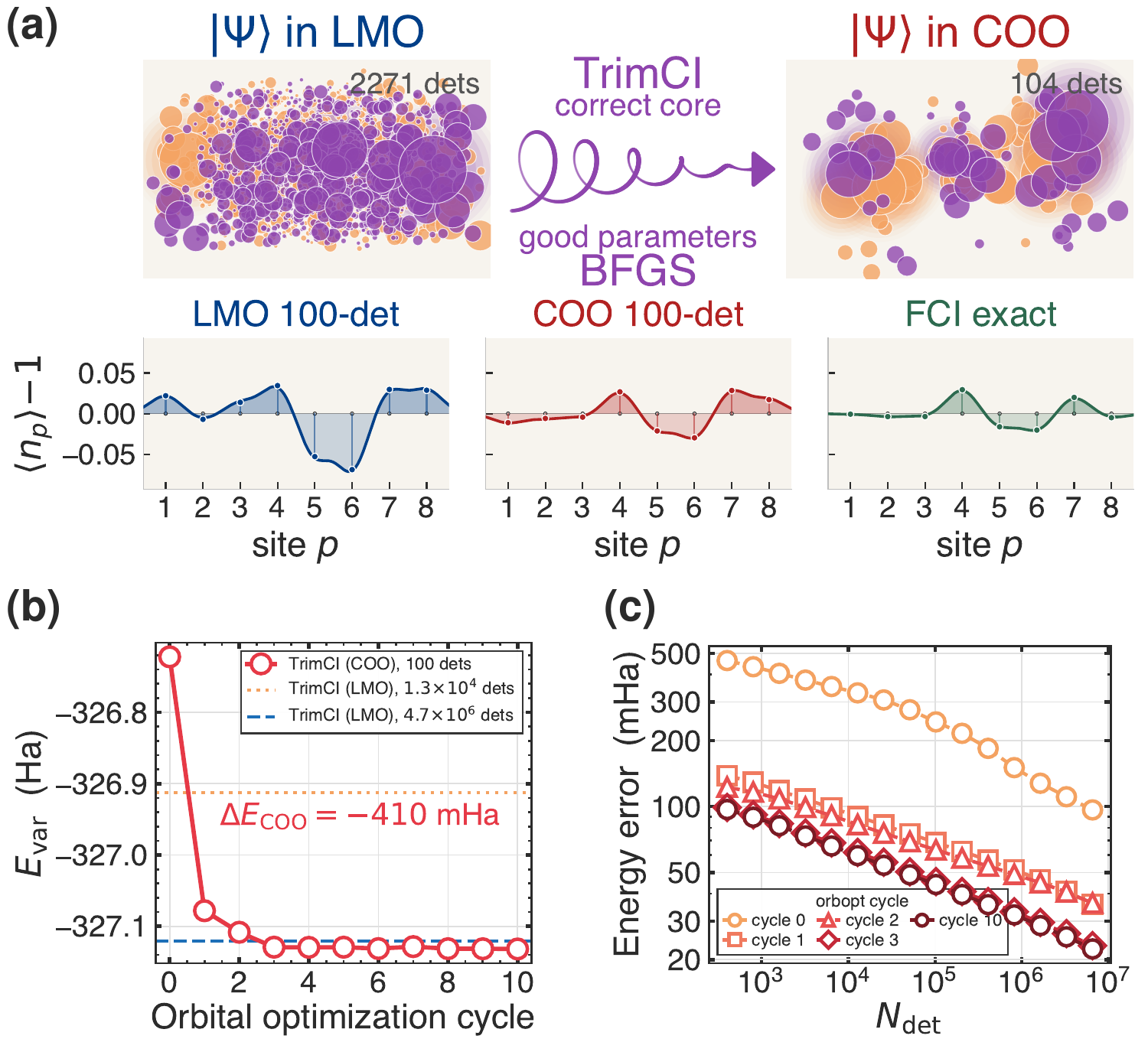}
\caption{Core Optimized Orbitals (COO).
\textbf{(a)}~Illustration of orbital rotation: the same
wavefunction $|\Psi\rangle$ (8-site Hubbard-on-graph
approximate ground state, Fig.~\ref{fig:mi}) needs 2271 dets
in LMO (left) but only 104 in COO (right). Each point is a
Slater determinant, placed by multidimensional scaling (MDS);
area $\propto |c_I|$, color by sign. Bottom row: site density
deviation $\langle n_p\rangle\!-\!1$ for the LMO and COO
100-det approximations vs the exact FCI.
\textbf{(b)}~Convergence of COO on a 100-determinant core for
\textrm{[Fe$_4$S$_4$]} (54e,\,36o), BS-1 (Fe$_1\!\uparrow$,
Fe$_2\!\uparrow$, Fe$_3\!\downarrow$, Fe$_4\!\downarrow$
broken-symmetry state).
\textbf{(c)}~Gain transfer: orbitals frozen at cycles 0, 1, 2, 3,
10 and expanded to $\sim\!10^{7}$ determinants at each fixed
orbital set. Both axes logarithmic; $\Delta E$ relative to the
BS-1 FCI energy $E_\mathrm{FCI} = -327.244$~Ha.}
\label{fig:coo}
\end{figure}

\begin{figure*}[!t]
\includegraphics[width=\textwidth]{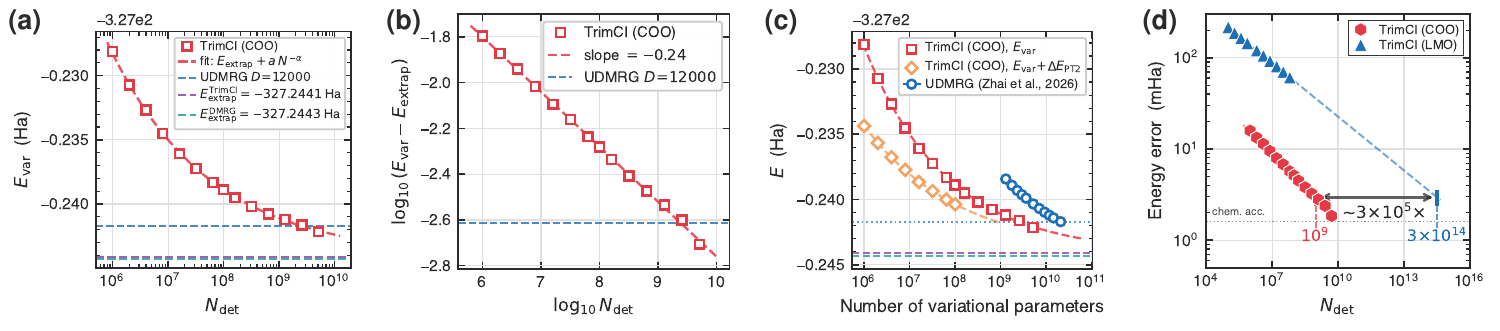}
\caption{Compression of the \textrm{[Fe$_4$S$_4$]} (54e,\,36o)
ground state.
\textbf{(a)}~$E_{\rm var}$ versus $N_{\rm det}$ for TrimCI~+~COO
with the power-law fit
$E(N_{\rm det}) = E_{\rm extrap} + a\,N_{\rm det}^{-\alpha}$
(dashed red); the UDMRG extrapolate of Ref.~\cite{Lee2026} is
shown for comparison.
\textbf{(b)}~Log--log form of (a).
\textbf{(c)}~Variational energy on a common variational-parameter
axis, with semistochastic PT2 corrections and the UDMRG benchmark.
\textbf{(d)}~$|E\!-\!E_{\rm FCI}|$ versus $N_{\rm det}$ for TrimCI
in COO orbitals (red hexagons) and in the starting LMO basis
(blue triangles), with power-law fits (dashed; COO slope $-0.24$,
LMO slope $-0.20$). Dotted line: chemical accuracy ($1.6$~mHa).}
\label{fig:dmrg}
\end{figure*}


\textit{COO method.}\label{sec:mechanism}---A CI wavefunction in an
orbital basis $\{\phi_p\}$ is
\begin{equation}
|\Psi\rangle \;=\; \sum_I c_I\,|D_I(\{\phi_p\})\rangle,
\label{eq:psi}
\end{equation}
with Slater determinants $|D_I\rangle$ built from those orbitals.
COO variationally optimizes both the amplitudes $\{c_I\}$ and the
orbitals. A change of orbital basis is a
unitary rotation,
\begin{equation}
\tilde\phi_p \;=\; \sum_q U_{qp}\,\phi_q,\qquad U\in U(n),
\label{eq:Urot}
\end{equation}
on the $n$ orbitals. $U$ can be parametrized as the matrix
exponential
\begin{equation}
U \;=\; e^{\kappa},
\label{eq:orbrot}
\end{equation}
of a real antisymmetric matrix $\kappa$ with $n(n{-}1)/2$ independent
entries $\kappa_{ai}$ ($a > i$); for simplicity we restrict to real
orbitals throughout. Figure~\ref{fig:coo}(a) illustrates this rotation: the same
wavefunction supported on thousands of LMO determinants
collapses to roughly a hundred in COO, the long tail
absorbed into the rotated orbitals.

However, the joint optimization of $\{c_I\}$ and $\kappa$ is
hard. The orbital rotation $e^{\kappa}$ acts on every determinant
of $|\Psi\rangle$, making the energy highly nonlinear in $\kappa$.
Combined with the $\{c_I\}\!\leftrightarrow\!\kappa$ coupling
noted above, naive alternation between the two parameter sets gets
stuck in self-consistent local minima where neither is improvable
on its own.

COO sidesteps this difficulty by anchoring the optimization on a
small, high-quality core wavefunction. TrimCI~\cite{TrimCI_paper}
produces this core by alternating two operations on the
Hamiltonian-connectivity graph of Slater determinants ---
\emph{expansion} (admit neighbors with strong $|H_{ij}|$) and
\emph{trim} (drop low-weight configurations after randomized-block
and global diagonalizations) --- starting from random determinants. For
the systems we study here, a $\sim\!100$-determinant core suffices, and
the closed-form orbital gradient
$\partial E/\partial \kappa_{ai}$ is computed in milliseconds from
the core's 1- and 2-RDMs.

COO alternates between sparse core search and
orbital rotation. TrimCI
finds a compact core in the current orbitals. With this core held
fixed, BFGS then optimizes $\kappa$ against the closed-form
gradient. The key design choice: each trial step in the BFGS line
search rotates the orbitals by $e^{\kappa}$ and re-diagonalizes the
projected Hamiltonian on the same core, letting $\{c_I\}$ relax to
the trial orbitals. The line-search energy then reflects the full
coupled $(\{c_I\}, \kappa)$ response, so the BFGS update sees the
right variational curvature rather than the noisier curvature of a
fixed-$\{c_I\}$ surrogate as in prior selected-CI orbital
optimization~\cite{YaoUmrigar2021}. This is where the
compact core pays off: a $100$-determinant Davidson takes
milliseconds, cheap enough to run inside every line-search trial,
so the line search probes the variational surface directly. We benchmarked every optimizer in
\texttt{scipy.optimize}~\cite{scipy} on the $\kappa$ parameters and
found BFGS leading in both speed and converged accuracy. The whole
$\kappa$ optimization runs in minutes per outer cycle on a single
workstation. Once it converges, the rotated integrals pass back to
TrimCI, which re-searches for a better core in the new basis.
Typically 5--10 outer cycles converge (see
Fig.~\ref{fig:coo}(b)): on \textrm{[Fe$_4$S$_4$]}, a
100-determinant core recovers $410$~mHa by cycle three --- already
lower than a fixed localized basis reaches at $4.7\!\times\!10^{6}$
determinants.


\textit{Gain transfer.}---A central test is whether orbital gains obtained from only $10^{2}$ determinants persist when the CI expansion is enlarged by many orders of magnitude. To test this, we freeze the orbitals at successive snapshots of the loop (cycles 0, 1, 2, 3, 10) and, for each fixed orbital set, run an independent TrimCI expansion up to $\sim\!10^{7}$ determinants [Fig.~\ref{fig:coo}(c)]. The result is a clear transfer of gain:
already after the first few orbital-optimization cycles, the energy
curve shifts left by several orders of magnitude relative to the
unrotated LMO basis. Most of this improvement is accumulated by
cycle~3, after which the curves change only weakly, indicating that
the orbitals are nearly converged. Overall, the cycle-10 orbitals
achieve the same accuracy as the starting LMO basis with roughly
$10^{4}$ fewer determinants. Thus a rotation learned from a
$10^{2}$-determinant core can transfer its compression gain across
five orders of magnitude of subsequent CI expansion.

This transferability has a simple physical origin. Every Slater
determinant in the expansion is built from the same single-particle
orbitals, so a single orbital rotation acts coherently on the entire
CI space. In second-quantized form, the same antisymmetric parameters
$\kappa_{ai}$ that define the orbital rotation matrix
$U=e^{\kappa}$ in Eq.~\eqref{eq:orbrot} also define the one-body operator
\begin{equation}
\hat{\kappa}
=
\sum_{a>i} \kappa_{ai}
(\hat a_a^\dagger \hat a_i-\hat a_i^\dagger \hat a_a),
\end{equation}
with
\begin{equation}
e^{\hat{\kappa}}\hat a_p^\dagger e^{-\hat{\kappa}}
=
\sum_q U_{qp}\hat a_q^\dagger .
\end{equation}

The compression depends on how far the optimized orbitals
move from the original ones. If the rotation is small, the induced
many-body matrix
$U_{JI}=\langle D_J|e^{\hat{\kappa}}|D_I\rangle$
is nearly diagonal, $U_{JI}\approx\delta_{JI}$, so each determinant
maps mostly onto itself, and the number of important determinants changes little. By contrast, when the optimized orbitals differ substantially
from the original basis, $U_{JI}$ becomes broadly distributed: a single
determinant in the rotated basis has weight on many determinants in the
original basis. Consequently, a compact COO wavefunction,
\begin{equation}
|\Psi_{\rm COO}\rangle
=
\sum_{I=1}^{10^2} c_I |\tilde D_I\rangle
=
e^{\hat{\kappa}}\sum_{I=1}^{10^2} c_I |D_I\rangle ,
\end{equation}
corresponds, when expressed back in the original orbital basis, to
\begin{equation}
|\Psi_{\rm COO}\rangle
=
\sum_J
\left(
\sum_{I=1}^{10^2} U_{JI}c_I
\right)
|D_J\rangle ,
\label{eq:eK_action}
\end{equation}
where many coefficients over $J$ can become non-negligible. In this sense,
the optimized orbitals absorb a large set of
determinants into the one-particle basis itself. This is why a rotation
learned from a $10^2$-determinant core can remain useful after the CI
expansion is enlarged by many orders of magnitude.

All results below use a three-phase TrimCI~+~COO workflow. In \textit{Phase~0 (Global Optimization)},
the joint core+orbital search just described starts from random
determinants and alternates TrimCI selection on a
$\sim\!100$-determinant core with BFGS rotation of $\kappa$ until
convergence. This stage provides most of the orbital
improvement. In \textit{Phase~1 (Local Refinement)}, starting from
the Phase~0 core, $N_{\rm det}$ grows slowly (growth factor
$\gamma\!\sim\!1.1$ per round) and $\kappa$ is re-rotated after
each expansion, fine-tuning the orbitals as the determinant space
broadens to $N_{\rm det}\!\sim\!10^{6}$. In \textit{Phase~2 (Fast
Expansion)}, the orbitals are frozen and $N_{\rm det}$ doubles per
round to the final $\sim\!10^{9}$ target, with optional
semistochastic PT2. Per-phase hyperparameters are listed in the
Supplemental Material~\cite{SM}; see also Fig.~S3 for the
full \textrm{[Fe$_4$S$_4$]} trajectory across all three phases.


\textit{Fe$_4$S$_4$ compression.}---We applied the full workflow to the challenging \textrm{[Fe$_4$S$_4$]}
(54e,\,36o) cluster in the BS-1 broken-symmetry state. The
resulting TrimCI~+~COO trajectory spans $10^{2}$ to
$5.12\!\times\!10^{9}$ determinants --- the largest variational
selected-CI calculation reported to date, $\sim\!2.5\times$ the
previous record~\cite{Sharma2017}.
Fitting a power-law ansatz
$E(N_{\rm det}) = E_{\rm extrap} + a\,N_{\rm det}^{-\alpha}$ to the
14 points from $10^{6}$ to $5.12\!\times\!10^{9}$
[Fig.~\ref{fig:dmrg}(a)] gives
$E^{\rm TrimCI}_{\rm extrap} = -327.2441$~Ha (R$^{2}$-scan
extrapolation, see SM), closely matching the independent UDMRG
extrapolate of Zhai \emph{et al.}~\cite{Lee2026},
$-327.2443$~Ha. We use $-327.244$~Ha as the FCI energy of BS-1 in
the error calculations below. In log--log
[Fig.~\ref{fig:dmrg}(b)] the 14 points closely follow a power law with
slope $-0.24$ over four decades. Fig.~\ref{fig:dmrg}(c) places CI
and DMRG on a common axis (total variational-parameter count).
TrimCI~+~COO reaches the largest UDMRG $D\!=\!12000$ benchmark
($-327.2417$~Ha, $2\!\times\!10^{10}$ parameters) with
$\sim\!8\times$ fewer parameters, $\sim\!25\times$ with
semistochastic PT2 corrections. Both TrimCI curves sit below UDMRG
at every matched budget. Our largest data point at
$5.12\!\times\!10^{9}$ determinants reaches $-327.2422$~Ha
($1.94$~mHa above the extrapolation), obtained with a purpose-built scalable distributed Davidson on 20 GPU
workers ($\sim\!30$~h wall time; see SM).

Fig.~\ref{fig:dmrg}(d) plots $|E\!-\!E_{\rm FCI}|$ versus
$N_{\rm det}$ for COO and for the same TrimCI expansion in the
starting localized basis (LMO). Both follow power laws, with COO
decaying faster (slope $-0.24$ vs $-0.20$), suggesting that the orbital rotation has absorbed part of the dynamical
correlation. Anchoring at $N_{\rm det}\!=\!10^{9}$, where
$\Delta E = 2.94$~mHa, the LMO power law extrapolates to
$\sim\!3\!\times\!10^{14}$ determinants for the same energy --- a
$\sim\!3\!\times\!10^{5}\!$ fold compression. This target remains below the full-CI dimension
$N_{\rm FCI}\!\approx\!10^{16}$,
so the comparison stays physically meaningful. By aligning the
orbitals with the ground state's dominant correlations, the
rotation $e^{\kappa}$ absorbs the dynamical correlation that an
LMO expansion would otherwise spread over $\sim\!10^{14}$
small-coefficient determinants --- effectively compressing a hundred-trillion-determinant LMO
description into a billion-determinant COO expansion.

\begin{figure}[t]
\includegraphics[width=\columnwidth]{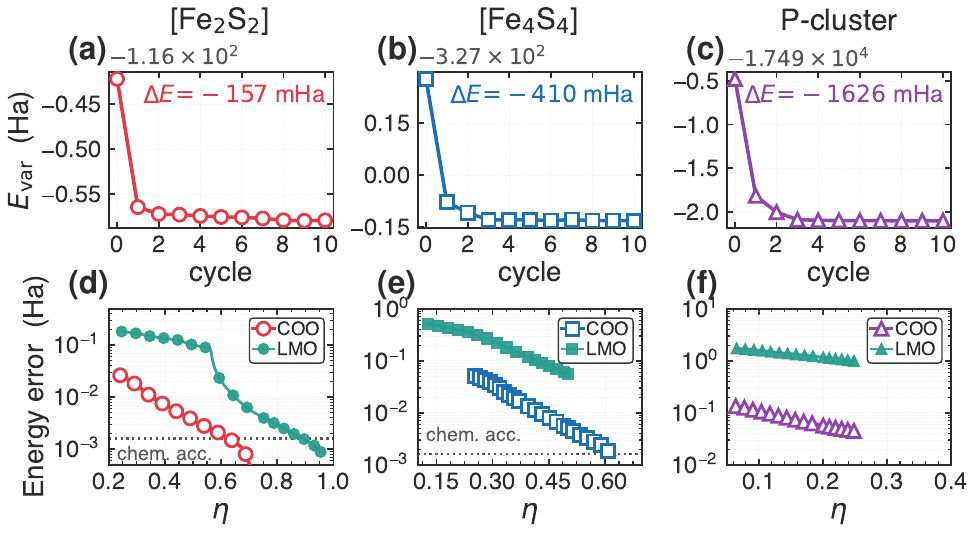}
\caption{\textbf{TrimCI~+~COO across the iron-sulfur series.}
Columns: Fe$_2$S$_2$ (30e,\,20o), Fe$_4$S$_4$ (54e,\,36o),
P-cluster (114e,\,73o).
\textbf{(a)--(c)}~Phase~0: variational energy versus
orbital-optimization cycle on a 100-determinant core.
\textbf{(d)--(f)}~Phase~1$+$2 expansion:
$|E_{\rm var} - E_{\rm FCI}|$ versus the log-normalized parameter
budget $\eta \equiv \log_{10} N_{\rm det} / \log_{10} N_{\rm FCI}$.
Hollow system-color markers: TrimCI~+~COO; filled teal: TrimCI+LMO.
Dotted line: chemical accuracy. FCI references and trajectory data
in the SM.}
\label{fig:fes_series}
\end{figure}

\textit{Generalization across the Fe-S series.}---We applied the
identical workflow to Fe$_2$S$_2$ (30e,\,20o) and the P-cluster
(114e,\,73o), in addition to the
\textrm{[Fe$_4$S$_4$]} (54e,\,36o) calculation already discussed
[Fig.~\ref{fig:fes_series}]. The $100$-determinant Phase~0 loop
reduces $E_{\rm var}$ by $157$--$1626$~mHa within ten
cycles~[(a)--(c)] (three to five already capture most of the
drop), and the gain scales monotonically with active-space size.
Continuing through Phases~1 and~2 to $10^{8}$ determinants
[(d)--(f)], TrimCI~+~COO uses $\sim\!10^{3}$--$10^{5}$$\times$
fewer determinants than the same TrimCI expansion without orbital
optimization (in the starting LMO basis) at matched energy
accuracy, across all three systems. The advantage widens with system size, showing that a rotation initialized from just $100$ determinants can deliver large compression across the Fe-S series.


\begin{figure}[t]
\includegraphics[width=\columnwidth]{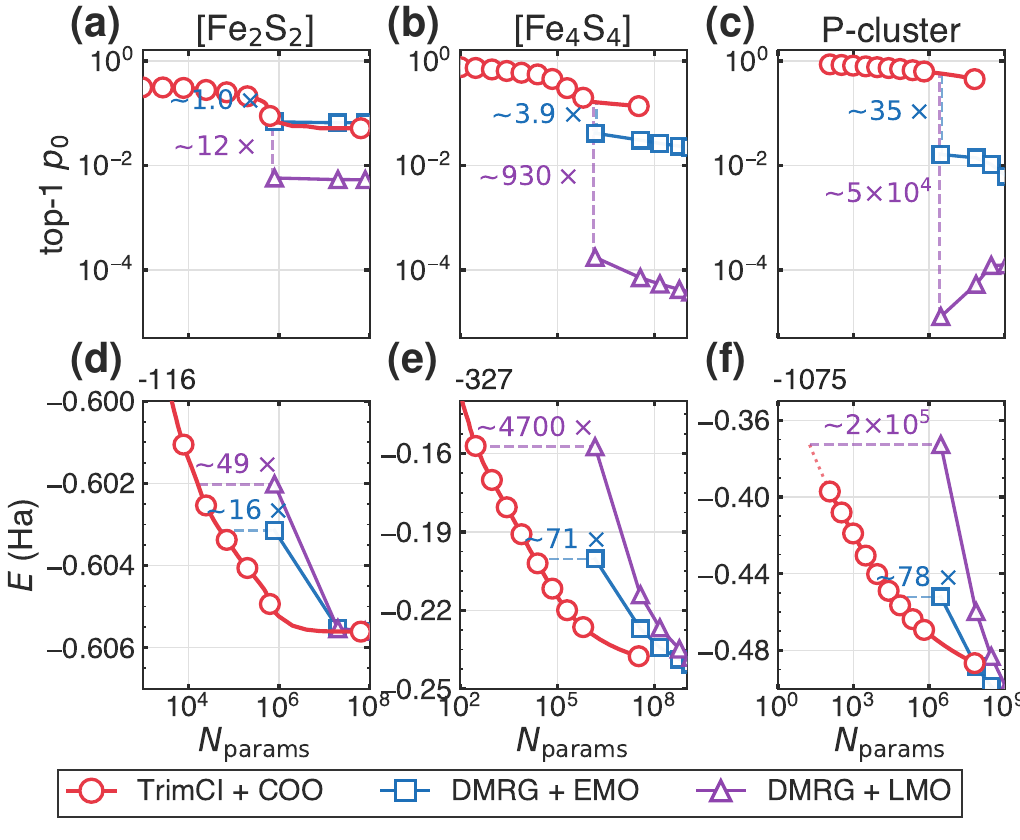}
\caption{\textbf{Compactness beyond entanglement minimization.}
Top row: dominant-determinant weight
$p_{0} \equiv |c_{\max}|^{2}$ versus parameter count. Bottom row:
absolute variational energy versus parameter count.
Columns: Fe$_{2}$S$_{2}$, Fe$_{4}$S$_{4}$, P-cluster. Red open
circles: TrimCI~+~COO; blue filled: DMRG in
entanglement-minimized orbitals (EMO) of Ref.~\cite{Li2025};
purple half-filled: DMRG in LMO. Parameter counts are defined
in SM Sec.~S4. Dashed connectors at $D\!=\!100$ mark
fixed-budget (top) or fixed-energy (bottom) comparisons.}
\label{fig:compactness}
\end{figure}

\textit{Compactness beyond entanglement minimization.}---Quantum
algorithms demand a trial state with high ground-state overlap at
low parameter count --- large ansätze are too costly to load. We
therefore compare at the smallest bond dimension reported in the
recent SU(2)-DMRG literature~\cite{Li2025}, $D\!=\!100$, where the
entanglement-minimized orbitals (EMO) of that work are themselves
optimized. We use the dominant-determinant weight
$p_{0} \equiv |c_{\max}|^{2}$ as the compactness diagnostic. At matched parameter budget
(Fig.~\ref{fig:compactness} and Table~\ref{tab:fig4_ratios}), COO concentrates weight on a single determinant up to five orders of
magnitude more than LMO and up to $35\times$ more than EMO across
the iron-sulfur series, with the largest separation on the P-cluster. COO therefore yields a
more compact wavefunction than EMO across the series, which may
indicate that EMO optimizes entanglement under the MPS structure
rather than absolute wavefunction compactness.

\begin{table}[t]
\caption{\textbf{TrimCI~+~COO vs DMRG at $D\!=\!100$.}
Top-1 $p_{0}$ ratio: COO/DMRG at
matched $N_{\rm det}\!=\!N^{\rm DMRG}$. $N_{\rm params}$ ratio:
$N^{\rm DMRG}/N^{\rm COO}$ at matched energy. DMRG data (EMO:
entanglement-minimized; LMO: localized molecular) from
Ref.~\cite{Li2025}. $^{\dagger}$Smallest COO point already below
LMO $D\!=\!100$; $N^{\rm COO}$ from power-law extrapolation
[Fig.~\ref{fig:compactness}(f)].}
\label{tab:fig4_ratios}
\begin{ruledtabular}
\begin{tabular}{l cc cc}
 & \multicolumn{2}{c}{top-1 $p_{0}$ ratio} &
   \multicolumn{2}{c}{$N_{\rm params}$ ratio} \\
\cline{2-3}\cline{4-5}
System & vs.\ EMO & vs.\ LMO & vs.\ EMO & vs.\ LMO \\
\hline
Fe$_{2}$S$_{2}$  & $1.0\times$ & $12\times$             & $16\times$ & $49\times$           \\
Fe$_{4}$S$_{4}$  & $3.9\times$ & $930\times$            & $71\times$ & $4700\times$         \\
P-cluster        & $35\times$  & $5\!\times\!10^{4}$    & $78\times$ & $1\!\times\!10^{5\,\dagger}$ \\
\end{tabular}
\end{ruledtabular}
\end{table}

At matched energy, TrimCI~+~COO uses $16$--$78\times$ fewer parameters
than DMRG+EMO at $D\!=\!100$, and up to five orders of
magnitude fewer than DMRG+LMO. Parameter count is the relevant metric because it controls classical storage, compute, and quantum state preparation costs. Whether measured by
compactness at fixed parameter count or by parameter count at
fixed accuracy, COO yields a better orbital basis than EMO.


\begin{figure*}[t]
\includegraphics[width=\textwidth]{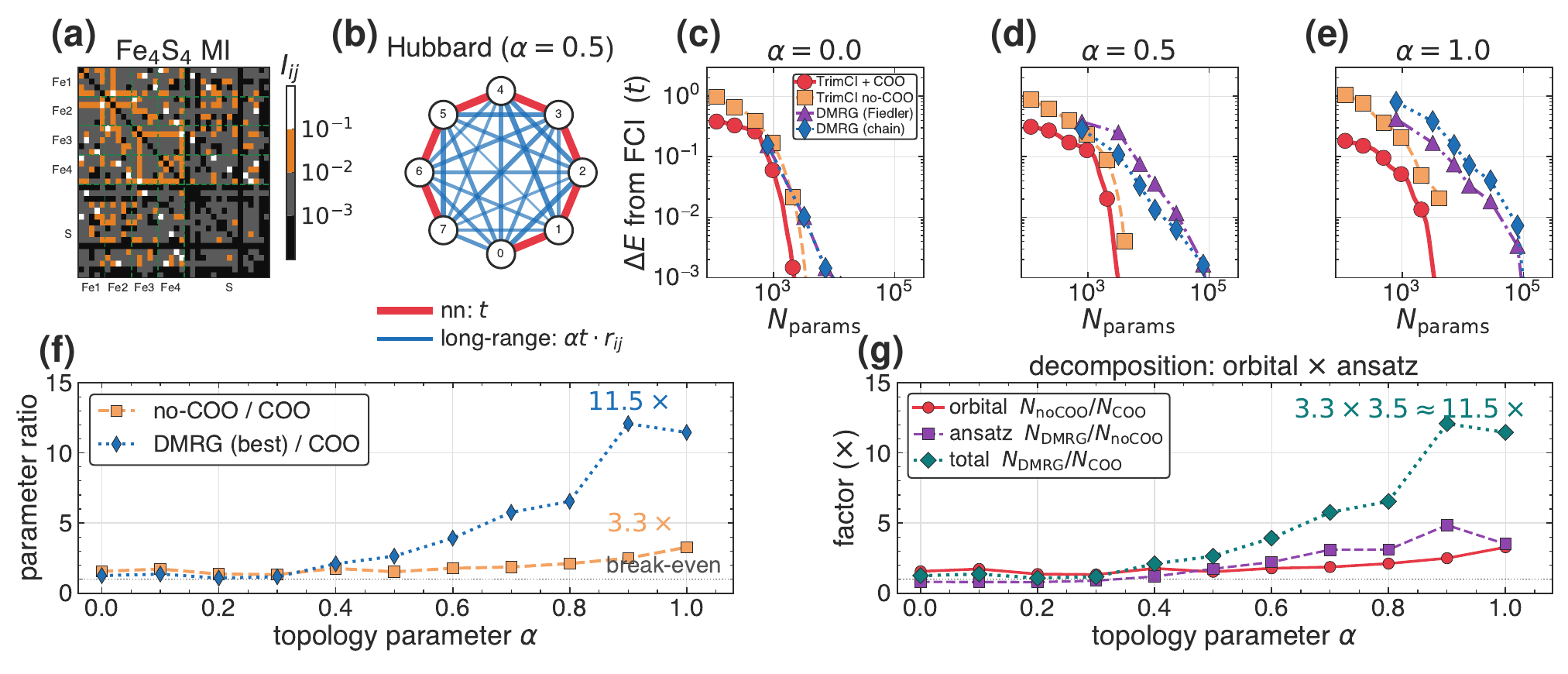}
\caption{\textbf{TrimCI~+~COO vs DMRG on a tunable Hubbard-on-graph
model.}
\textbf{(a)}~Orbital mutual information $I_{ij}$ of
\textrm{[Fe$_4$S$_4$]} in COO orbitals, re-ordered by Fe
localization (Fe1, Fe2, Fe3, Fe4, S; dashed green lines mark group
boundaries).
\textbf{(b)}~Hubbard-on-graph schematic at $\alpha\!=\!0.5$
($L\!=\!8$, $U/t\!=\!4$): nearest-neighbor edges $-t$ (red, open chain), non-nn edges
$-\alpha t\,r_{ij}$ with $r_{ij}\!\sim\!\mathcal{U}[0.5,1.5]$
(blue, width $\propto |t_{ij}|$).
\textbf{(c)}--\textbf{(e)}~$\Delta E$ from FCI versus
$N_{\rm params}$ for $\alpha\!=\!0,\,0.5,\,1.0$: TrimCI~+~COO
(red), TrimCI no-COO (orange), DMRG with chain (blue) and Fiedler
(purple) orderings.
\textbf{(f)}~Parameter ratio to reach $\Delta E\!<\!0.1\,t$:
DMRG (best ordering)/COO grows from $\sim\!1\times$ (chain) to
$\sim\!12\times$ (complete graph).
\textbf{(g)}~Decomposition of (f) into orbital and ansatz
factors.}
\label{fig:mi}
\end{figure*}

\textit{Why determinants outperform MPS here.}---The remaining question is why COO changes the comparison with MPS. The answer is that the
orbital rotation and the determinant ansatz compress different parts of the correlation structure. On \textrm{[Fe$_4$S$_4$]} in the converged COO basis, the orbital
mutual information $I_{ij}$ [Fig.~\ref{fig:mi}(a)] places all 18
strongly entangled orbital pairs ($I_{ij} > 0.1$) across distinct
atomic centers
(none intra-Fe; 8 direct Fe--Fe and 10 traversing bridging S;
see SM): COO absorbs the dynamical correlation into the orbitals,
exposing the multi-center structure that remains in the CI
expansion. One-dimensional ordering cannot
localize this graph: the Fiedler-reordered MI still spreads $95\%$
of its off-diagonal mass $\sum_{i\ne j} I_{ij}$ across a band of
half-width $k\!=\!15$, $5\times$ wider than for an H$_{36}$ chain
(see SM). An MPS must therefore carry this entanglement through a growing bond dimension.

TrimCI~+~COO captures this multi-center entanglement efficiently
through the combination of a sparse CI ansatz and the
core-optimized orbital basis. To separate how much each contributes to the
parameter-efficiency, we built a minimal multi-center model, a tunable
Hubbard-on-graph
[Fig.~\ref{fig:mi}(b)] with $L\!=\!8$ sites at half filling,
\begin{equation}
\hat H = -t \!\sum_{\langle ij\rangle_{\rm nn}} \hat c^\dagger_i \hat c_j
        -\alpha t \!\sum_{ij\notin{\rm nn}} r_{ij}\,\hat c^\dagger_i \hat c_j
        + U \sum_i \hat n_{i\uparrow}\hat n_{i\downarrow},
\label{eq:Hgraph}
\end{equation}
where $\langle ij\rangle_{\rm nn}$ runs over the $L\!-\!1$
nearest-neighbor pairs of an open $L$-site chain, the long-range
bond weights $r_{ij}$ are drawn uniformly from $[0.5, 1.5]$, and
the knob $\alpha\!\in\![0,1]$ interpolates from the bare 1D chain
to a complete graph at $U/t\!=\!4$ (see SM). The parameter budget
to reach an energy error $\Delta E < 0.1\,t$
[Fig.~\ref{fig:mi}(c)--(f)] yields a ratio $N_{\rm DMRG}/N_{\rm COO}$
that grows from unity on the chain to $\sim\!12\times$ on the
complete graph; relative to the site basis it factorizes
[Fig.~\ref{fig:mi}(g)] into an \emph{orbital} factor
($N_{\rm noCOO}/N_{\rm COO}\!\approx\!1.3$--$3.3$, the gain from
a better single-particle basis at fixed ansatz) and an
\emph{ansatz} factor
($N_{\rm DMRG}/N_{\rm noCOO}\!\approx\!1$--$3.5$, the gain from
a sparse CI over an MPS at fixed basis). The two factors
multiply on \textrm{[Fe$_4$S$_4$]} --- which lies in the
high-$\alpha$ regime --- to give the observed $10$--$100\times$
parameter advantage over DMRG variants at matched accuracy
(Fig.~\ref{fig:dmrg}, Table~\ref{tab:fig4_ratios}). The orbital
factor absorbs dynamical correlation into the basis; the ansatz
factor captures the multi-center entanglement that resists 1D
ordering. Together they make sparse CI in the COO basis more
compact than MPS.


\textit{Outlook.}---COO compresses sparse CI wavefunctions by
orders of magnitude by absorbing dynamical correlation into the
orbital basis, allowing sparse CI to outperform state-of-the-art
DMRG variants in the multi-center strongly correlated regime. This
suggests a reassessment of CI parameter efficiency relative to other
many-body ansätze. Natural next steps are to apply COO to larger
iron-sulfur targets, such as the FeMo cofactor of
nitrogenase~\cite{Reiher2017,Lee2026}; to assess COO in combination
with other many-body ansätze, including tensor networks, coupled
cluster, and neural quantum states; and to determine where the same
compression mechanism persists beyond multi-center systems.

\section*{Data and Code Availability}
The data that support the findings of this study are presented
in the Supplemental Material~\cite{SM}. The orbital optimization driver
and the complete TrimCI workflow are available at
\url{https://github.com/hao-zhang-quantum/TrimCI}, packaged as a
Python interface with an efficient C++ backend.

\section*{Acknowledgments}
This work was supported by the NSF QLCI for Hybrid Quantum
Architectures and Networks (NSF award 2016136). We gratefully
acknowledge computing resources provided by the Center for High
Throughput Computing (CHTC) at UW--Madison.

\putbib[coo-ref]
\end{bibunit}

\beginsupplement
\begin{bibunit}[apsrev4-2]

\makeatletter
\renewcommand*\l@section{\@dottedtocline{1}{0em}{2.2em}}
\renewcommand*\l@subsection{\@dottedtocline{2}{2.2em}{3.2em}}
\renewcommand*\l@subsubsection{\@dottedtocline{3}{5.4em}{4.0em}}
\makeatother

\begin{center}
{\large\bfseries Supplemental Material: \\[2pt]
Absorbing Many-Body Correlations into Core-Optimized Orbitals}
\end{center}
\vspace{1ex}

\linespread{1.667}\selectfont

\begingroup
\setlength{\parskip}{0pt}
\hypersetup{linkcolor=black}
\begin{center}{\large\bfseries Contents}\end{center}
\vspace{0.5em}
\begin{enumerate}[label=S\arabic*.,leftmargin=2.2em,itemsep=2pt]
  \item \textit{Complete TrimCI~+~COO workflow}\dotfill \pageref{sec:workflow}
  \item \textit{Orbital optimization algorithm}\dotfill \pageref{sec:algorithm}
  \item \textit{Energy basins and convergence robustness}\dotfill \pageref{sec:robustness}
  \item \textit{Parameter counting}\dotfill \pageref{sec:parameter_counting}
  \item \textit{Data tables for the main-text figures}\dotfill \pageref{sec:data_tables}
  \item \textit{Orbital mutual information analysis}\dotfill \pageref{sec:strict_mi}
  \item \textit{Multi-center excitation analysis}\dotfill \pageref{sec:multicenter}
  \item \textit{Hubbard-on-graph details}\dotfill \pageref{sec:hubbard_graph}
  \item \textit{Distributed Davidson via mini-task bundles: two-axis ($K\!\times\!Z$) scalability}\dotfill \pageref{sec:gpu_davidson_5120m}
  \item \textit{Finding the P-cluster ground-state spin pattern from random initial determinants}\dotfill \pageref{sec:pcluster_spin}
\end{enumerate}
\endgroup
\clearpage

\makeatletter
\renewcommand{\section}{\@startsection{section}{1}{\z@}%
    {-2.0ex \@plus -0.5ex \@minus -0.2ex}%
    {1.0ex \@plus 0.2ex}%
    {\normalfont\large\bfseries}}
\makeatother

\section{Complete TrimCI~+~COO workflow}
\label{sec:workflow}

\begin{figure}[!ht]
\centering
\includegraphics[width=0.85\textwidth]{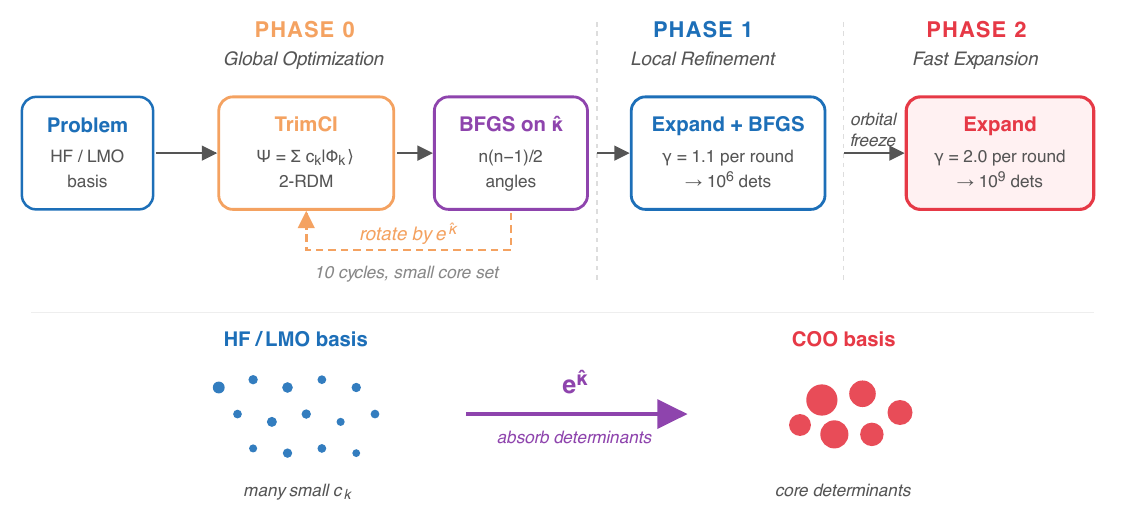}
\caption{Full TrimCI~+~COO workflow.
\textit{Phase~0 (Global Optimization)} alternates TrimCI core
search with BFGS optimization of the $n(n{-}1)/2$ orbital angles
$\kappa_{ai}$ on a small core set (typically $100$ dets), until
both the core and the orbitals converge;
\textit{Phase~1 (Local Refinement)} grows the determinant space
($\gamma\!=\!1.1$ per round) while continuing the per-round BFGS
update; \textit{Phase~2 (Fast Expansion)} freezes the orbitals and
expands aggressively ($\gamma\!=\!2.0$ per round) to the final
target. The bottom panel cartoons the basis transformation: a
single orbital rotation $e^{\hat\kappa}$ absorbs the wavefunction's
many small-coefficient determinants in the HF/LMO basis into a
compact set of core determinants in the COO basis.}
\label{fig:workflow}
\end{figure}

The end-to-end calculation is organized into three phases
(Fig.~\ref{fig:workflow}): \textit{Phase~0 (Global Optimization)}
jointly optimizes a small TrimCI core and its orbitals starting from
random initial determinants; \textit{Phase~1 (Local Refinement)}
grows the determinant space while continuing to refine the orbitals
each round; \textit{Phase~2 (Fast Expansion)} freezes the orbitals
and pushes the expansion to its final target.

\paragraph{Phase~0 (Global Optimization).}
The goal of Phase~0 is to discover a small core set of
$N_{\rm det}^{0}$ determinants together with the orbitals adapted to
it (default $N_{\rm det}^{0}=100$, set by \texttt{max\_final\_dets}).
It needs no prior knowledge of the ground-state wavefunction:
starting from any reasonable orbital guess (e.g.\ HF or LMO) and a
pool of random initial determinants, the core is built by TrimCI
while BFGS rotates the orbitals. Each \emph{cycle} runs three steps
in sequence:

\begin{enumerate}
\item \textbf{Core determinants search.} A single TrimCI run, given
a set of random initial determinants, iteratively expands the
current core $C$ to a pool $P$ via large Hamiltonian couplings
($|H_{ij}c_j| > \theta$), then trims $P$ in two passes --- local
diagonalization of randomized sub-blocks, followed by a global
diagonalization on the survivors --- to produce the next core,
until $N_{\rm det}^{0}$ is reached~\cite{TrimCI_paper}. To
thoroughly explore the global energy landscape, $N_{\rm runs}$
such runs are launched in parallel with different random seeds; we
keep the lowest-energy run as the cycle output core set, or in
broken-symmetry studies retain the basin of interest
(Sec.~\ref{sec:robustness}).
\item \textbf{BFGS orbital rotation.} The two-body reduced density
matrix from the output core set feeds BFGS, which optimizes all
$n(n{-}1)/2$ rotation angles $\kappa_{ai}$. The core determinants
themselves are kept fixed throughout BFGS; at every line-search
trial step we re-solve the projected CI for fresh coefficients
$\{c_I\}$, so the line-search energy is the exact variational
energy at the trial orbitals --- a much sharper signal than the
fixed-coefficient approximation that BFGS would otherwise use
(details in Sec.~\ref{sec:algorithm}).
\item \textbf{Basis rotation.} The integrals are rotated by
$e^{\hat\kappa}$, and the next cycle re-runs all $N_{\rm runs}$
TrimCI runs in the new basis.
\end{enumerate}

After ten cycles the orbitals typically have converged --- 3--5
cycles already capture most of the gain, with robustness across
basins and seeds documented in Sec.~\ref{sec:robustness}.

\begin{table}[h]
\caption{Phase~0 hyperparameters. $N_{\rm runs}\!\times\!$cycles is
the dominant cost; \texttt{max\_final\_dets} controls how expressive
the core that drives orbital optimization is.}
\label{tab:phase0_config}
\begin{tabular}{lll}
\toprule
parameter & default & \parbox[t]{4in}{\raggedright meaning} \\
\midrule
\texttt{num\_runs}                & 64                & \parbox[t]{4in}{\raggedright independent random starts per cycle} \\
\texttt{cycles}                   & 10                & \parbox[t]{4in}{\raggedright orbital-optimization-then-CI iterations} \\
\texttt{max\_final\_dets}         & 100               & \parbox[t]{4in}{\raggedright target core size $N_{\rm det}^{0}$} \\
\texttt{initial\_dets\_dict}      & \texttt{HF=1, rand=[1,10000]} & \parbox[t]{4in}{\raggedright initial determinants per run: 1 HF reference + 10000 random dets, each with weight 1} \\
\texttt{first\_cycle\_keep\_size} & 10                & \parbox[t]{4in}{\raggedright dets retained after the first expand-trim inside a TrimCI run, before further expansion} \\
\texttt{threshold} $\theta$       & $10^{-2}$         & \parbox[t]{4in}{\raggedright initial heat-bath screening $|H_{ij}c_j| > \theta$} \\
\texttt{pool\_core\_ratio}        & 40                & \parbox[t]{4in}{\raggedright candidate pool / core ratio} \\
\texttt{core\_set\_ratio}         & $[1.0, 1.1]$      & \parbox[t]{4in}{\raggedright inter-round core-size growth} \\
\texttt{num\_groups}              & 20                & \parbox[t]{4in}{\raggedright number of TRIM sub-divisions} \\
\texttt{local\_trim\_keep\_ratio} & 4                 & \parbox[t]{4in}{\raggedright dets retained by local trim, relative to core size} \\
\texttt{max\_rounds}              & 4                 & \parbox[t]{4in}{\raggedright number of excitation hops during pool build} \\
\texttt{pool\_build\_strategy}    & \texttt{heat\_bath} & \parbox[t]{4in}{\raggedright candidate generator} \\
\bottomrule\end{tabular}
\end{table}

\begin{table}[h]
\caption{Orbital-optimizer settings used inside each Phase~0 cycle.
The same BFGS algorithm is invoked per round in Phase~1 with a
tighter iteration cap (Table~\ref{tab:phase1_config}); see
Sec.~\ref{sec:algorithm} for the line-search re-diagonalization
protocol.}
\label{tab:bfgs_config}
\begin{tabular}{lll}
\toprule
parameter              & default          & \parbox[t]{4in}{\raggedright meaning} \\
\midrule
\texttt{optimizer}     & \texttt{bfgs}    & \parbox[t]{4in}{\raggedright BFGS optimizer for the rotation angles $\kappa_{ai}$} \\
\texttt{maxiter}       & 100              & \parbox[t]{4in}{\raggedright maximum BFGS iterations per cycle} \\
\texttt{ftol}          & $10^{-8}$        & \parbox[t]{4in}{\raggedright BFGS convergence tolerance on $|\Delta E|$} \\
\texttt{davidson\_tol} & $10^{-7}$        & \parbox[t]{4in}{\raggedright Davidson tolerance inside line-search re-diagonalization} \\
\texttt{tracking\_dets}& \texttt{False}   & \parbox[t]{4in}{\raggedright retain the previous cycle's dominant determinants when re-running CI in the rotated basis} \\
\texttt{loaded\_dets\_randomness} & $0.0$ & \parbox[t]{4in}{\raggedright when re-running CI from a saved wavefunction (\texttt{initial\_dets\_path}, or via \texttt{tracking\_dets}), this fraction of the loaded determinants is replaced by random ones each cycle; setting to $0.1$ lets the search explore a wider neighbourhood of the seed basin} \\
gradient               & analytic         & \parbox[t]{4in}{\raggedright closed form from 2-RDM, $O(n_{\rm orb}^{4})$ per BFGS step} \\
\bottomrule\end{tabular}
\end{table}

\paragraph{Phase~1 (Local Refinement).}
Starting from the converged Phase~0 result, the determinant space
grows slowly while the orbitals continue to refine. Each round
expands $N_{\rm det}$ by \texttt{growth\_factor}, runs Davidson with
warm-started CI vectors, and invokes BFGS on the rotation angles;
the integrals are rotated by $e^{\hat\kappa}$ before the next
expansion step. The Davidson matvec is accelerated by a
\emph{connection cache} that precomputes the nonzero $H_{ij}$
couplings between determinants in the current core, trading
$O(N_{\rm det}\!\times\!\overline{\rm nnz})$ memory (where
$\overline{\rm nnz}$ is the average number of connected determinants
per row) for fast matvecs. Davidson uses a loose energy tolerance
because round-to-round orbital-optimization steps already sit above
this scale. Phase~1 ends at \texttt{max\_n\_dets}, typically
$N_{\rm det}\!\sim\!10^{6}$ (e.g.\ for \textrm{[Fe$_4$S$_4$]}
(54e,\,36o)). At this scale the orbitals are sufficiently adapted
that Phase~2 can freeze them without loss of variational quality.

\begin{table}[h]
\caption{Phase~1 expansion + orbital-optimization hyperparameters.}
\label{tab:phase1_config}
\begin{tabular}{lll}
\toprule
parameter                          & default        & \parbox[t]{4in}{\raggedright meaning} \\
\midrule
\texttt{max\_n\_dets}              & $10^{6}$        & \parbox[t]{4in}{\raggedright Phase~1 endpoint} \\
\texttt{growth\_factor}            & 1.1            & \parbox[t]{4in}{\raggedright $N_{\rm det}$ growth ratio per expansion round} \\
\texttt{orbital\_optimization}     & \texttt{True}  & \parbox[t]{4in}{\raggedright BFGS each round} \\
\texttt{orbital\_opt\_max\_iter}   & 50             & \parbox[t]{4in}{\raggedright per-round BFGS iteration cap} \\
\texttt{use\_connection\_cache}    & \texttt{True}  & \parbox[t]{4in}{\raggedright cache nonzero $H_{ij}$ couplings between dets} \\
\texttt{davidson.energy\_tol}      & $10^{-4}$      & \parbox[t]{4in}{\raggedright Davidson stop on $|\Delta E_{\rm iter}|$} \\
\bottomrule\end{tabular}
\end{table}

\paragraph{Phase~2 (Fast Expansion).}
Orbitals are frozen at the Phase~1 result and the determinant space
grows fast --- typically doubling per round
(\texttt{growth\_factor}=2) --- toward the final target. An optional
semistochastic PT2 correction may be applied per round for two
purposes: to lower the total variational$+$PT2 energy estimate, and
to gauge the residual distance to the variational limit. Davidson
tightens its energy tolerance relative to Phase~1, since Phase~2 is
the final stage. The connection cache is disabled at this scale
because its memory cost becomes prohibitive; GPU-accelerated
Davidson handles the large-scale matvecs instead. On
\textrm{[Fe$_4$S$_4$]}, Phase~2 grows
$10^{6}\!\rightarrow\!5\!\times\!10^{9}$ in thirteen doubling rounds.

\begin{table}[h]
\caption{Phase~2 frozen-orbital expansion hyperparameters.}
\label{tab:phase2_config}
\begin{tabular}{lll}
\toprule
parameter                     & default            & \parbox[t]{4in}{\raggedright meaning} \\
\midrule
\texttt{max\_n\_dets}         & $10^{8}$           & \parbox[t]{4in}{\raggedright target det count (problem-dependent; up to $5\!\times\!10^{9}$ for \textrm{[Fe$_4$S$_4$]})} \\
\texttt{growth\_factor}       & 2.0                & \parbox[t]{4in}{\raggedright doubling per round} \\
\texttt{pt2\_correction}      & \texttt{True}      & \parbox[t]{4in}{\raggedright semistochastic PT2, per round} \\
\texttt{use\_connection\_cache} & \texttt{False}   & \parbox[t]{4in}{\raggedright disabled to reduce memory at large $N_{\rm det}$} \\
\texttt{davidson.energy\_tol} & $10^{-5}$          & \parbox[t]{4in}{\raggedright Davidson stop tolerance (final stage)} \\
\bottomrule\end{tabular}
\end{table}

Phase~0 and Phase~1 run comfortably on a single CPU node
(64--128 cores). A complete calculation to $N_{\rm det}\!\sim\!10^{8}$
typically takes $\sim\!20$--$40$~h wall-time; only Phase~2 at
$N_{\rm det} \gtrsim 10^{8}$ requires multi-node or GPU resources.

\section{Orbital optimization algorithm}
\label{sec:algorithm}

Three implementation choices distinguish the COO orbital
optimization. (i)~The compact wavefunction comes from
TrimCI~\cite{TrimCI_paper}: a $\sim\!100$-determinant core that
already captures the dominant ground-state weight, making both the
2-RDM and the projected-Hamiltonian Davidson cheap. (ii)~BFGS
optimizes all $n_{\rm orb}(n_{\rm orb}{-}1)/2$ rotation angles
$\kappa_{ai}$ simultaneously. (iii)~The projected Hamiltonian is
re-diagonalized inside every BFGS line-search trial step, so the
line-search energy is the exact variational energy at the trial
orbitals rather than a fixed-CI approximation. This section
documents the algorithm.

\paragraph{Setup.} The optimization variables are the
$n_{\rm orb}(n_{\rm orb}{-}1)/2$ independent entries $\kappa_{ai}$
of an antisymmetric matrix $\kappa$. This matrix enters the
algorithm in two equivalent forms. The matrix exponential
$U = e^{\kappa}$ is a unitary that rotates the orbital coefficients,
and is what we apply to the integrals via $h \to U^{T} h\, U$ and
the analogous four-index transform of $V$. The corresponding
one-body operator
$\hat\kappa = \sum_{pq} \kappa_{pq}\, a^{\dagger}_p a_q$
generates the unitary on Fock space and rotates the Hamiltonian
itself, $\hat H \to e^{-\hat\kappa}\hat H e^{\hat\kappa}$; this form
is what makes the variational energy an analytic function of
$\kappa$ and yields the closed-form gradient below. In practice we implement the matrix form: rotate the integrals once
per trial step, then re-diagonalize. We work on a fixed small core $\mathcal{C}$
(typically $|\mathcal{C}|\sim 100$ determinants) using the integrals
$(h_{pq}, V_{pqrs})$ in the current orbital basis. They define the
second-quantized Hamiltonian
\begin{equation}
\hat H = \sum_{pq} h_{pq}\, a^{\dagger}_p a_q
       + \tfrac{1}{2}\!\sum_{pqrs} V_{pqrs}\, a^{\dagger}_p a^{\dagger}_q a_s a_r .
\end{equation}
At BFGS iteration $t$ the projected Hamiltonian
$H_{\mathcal{C}} = P_{\mathcal{C}} \hat H P_{\mathcal{C}}$ is
diagonalized by Davidson to give CI coefficients $\{c_I\}$ and
variational energy $E_t = \langle\Psi|\hat H|\Psi\rangle$. From this
wavefunction we form the one- and two-body reduced density matrices
on $\mathcal{C}$, and from those the closed-form orbital gradient
$g_t = \partial E / \partial \kappa$. The BFGS Hessian approximation
$B_t$ is initialized from the diagonal of the orbital Hessian and
updated by the standard formula
\begin{equation}
B_{t+1} = B_t + \frac{y\,y^{T}}{y^{T}\!s} -
                  \frac{B_t\, s\, s^{T} B_t}{s^{T} B_t\, s},
\end{equation}
with $s = \kappa_{t+1} - \kappa_t$, $y = g_{t+1} - g_t$. The
curvature condition $y^{T}\!s > \varepsilon$
($\varepsilon = 10^{-6}$) is required to preserve positive
definiteness; updates that fail it are skipped. A small ridge
$10^{-3} I$ is added to $B_t$ before inversion to guarantee a
descent direction even when the unshifted $B_t$ has near-zero
eigenvalues.

\paragraph{Inner loop: BFGS step with line-search re-diagonalization.}
Each iteration proposes a search direction
$d_t = -(B_t + 10^{-3} I)^{-1} g_t$ and then runs the following
line search:

\begin{enumerate}
\item Set $\alpha \leftarrow 1$ and store $E_{\rm prev} = E_t$.
\item Line-search trial loop:
\begin{enumerate}
\item Compute trial parameters $\kappa_{\rm try} = \kappa_t + \alpha\, d_t$.
\item Build the rotation matrix $U_{\rm try} = e^{\kappa_{\rm try}}$ and rotate the integrals: $h_{\rm try}, V_{\rm try} = U_{\rm try}^{T}(h, V)\, U_{\rm try}$.
\item Re-diagonalize $H_{\mathcal C}$ via Davidson with $(h_{\rm try}, V_{\rm try})$, warm-started from $\{c_I\}$, giving $\{c_I^{\rm try}\}$ and $E_{\rm try}$.
\item If $E_{\rm try} \le E_{\rm prev} + \delta$, accept: set $\kappa_{t+1} \leftarrow \kappa_{\rm try}$, $E_{t+1} \leftarrow E_{\rm try}$, and break.
\item Else shrink $\alpha \leftarrow \alpha / 2$ and continue (up to a max of $\sim\!10$ trials).
\end{enumerate}
\item Update the BFGS Hessian $B_{t+1}$ from $(s, y)$.
\end{enumerate}

The acceptance tolerance $\delta = \max(\delta_{\rm tol},
10^{-12} |E_{\rm prev}|)$ is a floating-point slack: round-off in
the energy comparison can flip the sign of a true descent step of
order $\sim\!10^{-12}|E_{\rm prev}|$, and $\delta$ absorbs that
noise so we do not reject genuine progress at machine precision.
The key property of this loop is that every accepted step
$\kappa_t \to \kappa_{t+1}$ corresponds to a strict decrease of the
\emph{exact variational energy} on the core, with $\{c_I\}$
self-consistently re-optimized at the new orbitals --- not the
fixed-$\{c_I\}$ approximation that BFGS would otherwise use.

\paragraph{Why the small core matters.} A line-search trial that
re-diagonalizes CI is only practical when the diagonalization is
cheap. With $|\mathcal{C}| \sim 10^{2}$ determinants the
projected-Hamiltonian Davidson takes milliseconds (warm-started
from the previous trial's eigenvector). Each BFGS step then costs
several Davidsons across its line-search trials plus one 2-RDM
contraction. The whole
inner loop runs in seconds; an outer cycle in minutes. By contrast,
running the same line search on a $10^{6}$-determinant core (the
standard regime for selected-CI orbital
optimization~\cite{YaoUmrigar2021}) would cost
$\sim\!10^{4}\!\times$ more per Davidson, making the line-search
re-diagonalization infeasible. Cheap 2-RDM evaluation and feasible
per-trial line-search re-diagonalization are therefore two virtues
of the same compact-core design.

\paragraph{Optimizer choice.} We benchmarked the full
\texttt{scipy.optimize.minimize} family (BFGS, L-BFGS-B, CG,
Newton-CG, trust-ncg, Nelder-Mead, Powell, COBYLA, and others) on
the orbital optimization problem and adopted BFGS for production.
The dimensionality ($n(n{-}1)/2 = 630$ for $n\!=\!36$) is small
enough that BFGS's dense Hessian fits comfortably in memory and
the limited-memory variant L-BFGS offers no advantage; gradient-free
methods fail to converge in this many dimensions; Newton-CG and
trust-region methods would require Hessian--vector products that
are no cheaper than the BFGS approximation maintains.

\paragraph{Comparison with prior selected-CI orbital optimization.}
The closest published scheme is Yao--Umrigar's BFGS for SHCI orbital
optimization~\cite{YaoUmrigar2021}. The schemes differ on three
methodological points, which compound:

\begin{enumerate}
\item \textbf{Working wavefunction --- size and quality.} Y\&U
  optimize on a $\sim\!10^{6}$-determinant SHCI expansion built on
  top of HF. On strongly-correlated systems where HF is far from
  the ground state, the expansion misses key parts of the
  ground-state structure, so the 2-RDM gradient signal is noisy. We
  instead optimize on a $\sim\!10^{2}$-determinant TrimCI core that
  already concentrates on the dominant ground-state
  weight~\cite{TrimCI_paper}: despite being four orders of magnitude
  smaller, it provides a sharper gradient signal, and its 2-RDM is
  correspondingly cheaper.

\item \textbf{Line search with adaptive $\{c_I\}$.} The Y\&U scheme
  runs no line search: each BFGS step is committed under a
  norm-clamp on $\|\Delta\kappa\|$, and CI is re-diagonalized only
  at the start of the next outer iteration. We instead run a
  backtracking line search inside each BFGS iteration,
  re-diagonalizing the projected CI at every trial step. Letting
  $\{c_I\}$ relax at each trial means the trial energy reflects the
  full coupled $(\{c_I\}, \kappa)$ response to the orbital
  perturbation, so each BFGS update sees the right curvature and
  the Hessian approximation (and the search direction it generates)
  is sharper; steps that fail to decrease the variational energy are
  rejected. This is practical only because the small core makes each
  trial Davidson take milliseconds.

\item \textbf{Global search in the new orbital basis.} At every
  outer orbital-optimization iteration, SHCI rebuilds its
  variational space starting from the HF reference and selecting
  deterministically through the $|H_{ij} c_j| > \theta$ couplings.
  A better orbital basis sharpens the resulting expansion, but the
  search remains anchored to the HF basin: the new dets are always
  those reachable from HF by repeated single/double excitation,
  never a globally different ground-state structure. We instead
  start TrimCI from independent random initial determinants at each
  Phase~0 cycle. When the orbitals improve, the random restart
  gives the core search a fresh shot at a globally better basin ---
  often a large-amplitude improvement on strongly correlated systems
  where HF is far from the ground state.
\end{enumerate}

These three differences compound. On standard benchmarks where HF
is already a good starting point, prior selected-CI orbital
optimization delivers a useful but moderate improvement; on
strongly-correlated systems where HF is far from the ground state,
the same schemes stall, and can sit hundreds of mHa above the
ground state --- for example, on \textrm{[Fe$_4$S$_4$]}. The
compact wavefunction, the adaptive line-search signal, and the
global core search of TrimCI~+~COO compound to a qualitative shift:
systems such as \textrm{[Fe$_4$S$_4$]}, previously out of reach for
selected-CI orbital optimization, are now solved on a single CPU
node (with GPU acceleration recovering the last $\lesssim\!4$\,mHa
of variational accuracy).

\paragraph{Relation to CASSCF and DMRG-SCF orbital optimization.}
The same picture extends to traditional CASSCF and to DMRG-SCF. In
the second-order MCSCF/CASSCF framework of
Werner--Knowles~\cite{Roos1980,Werner1985,Legeza2025OrbOpt} and its
modern variants, a fresh CI on the original Hamiltonian is run only
at macro-iteration boundaries, where the integral re-transformation
has just been done. DMRG-SCF~\cite{Zgid2008} inherits the same
framework with the DMRG sweep replacing CI; a per-microstep DMRG
re-run is prohibitive in cost. The implicit assumption shared by
all these schemes is that re-solving the CI/DMRG problem on the
original Hamiltonian inside an inner step is too expensive to
consider, so the inner step makes do with stale, linearly-corrected,
or quasi-Newton-eliminated CI coefficients --- never a fresh
diagonalization. Our line-search-with-rediagonalization becomes
possible only because the $\sim\!100$-determinant TrimCI core has
two simultaneous properties: it already captures the dominant
ground-state correlation, so the variational energy on the core is
a meaningful signal at every line-search trial; and it is small
enough that the projected-Hamiltonian Davidson takes milliseconds,
making per-trial re-diagonalization affordable.

\section{Energy basins and convergence robustness}
\label{sec:robustness}

\begin{figure}[t]
\centering
\includegraphics[width=0.92\textwidth]{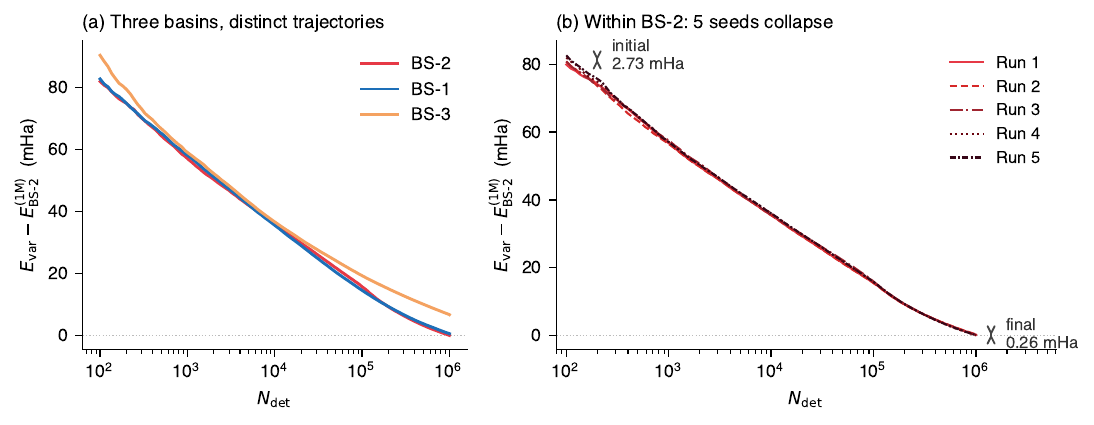}
\caption{Robustness of the Phase~1 endpoint on
\textrm{[Fe$_4$S$_4$]} (54e,\,36o). All seven trajectories use
the standard Phase~1 hyperparameters of Sec.~\ref{sec:workflow}
($\gamma\!=\!1.1$, orbital optimization ON, Davidson
$\mathrm{tol}\!=\!10^{-4}$); energies are referenced to the BS-2
endpoint at $N_{\rm det}=10^{6}$. (a) One representative
trajectory per basin, full N range. BS-1 closes from
$+2.85$\,mHa at $100$ dets to $+0.56$\,mHa at $10^{6}$
(near-degenerate with BS-2); BS-3 stays $\sim\!6.7$\,mHa above
throughout. (b) Within the BS-2 basin, five trajectories from
different random-seed Phase~0 starts (Run~1--5, ordered by
$100$-det energy) compress an initial $2.73$\,mHa spread to a
final $0.26$\,mHa spread at $N_{\rm det}=10^{6}$.}
\label{fig:basin_convergence}
\end{figure}

\paragraph{TrimCI sampling discovers low-energy broken-symmetry
basins.}
The energy landscape on a strongly-correlated multi-reference
system in general contains multiple energy basins, one for each
distinct broken-symmetry configuration of the open-shell electrons.
Running TrimCI Phase~0 many times from independent random initial
determinants samples this landscape: each run converges to one of
the basins, and the collection of converged outputs maps the basin
structure. On \textrm{[Fe$_4$S$_4$]} (54e,\,36o), this sampling
identifies exactly three energy basins
(Fig.~\ref{fig:basin_convergence}a), corresponding to the three
antiferromagnetic spin orderings of the four iron centers: each Fe
carries five unpaired $d$-electrons polarized either ``up'' (U) or
``down'' (D), and the four-Fe cluster admits three distinct AF
patterns --- UUDD\,/\,DDUU, UDUD\,/\,DUDU, and UDDU\,/\,DUUD ---
labelled BS-1, BS-2, and BS-3 following the standard 4Fe
broken-symmetry nomenclature (Table~\ref{tab:basin_landscape}).
Two independent sampling experiments --- Experiment~A
($9$ independent Phase~0 jobs) and Experiment~B ($40$ independent
Phase~0 jobs, on a different computing cluster) --- both identify the same
three basins. Within each experiment we keep the lowest-energy
Phase~0 job per basin (i.e.\ the best-of-many output); these
basin-best energies agree between the two experiments to within
$\sim\!0.02$\,mHa (Table~\ref{tab:basin_landscape}), confirming
that they reflect physical broken-symmetry states rather than
numerical artifacts.
For a quick landscape survey alone, the Phase~0 cost can be cut
further by disabling orbital optimization or limiting to $1$--$2$
cycles. The
convergence robustness check below is performed in BS-2; the
main-text studies focus on BS-1.

\paragraph{Convergence robustness within a basin.}
Different random seeds across independent Phase~0 runs produce
slightly different $100$-determinant core sets within the same
basin, but Phase~1 expansion converges all of them to the same
physical state to sub-mHa precision. On \textrm{[Fe$_4$S$_4$]}
(54e,\,36o), we took five independent Phase~0 outputs that
landed in BS-2 (Run~1--5 in Fig.~\ref{fig:basin_convergence}b);
their $100$-det Phase~0 energies span $2.73$\,mHa. Expanded under
the same standard Phase~1 protocol
($\gamma\!=\!1.1$, orbital optimization ON, Davidson
$\mathrm{tol}\!=\!10^{-4}$) their $N_{\rm det}=10^{6}$ endpoints
collapse to a $0.26$\,mHa spread --- a tenfold compression and
well below chemical accuracy ($1.6$\,mHa). The Phase~1 expansion is
therefore strongly basin-correcting at the seed level: every seed
within a basin converges to essentially the same final energy.

\begin{table}[h]
\caption{Phase~0 basin energies on \textrm{[Fe$_4$S$_4$]}
($100$-determinant core). Two independent sampling experiments ---
Experiment~A ($9$ independent Phase~0 jobs) and Experiment~B
($40$ independent Phase~0 jobs, different computing cluster) --- each
identify the same three basins; the column reports the
\emph{lowest-energy} Phase~0 job per basin in each experiment.
The two experiments agree at the $\sim\!0.02$\,mHa level on every
basin.}
\label{tab:basin_landscape}
\centering
\begin{tabular}{l l c c c c}
\toprule
Basin & Spin pattern & Best of Exp.\ A & Best of Exp.\ B & $|$A $-$ B$|$ & vs.\ BS-2 (A) \\
\midrule
BS-2 & DUDU\,/\,UDUD & $-327.149087$\,Ha & $-327.149104$\,Ha & $0.017$\,mHa & $0$ \\
BS-1 & UUDD\,/\,DDUU & $-327.146237$\,Ha & $-327.146228$\,Ha & $0.009$\,mHa & $+2.85$\,mHa \\
BS-3 & DUUD\,/\,UDDU & $-327.142197$\,Ha & $-327.142182$\,Ha & $0.015$\,mHa & $+6.89$\,mHa \\
\bottomrule
\end{tabular}
\end{table}

\section{Parameter counting}
\label{sec:parameter_counting}

\paragraph{Selected CI.}
Each determinant carries one variational CI coefficient. The number of
parameters equals the number of selected determinants $N_{\rm det}$.
Orbital rotation parameters ($n(n{-}1)/2 = 630$ for $n\!=\!36$) are optimized
on a fixed core set and frozen during expansion; they are not counted as
variational parameters of the final wavefunction.

\paragraph{DMRG (matrix product states).}
The MPS for $n$ sites with local dimension $d$ and bond dimension $D$ has
$n \cdot d \cdot D^2$ tensor elements (neglecting boundary corrections).
With $d\!=\!4$ (empty, $\uparrow$, $\downarrow$, $\uparrow\downarrow$):
$N_{\rm MPS} = 4nD^2 = 4LD^2$.
For \mbox{[Fe$_4$S$_4$]} ($n\!=\!36$) at $D\!=\!12{,}000$:
$N_{\rm MPS} = 2.07 \times 10^{10}$.

\paragraph{SU$_2$-adapted DMRG.}
For spin-adapted DMRG we use the same nominal $N_{\rm MPS} = 4LD^2$,
with $D$ now interpreted as the multiplet bond dimension. The actual
parameter count is structurally smaller than this (Wigner--Eckart
factorization separates reduced matrix elements from Clebsch--Gordan
coefficients), but the precise reduction depends on the multiplet
population at each bond and is not easily estimated. We do not apply a SU$_2$ correction here.
SU$_2$ adaptation is an orthogonal symmetry-exploitation step that
applies in principle to either ansatz: a spin-adapted selected-CI
scheme would compress the determinant count comparably. Withholding
the SU$_2$ reduction from both sides --- reporting $4LD^2$ for DMRG
and $N_{\rm det}$ for selected CI --- therefore isolates the
comparison to ansatz structure and orbital basis, rather than to
SU$_2$ symmetry handling.

\paragraph{DMRG parameter counting for the Hubbard-on-graph results.}
The Hubbard-on-graph calculations in Fig.~5 of the main text use the
$S_z$ DMRG driver of block2. We report the same
$N_{\rm MPS} = 4LD^2$ for each bond dimension $D$, consistent with
the convention above.

\section{Data tables for the main-text figures}
\label{sec:data_tables}

Sections~\ref{sec:fig1_data}--\ref{sec:fig4_data} below tabulate the raw point-by-point data underlying Figs.~1--4 of the main text.

\subsection{Fig.~1 data}
\label{sec:fig1_data}
Table~\ref{tab:orbopt_trajectory} lists the cycle-by-cycle energies
underlying Fig.~1(b), and Table~\ref{tab:gain_transfer} lists the
per-round energies underlying Fig.~1(c). They are from the
\textrm{[Fe$_4$S$_4$]} (54e,\,36o) BS-1 state
(${\uparrow}{\uparrow}{\downarrow}{\downarrow}$
broken-symmetry configuration; basin taxonomy in
Sec.~\ref{sec:robustness}) with the full hyperparameter set in
Sec.~\ref{sec:workflow}. The particular Phase~0 seed shown here
lands at $-327.132$\,Ha at cycle~10, about $14$\,mHa above the
BS-1 basin-best of Table~\ref{tab:basin_landscape}; this gap is
ordinary intra-basin seed-level scatter at the $100$-determinant
Phase~0 stage, and is closed by Phase~1 expansion.

\begin{table}[h]
\caption{COO trajectory on the 100-determinant BS-1 core set
[Fig.~1(b) of the main text]. $E_\mathrm{BFGS}$ is the variational
energy at the end of the cycle's BFGS orbital optimization (the
$100$-determinant core is fixed across this step; CI is re-diagonalized
at every line-search trial, see Sec.~\ref{sec:algorithm});
$E_\mathrm{CI}$ is the energy after re-running the TrimCI search
(new $100$-determinant selection) in the rotated basis.
$\Delta E_\mathrm{COO}$ is measured relative to the initial energy
$E_0 = -326.722135$~Ha. By cycle~3 the orbital optimization has recovered
$\sim\!99\%$ of the final gain.}
\label{tab:orbopt_trajectory}
\begin{ruledtabular}
\begin{tabular}{cccc}
cycle & $E_\mathrm{BFGS}$ (Ha) & $E_\mathrm{CI}$ (Ha) & $\Delta E_\mathrm{COO}$ (mHa) \\
\midrule
0  &       ---         & $-326.722135$ &     $0.00$ \\
1  & $-327.052937$     & $-327.078142$ & $-356.01$ \\
2  & $-327.102801$     & $-327.108337$ & $-386.20$ \\
3  & $-327.125327$     & $-327.129260$ & $-407.13$ \\
4  & $-327.130326$     & $-327.129505$ & $-407.37$ \\
5  & $-327.131395$     & $-327.129278$ & $-407.14$ \\
6  & $-327.130065$     & $-327.131226$ & $-409.09$ \\
7  & $-327.131850$     & $-327.128472$ & $-406.34$ \\
8  & $-327.129166$     & $-327.131371$ & $-409.24$ \\
9  & $-327.131888$     & $-327.131462$ & $-409.33$ \\
10 & $-327.132071$     & $-327.131776$ & $-409.64$ \\
\end{tabular}
\end{ruledtabular}
\end{table}

\begin{table}[h]
\caption{Gain-transfer data [Fig.~1(c) of the main text]. Each entry is the
energy error $E_\mathrm{var} - E_\mathrm{FCI}$ in mHa, with
$E_\mathrm{FCI} = -327.244$~Ha (BS-1 FCI energy estimated from the
TrimCI~+~COO power-law extrapolation, Sec.~\ref{sec:fig2_data}). Columns are orbital snapshots (cycles 0,
1, 2, 3, 10 of the COO loop); rows are the per-round state of a frozen-
orbital, variational TrimCI expansion started from the corresponding
100-determinant core. Growth factor~$\gamma = 2$ per round.
The ``init'' row reports the energy of the $100$-determinant core
itself (cycle~0: the LMO seed; cycle~$N\!\geq\!1$: the TrimCI
re-searched core in cycle-$N$'s rotated basis, equal to
$E_\mathrm{CI}$ at cycle~$N$ in Table~\ref{tab:orbopt_trajectory}).}
\label{tab:gain_transfer}
\begin{ruledtabular}
\begin{tabular}{rrrrrrr}
round & $N_\mathrm{det}$ & c=0 & c=1 & c=2 & c=3 & c=10 \\
\midrule
init &         100  & $521.87$ & $165.86$ & $135.66$ & $114.74$ & $112.22$ \\
 0  &         200  & $489.46$ & $148.53$ & $127.86$ & $106.87$ & $104.43$ \\
 1  &         400  & $463.78$ & $138.41$ & $122.54$ & $ 99.07$ & $ 96.90$ \\
 2  &         800  & $434.60$ & $129.02$ & $116.36$ & $ 91.65$ & $ 89.36$ \\
 3  &       1\,600 & $405.55$ & $118.07$ & $108.48$ & $ 83.58$ & $ 81.75$ \\
 4  &       3\,200 & $377.63$ & $108.08$ & $ 99.16$ & $ 75.70$ & $ 73.42$ \\
 5  &       6\,400 & $353.01$ & $ 98.72$ & $ 90.87$ & $ 68.25$ & $ 66.05$ \\
 6  &      12\,800 & $330.31$ & $ 90.00$ & $ 82.76$ & $ 61.69$ & $ 59.71$ \\
 7  &      25\,600 & $306.55$ & $ 82.19$ & $ 75.50$ & $ 55.72$ & $ 53.92$ \\
 8  &      51\,200 & $276.44$ & $ 75.03$ & $ 69.17$ & $ 50.34$ & $ 48.80$ \\
 9  &     102\,400 & $243.42$ & $ 68.48$ & $ 63.38$ & $ 45.44$ & $ 44.04$ \\
10  &     204\,800 & $215.85$ & $ 62.48$ & $ 58.16$ & $ 40.92$ & $ 39.66$ \\
11  &     409\,600 & $184.29$ & $ 56.88$ & $ 53.47$ & $ 36.82$ & $ 35.67$ \\
12  &     819\,200 & $150.51$ & $ 51.51$ & $ 49.07$ & $ 33.06$ & $ 32.02$ \\
13  &   1\,638\,400 & $127.73$ & $ 46.25$ & $ 44.87$ & $ 29.58$ & $ 28.62$ \\
14  &   3\,276\,800 & $110.79$ & $ 41.00$ & $ 40.71$ & $ 26.29$ & $ 25.43$ \\
15  &   6\,553\,600 & $ 96.51$ & $ 35.79$ & $ 36.24$ & $ 23.18$ & $ 22.39$ \\
\end{tabular}
\end{ruledtabular}
\end{table}

The compression follows the three-regime pattern reported in the main
text. From cycle~0 to cycles~1--2 the determinant count required for
$\Delta E\!\approx\!97$\,mHa drops from $6.5\!\times\!10^{6}$ to
$\sim\!8\!\times\!10^{3}$, a $\sim\!10^{3}\times$ compression purely
from the basis change. Cycle~3 contributes another $\sim\!14\times$
over cycle~2 (at $\Delta E\!\approx\!36$\,mHa: $4.5\!\times\!10^{5}$ vs
$6.5\!\times\!10^{6}$ determinants). Beyond cycle~3 the orbitals are
essentially saturated (cycle~10 is only $\sim\!1.2\times$ tighter than
cycle~3 at $\Delta E\!\approx\!23$\,mHa). The total cycle-0 to
cycle-10 determinant compression at fixed accuracy is $\sim\!10^{4}\times$.

\subsection{Fig.~2 data}
\label{sec:fig2_data}
Tables~\ref{tab:trimci_fe4s4}, \ref{tab:udmrg_fe4s4}, and
\ref{tab:extrap_fe4s4} list the raw data underlying Fig.~2 of the main
text. All calculations are on \textrm{[Fe$_4$S$_4$]} (54e,\,36o) in
the BS-1 state
(${\uparrow}{\uparrow}{\downarrow}{\downarrow}$
broken-symmetry configuration; basin taxonomy in
Sec.~\ref{sec:robustness}) with TrimCI hyperparameters as in
Sec.~\ref{sec:workflow}.

\begin{figure}[h]
\includegraphics[width=\columnwidth]{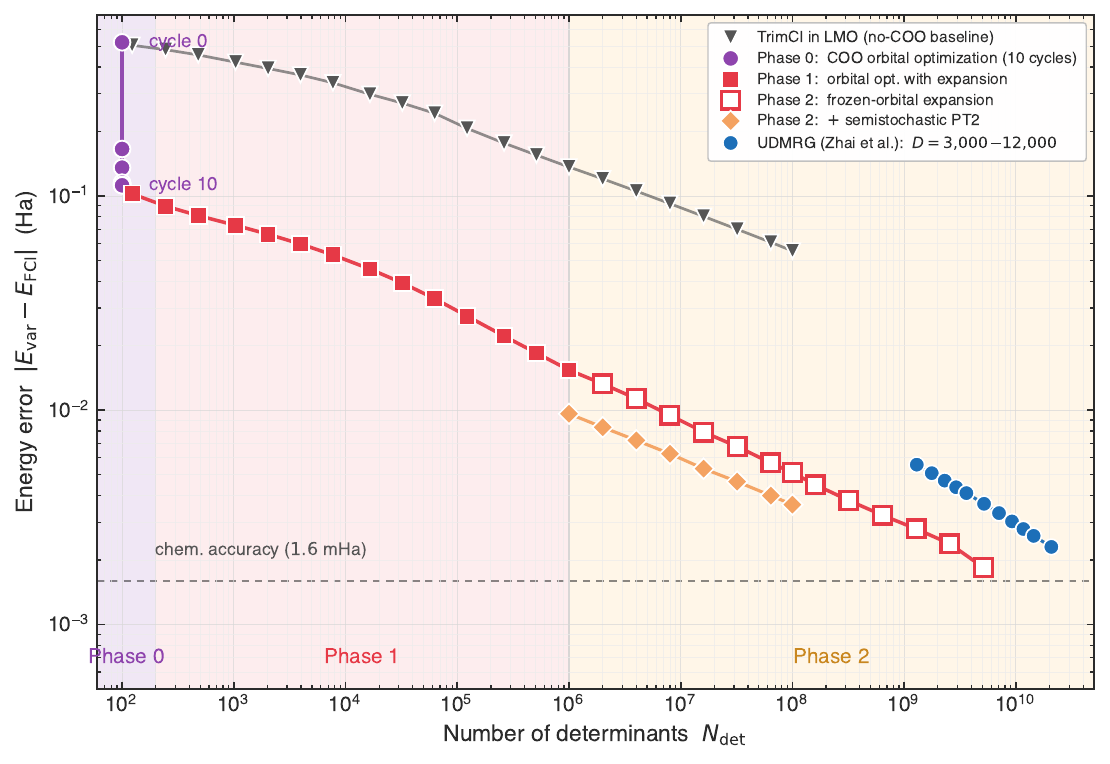}
\caption{Full TrimCI~+~COO trajectory on \textrm{[Fe$_4$S$_4$]}
(54e,\,36o) BS-1, all three phases on a common
energy-error-vs-$N_{\rm det}$ plot. Phase~0
(purple): 100-det COO orbital optimization, cycles 0--10. Phase~1
(filled red squares): orbital optimization with expansion from
$N_{\rm det}\!=\!100$ to $\sim\!10^6$ determinants. Phase~2 (open
red squares): frozen-orbital expansion to
$5.12\!\times\!10^9$ determinants. Orange diamonds: variational
energies with semistochastic PT2 correction. Gray line: TrimCI in
LMO (no-COO baseline) for reference. Blue circles: UDMRG reference
$D = 3000$--$12000$ from Ref.~\cite{Lee2026}. Dashed
gray line: chemical accuracy ($1.6$~mHa).}
\label{fig:phase012_overview}
\end{figure}

Figure~\ref{fig:phase012_overview} places the entire trajectory on
a single plot of energy error against $N_{\rm det}$. The three
phases are: Phase~0 (orbital optimization at a fixed 100-determinant
core); Phase~1 (slow expansion with $\gamma\!=\!1.1$ per round while
orbitals continue to refine, reaching $\sim\!10^6$ determinants);
and Phase~2 (frozen-orbital expansion with $\gamma\!=\!2$ to
$5.12\!\times\!10^9$ determinants). Phase~0 sits at fixed
$N_{\rm det}\!=\!100$ and lowers $\Delta E$ by $\sim\!5\!\times$ over
ten orbital cycles. Phase~1 carries the trajectory across four
decades of $N_{\rm det}$ with orbital optimization still in the
loop. Phase~2 then expands four further decades along a clean
power law with the orbitals frozen.

Each phase prepares the next: Phase~0 establishes high-quality
orbitals at the smallest scale, Phase~1 carries that quality
forward as the determinant space grows, and Phase~2 then expands
rapidly with the orbitals held fixed.

\begin{table}[h]
\caption{TrimCI~+~COO variational energies on
\textrm{[Fe$_4$S$_4$]} along the Phase~2 expansion (frozen
orbitals, growth $\gamma\!=\!2$ per round). The ``Source'' column
records whether the Davidson eigensolve at that point ran on CPU or
GPU. $\Delta E_{\rm PT2}$ is the semistochastic PT2 correction; we
did not compute it for $N_{\rm det}\!>\!10^{8}$.}
\label{tab:trimci_fe4s4}
\begin{ruledtabular}
\begin{tabular}{rrrrl}
$N_\mathrm{det}$ & $E_\mathrm{var}$ (Ha) & $\Delta E_\mathrm{PT2}$ (mHa) & $E_\mathrm{var}\!+\!\Delta E_\mathrm{PT2}$ (Ha) & Source \\
\midrule
$1.0\!\times\!10^{6}$  & $-327.228131$ & $-6.23$ & $-327.234365$ & CPU \\
$2.0\!\times\!10^{6}$  & $-327.230715$ & $-4.95$ & $-327.235665$ & CPU \\
$4.0\!\times\!10^{6}$  & $-327.232666$ & $-4.10$ & $-327.236762$ & CPU \\
$8.0\!\times\!10^{6}$  & $-327.234513$ & $-3.22$ & $-327.237731$ & CPU \\
$1.6\!\times\!10^{7}$  & $-327.236084$ & $-2.58$ & $-327.238659$ & CPU \\
$3.2\!\times\!10^{7}$  & $-327.237229$ & $-2.13$ & $-327.239356$ & CPU \\
$6.4\!\times\!10^{7}$  & $-327.238319$ & $-1.68$ & $-327.240001$ & CPU \\
$1.0\!\times\!10^{8}$  & $-327.238886$ & $-1.49$ & $-327.240372$ & CPU \\
$1.6\!\times\!10^{8}$  & $-327.239507$ & ---      & ---            & GPU \\
$3.2\!\times\!10^{8}$  & $-327.240207$ & ---      & ---            & GPU \\
$6.4\!\times\!10^{8}$  & $-327.240764$ & ---      & ---            & GPU \\
$1.28\!\times\!10^{9}$ & $-327.241205$ & ---      & ---            & GPU \\
$2.56\!\times\!10^{9}$ & $-327.241618$ & ---      & ---            & GPU \\
$5.12\!\times\!10^{9}$ & $-327.242156$ & ---      & ---            & GPU \\
\end{tabular}
\end{ruledtabular}
\end{table}

\begin{table}[h]
\caption{Unrestricted DMRG reference data on the BS-1 state of
\textrm{[Fe$_4$S$_4$]}, copied from Zhai \textit{et al.}\
(arXiv:2601.04621, Ref.~\cite{Lee2026}). Parameter count
$N_\mathrm{params} = n_\mathrm{orb}\cdot d\cdot D^{2} = 144\,D^{2}$
($n_\mathrm{orb}\!=\!36$, $d\!=\!4$).}
\label{tab:udmrg_fe4s4}
\begin{ruledtabular}
\begin{tabular}{rrr}
$D$ & $N_\mathrm{params}$ & $E_\mathrm{var}$ (Ha) \\
\midrule
$3\,000$  & $1.30\!\times\!10^{9}$  & $-327.238426$ \\
$3\,500$  & $1.76\!\times\!10^{9}$  & $-327.238913$ \\
$4\,000$  & $2.30\!\times\!10^{9}$  & $-327.239304$ \\
$4\,500$  & $2.92\!\times\!10^{9}$  & $-327.239627$ \\
$5\,000$  & $3.60\!\times\!10^{9}$  & $-327.239894$ \\
$6\,000$  & $5.18\!\times\!10^{9}$  & $-327.240340$ \\
$7\,000$  & $7.06\!\times\!10^{9}$  & $-327.240688$ \\
$8\,000$  & $9.22\!\times\!10^{9}$  & $-327.240971$ \\
$9\,000$  & $1.17\!\times\!10^{10}$ & $-327.241208$ \\
$10\,000$ & $1.44\!\times\!10^{10}$ & $-327.241407$ \\
$12\,000$ & $2.07\!\times\!10^{10}$ & $-327.241697$ \\
\end{tabular}
\end{ruledtabular}
\end{table}

\textbf{R$^{2}$-scan power-law extrapolation.} Assuming the
power-law ansatz $E(N_{\rm det}) = E_{\rm extrap} + a\,N_{\rm det}^{-\alpha}$,
we take logs to obtain
$\log(E - E_{\rm extrap}) = \log a - \alpha \log N_{\rm det}$
and scan $5{,}000$ candidate values of $E_{\rm extrap}$ below
$E_{\rm min}$. For each candidate, the remaining $(a, \alpha)$ are
determined by a linear least-squares fit in log--log coordinates; the
$E_{\rm extrap}$ value giving the highest $R^{2}$ is selected.
Uncertainty is estimated by a nonparametric bootstrap. The
bootstrap is a standard statistical procedure~\cite{Efron1979}
that simulates having many independent data sets by repeatedly
resampling the existing one, so the variability of a fitted
quantity can be assessed without acquiring fresh data.
Concretely: from the $14$ data points of
Table~\ref{tab:trimci_fe4s4}, we draw $14$ points uniformly at
random \emph{with replacement} --- some points appear multiple
times, others may be missing --- to form a bootstrap replicate,
refit it with the same $R^{2}$-scan procedure, and record the
resulting $E_{\rm extrap}$. Repeating this $500$ times yields a
distribution of fits; the standard deviation across the $500$
replicates is the quoted uncertainty
(Table~\ref{tab:extrap_fe4s4}). The UDMRG extrapolate is taken
directly from the published fit in Ref.~\cite{Lee2026}, not refit
here.

\begin{table}[h]
\caption{Power-law extrapolation results for the BS-1 state of
\textrm{[Fe$_4$S$_4$]}. TrimCI (COO) values are from the
R$^{2}$-scan fit to the $14$ points in Table~\ref{tab:trimci_fe4s4};
$\sigma(E_{\rm extrap})$ is the standard deviation across the $500$
bootstrap replicates, and the $90\%$ confidence interval (c.i.)
is the $[5\text{th},\,95\text{th}]$ percentile of the same
bootstrap distribution.}
\label{tab:extrap_fe4s4}
\centering
\begin{tabular}{lcc}
\toprule
Quantity & TrimCI (COO) & UDMRG~\cite{Lee2026} \\
\midrule
$E_{\rm extrap}$ (Ha)          & $-327.2441$     & $-327.2443$ \\
$\sigma(E_{\rm extrap})$ (mHa) & $0.26$          & --- \\
90\% c.i.\ (Ha)                & $[-327.2443, -327.2436]$ & --- \\
$\alpha$                        & $0.24 \pm 0.01$ & --- \\
$a$                             & $0.44 \pm 0.04$ & --- \\
$R^{2}$                         & $0.9992$        & --- \\
\bottomrule
\end{tabular}
\end{table}

The agreement between the two independent extrapolations
(gap $= 0.19$~mHa, well within the TrimCI bootstrap uncertainty
$\sigma\!=\!0.26$\,mHa) provides cross-validation of the full-CI
limit for this system.

\textbf{Hundred-trillion compression (panel (d) of Fig.~2 of the main text).}
From the R$^{2}$-scan fit above we have $\alpha_{\rm COO} = 0.24$
(Table~\ref{tab:extrap_fe4s4}), and the corresponding late-window
fit on the LMO trajectory of Table~\ref{tab:fig3_fe4s4} below
($N_\mathrm{det} \ge 10^{6}$) gives $\alpha_{\rm LMO} = 0.196$;
varying the lower $N_\mathrm{det}$ bound of the fit window
between $10^{5}$ and $10^{7}$ shifts the slope by less than $0.01$. Anchoring at $N_\mathrm{det} = 10^{9}$
on the COO trajectory, where $\Delta E = 2.94$~mHa, the LMO power
law extrapolates to $N_\mathrm{LMO} = 3.23 \times 10^{14}$ for the
same energy. COO therefore reaches this energy with
$3.2 \times 10^{5}\!\times$ fewer determinants than LMO. This
extrapolation sits at $3.65\%$ of the full-CI dimension
$N_\mathrm{FCI} = \binom{36}{27}^{2} = 8.86 \times 10^{15}$, so the
comparison stays comfortably below the physical ceiling.

\textbf{Semistochastic PT2 correction and origin of the
$\boldsymbol{25\times}$ ratio.} The $\Delta E_{\rm PT2}$ column of
Table~\ref{tab:trimci_fe4s4} is the Epstein--Nesbet second-order
correction to the variational TrimCI energy,
\begin{equation}
\Delta E_{\rm PT2}
  = \sum_{a \notin \mathcal{V}}
    \frac{|\langle a | \hat{H} | \Psi_{\rm var} \rangle|^{2}}
         {E_{\rm var} - H_{aa}} \,,
\label{eq:pt2}
\end{equation}
where the sum runs over all determinants outside the variational
space $\mathcal{V}$ that are connected to $|\Psi_{\rm var}\rangle$
by a one- or two-electron excitation. We evaluate this sum semistochastically with a modified SHCI-style
screening~\cite{Holmes2016,Sharma2017}: the variational core is
sorted by $|c_i|$ and partitioned into a deterministic block (top
dets capturing $\geq\!99\%$ of $\sum_i|c_i|^{2}$) and an
importance-sampled stochastic remainder for the rest. Both phases
use the same SHCI-style heat-bath cutoff $\varepsilon_{Hc}$ ---
couplings with $|H_{ij}c_{j}| < \varepsilon_{Hc}$ are skipped from
the enumeration --- and $\varepsilon_{Hc}$ is adaptively tightened
until the round-to-round change in $\Delta E_{\rm PT2}$ falls
below $3\%$.
Compared to a naive enumeration, the implementation reduces the
peak working-set memory by chunking the deterministic block and
by streaming external determinants without ever materializing the
full connected space; full algorithmic and numerical detail, including the key
innovations, is deferred to a separate paper. We computed PT2 up to
$N_{\rm det}\!=\!10^{8}$ in Table~\ref{tab:trimci_fe4s4}.

A power-law fit to all eight PT2-corrected energies of
Table~\ref{tab:trimci_fe4s4} (same $R^{2}$-scan procedure as for the
variational fit above) gives $\alpha = 0.20$,
$E_{\rm extrap}^{\rm PT2} = -327.2443$~Ha, $R^{2}=0.9999$,
consistent with the variational extrapolate $-327.2441$~Ha
(Table~\ref{tab:extrap_fe4s4}). Solving the fitted curve for the
largest UDMRG energy $E = -327.2417$~Ha ($D\!=\!12{,}000$,
$N_{\rm UDMRG} = 2.07\!\times\!10^{10}$ parameters) gives
$N_{\rm det}^{\rm match} \approx 7.7\!\times\!10^{8}$, i.e.\ a
parameter-count ratio of $2.07\!\times\!10^{10} / 7.7\!\times\!10^{8}
\approx 27\times$. The same calculation restricted to the last
$5$--$6$ points gives $21$--$24\times$; we report
``$\sim\!25\times$'' in the abstract and main text as the central
value across these fit windows. The variational-only
``$\sim\!8\times$'' compression follows from the same
variational fit of Table~\ref{tab:extrap_fe4s4}:
$E_{\rm var} = -327.2417$~Ha is reached at
$N_{\rm det} \approx 2.6\!\times\!10^{9}$, giving
$2.07\!\times\!10^{10} / 2.6\!\times\!10^{9} \approx 8\times$.

\subsection{Fig.~3 data}
\label{sec:fig3_data}
Tables~\ref{tab:fig3_phase0}--\ref{tab:fig3_pcluster} underlie
Fig.~3 of the main text. Table~\ref{tab:fig3_phase0} lists the
cycle-by-cycle Phase~0 variational energies for the three Fe-S systems
(Rows~(a)--(c) of Fig.~3). Tables~\ref{tab:fig3_fe2s2}--\ref{tab:fig3_pcluster}
sample the Phase~1$+$2 expansion trajectories with COO and LMO at
every doubling of $N_{\rm det}$, with rows from the two orbital
sets aligned on the same $N_{\rm det}$ where data are available
(Rows~(d)--(f) of Fig.~3).
Reference $E_{\rm FCI}$ values (electronic, Ha):
$-116.6056$ (Fe$_2$S$_2$, Li \emph{et al.}~\cite{Li2025} converged EMO/LMO
at $D=3000$, and confirmed by our TrimCI COO trajectory at
$N_{\rm det}=10^{8}$, Table~\ref{tab:fig3_fe2s2}), $-327.244$ (Fe$_4$S$_4$ BS-1, TrimCI power-law extrapolation
in agreement with UDMRG~\cite{Lee2026} to $0.19$~mHa,
Sec.~\ref{sec:fig2_data}), $-1075.530$ (P-cluster, electronic; corresponds to the total
$E_{\rm tot}^{\rm FCI} = -17{,}492.236408$~Ha of Xiang
\emph{et al.}~\cite{xiangDistributedMultiGPUInitio2024} minus
the FCIDUMP nuclear-repulsion energy
$E_{\rm nuc} = -16{,}416.706473$~Ha).

\begin{table}[h]
\caption{Phase~0 orbital-optimization trajectory. Variational
energy $E_{\rm var}$ (Ha) versus cycle for the three Fe-S systems
at a $100$-determinant core, corresponding to panels (a)--(c) of
Fig.~3. The descent is monotone in trend but oscillates by
$\lesssim\!1$~mHa near convergence (sub-mHa late-cycle wobbles
are a normal feature at this scale); the protocol takes
the lowest-energy cycle as the Phase~0 endpoint, which is not
necessarily cycle~10.
P-cluster energies are total energies
(electronic $+\ E_{\rm core}$, with $E_{\rm core} = -16416.706473$~Ha).}
\label{tab:fig3_phase0}
\begin{ruledtabular}
\begin{tabular}{cccc}
Cycle & Fe$_2$S$_2$ & Fe$_4$S$_4$ & P-cluster \\
\midrule
 0 & $-116.421800$ & $-326.722135$ & $-17490.476175$ \\
 1 & $-116.564568$ & $-327.078142$ & $-17491.815324$ \\
 2 & $-116.572244$ & $-327.108337$ & $-17492.004667$ \\
 3 & $-116.573003$ & $-327.129260$ & $-17492.089451$ \\
 4 & $-116.574414$ & $-327.129505$ & $-17492.097749$ \\
 5 & $-116.575527$ & $-327.129278$ & $-17492.100646$ \\
 6 & $-116.576039$ & $-327.131226$ & $-17492.101228$ \\
 7 & $-116.577189$ & $-327.128472$ & $-17492.101952$ \\
 8 & $-116.579164$ & $-327.131371$ & $-17492.102214$ \\
 9 & $-116.579768$ & $-327.131462$ & $-17492.102115$ \\
10 & $-116.579256$ & $-327.131776$ & $-17492.102528$ \\
\end{tabular}
\end{ruledtabular}
\end{table}

\begin{table}[h]
\caption{Fe$_2$S$_2$ Phase~1$+$2 expansion trajectories, sampled
at every doubling of $N_{\rm det}$ from the round-by-round logs.
Columns: $E_{\rm var}^{\rm COO}$ (starting from the Phase~0 COO
100-determinant core), $E_{\rm var}^{\rm LMO}$
(starting from the LMO 100-determinant core; identical expansion
with orbital optimization disabled). Both trajectories share the
same Phase~1 growth schedule, so rows are aligned at the same
$N_{\rm det}$ in both columns. The $N_{\rm det}=100$ row is the
starting core of each trajectory (Phase~0 cycle~10 endpoint for
COO; LMO 100-determinant core for LMO).
FCI reference (electronic): $-116.6056$~Ha.}
\label{tab:fig3_fe2s2}
\begin{ruledtabular}
\begin{tabular}{rrr}
$N_{\rm det}$ & $E_{\rm var}^{\rm COO}$ (Ha) & $E_{\rm var}^{\rm LMO}$ (Ha) \\
\midrule
$100$ & $-116.579256$ & $-116.421800$ \\
$115$ & $-116.580521$ & $-116.422869$ \\
$207$ & $-116.585205$ & $-116.430653$ \\
$408$ & $-116.590437$ & $-116.442363$ \\
$798$ & $-116.594897$ & $-116.456759$ \\
$1\,558$ & $-116.597459$ & $-116.466219$ \\
$3\,346$ & $-116.599410$ & $-116.475284$ \\
$6\,526$ & $-116.600741$ & $-116.482483$ \\
$12\,721$ & $-116.601689$ & $-116.501403$ \\
$24\,796$ & $-116.602538$ & $-116.511285$ \\
$53\,158$ & $-116.603178$ & $-116.526909$ \\
$103\,595$ & $-116.603643$ & $-116.584551$ \\
$201\,882$ & $-116.604062$ & $-116.593224$ \\
$393\,416$ & $-116.604462$ & $-116.597520$ \\
$843\,328$ & $-116.605153$ & $-116.600066$ \\
$2\,000\,000$ & $-116.605465$ & $-116.601623$ \\
$4\,000\,000$ & $-116.605560$ & $-116.602421$ \\
$8\,000\,000$ & $-116.605594$ & $-116.603099$ \\
$16\,000\,000$ & $-116.605604$ & $-116.603644$ \\
$32\,000\,000$ & $-116.605607$ & $-116.604040$ \\
$64\,000\,000$ & $-116.605607$ & $-116.604426$ \\
$100\,000\,000$ & $-116.605607$ & $-116.604715$ \\
\end{tabular}
\end{ruledtabular}
\end{table}

\begin{table}[h]
\caption{Fe$_4$S$_4$ Phase~1$+$2 expansion trajectories, sampled at
every doubling of $N_{\rm det}$ from the round-by-round logs.
$E_{\rm var}^{\rm COO}$ starts from the Phase~0 COO
100-determinant core and continues to
$5.12\!\times\!10^{9}$~dets via GPU Davidson (cf.\
Fig.~2 of the main text); $E_{\rm var}^{\rm LMO}$ starts from the
LMO 100-determinant core (identical expansion with orbital
optimization disabled) and is logged to $10^{8}$~dets. The
$N_{\rm det}=100$ row is the starting core of each trajectory.
FCI reference (electronic): $-327.244$~Ha.}
\label{tab:fig3_fe4s4}
\begin{ruledtabular}
\begin{tabular}{rrr}
$N_{\rm det}$ & $E_{\rm var}^{\rm COO}$ (Ha) & $E_{\rm var}^{\rm LMO}$ (Ha) \\
\midrule
$100$ & $-327.131776$ & $-326.722135$ \\
$111$ & $-327.139084$ & $-326.732731$ \\
$201$ & $-327.150547$ & $-326.755144$ \\
$397$ & $-327.160535$ & $-326.780400$ \\
$778$ & $-327.168386$ & $-326.809110$ \\
$1\,674$ & $-327.175772$ & $-326.842476$ \\
$3\,266$ & $-327.182431$ & $-326.868941$ \\
$6\,370$ & $-327.189027$ & $-326.895704$ \\
$12\,417$ & $-327.195645$ & $-326.931867$ \\
$26\,622$ & $-327.202911$ & $-326.965099$ \\
$51\,883$ & $-327.209163$ & $-326.990966$ \\
$101\,110$ & $-327.214977$ & $-327.027806$ \\
$197\,040$ & $-327.219942$ & $-327.057072$ \\
$422\,379$ & $-327.224544$ & $-327.082938$ \\
$823\,100$ & $-327.227761$ & $-327.102143$ \\
$2\,000\,000$ & $-327.230715$ & $-327.123672$ \\
$4\,000\,000$ & $-327.232666$ & $-327.138511$ \\
$8\,000\,000$ & $-327.234513$ & $-327.151816$ \\
$16\,000\,000$ & $-327.236084$ & $-327.163623$ \\
$32\,000\,000$ & $-327.237229$ & $-327.174023$ \\
$64\,000\,000$ & $-327.238319$ & $-327.183151$ \\
$100\,000\,000$ & $-327.238886$ & $-327.188557$ \\
$160\,000\,000$ & $-327.239507$ & --- \\
$320\,000\,000$ & $-327.240207$ & --- \\
$640\,000\,000$ & $-327.240764$ & --- \\
$1\,280\,000\,000$ & $-327.241205$ & --- \\
$2\,560\,000\,000$ & $-327.241618$ & --- \\
$5\,120\,000\,000$ & $-327.242156$ & --- \\
\end{tabular}
\end{ruledtabular}
\end{table}

\begin{table}[h]
\caption{P-cluster Phase~1$+$2 expansion trajectories, sampled at
every doubling of $N_{\rm det}$ from the round-by-round logs.
$E_{\rm var}^{\rm COO}$ starts from the Phase~0 COO
100-determinant core; $E_{\rm var}^{\rm LMO}$ starts
from the LMO 100-determinant core (identical expansion with
orbital optimization disabled). The $N_{\rm det}=100$ row is the
starting core of each trajectory. All energies are electronic
(total = electronic + $E_{\rm nuc}$, with
$E_{\rm nuc} = -16{,}416.706473$~Ha from the corresponding FCIDUMP). FCI
reference (electronic): $-1075.530$~Ha, corresponding to the
total $-17{,}492.236408$~Ha of Xiang \emph{et
al.}~\cite{xiangDistributedMultiGPUInitio2024}.}
\label{tab:fig3_pcluster}
\begin{ruledtabular}
\begin{tabular}{rrr}
$N_{\rm det}$ & $E_{\rm var}^{\rm COO}$ (Ha) & $E_{\rm var}^{\rm LMO}$ (Ha) \\
\midrule
$100$ & $-1075.396055$ & $-1073.769702$ \\
$115$ & $-1075.397097$ & $-1073.785446$ \\
$207$ & $-1075.403184$ & $-1073.830548$ \\
$408$ & $-1075.410157$ & $-1073.858624$ \\
$798$ & $-1075.417169$ & $-1073.896982$ \\
$1\,558$ & $-1075.423997$ & $-1073.936969$ \\
$3\,346$ & $-1075.431463$ & $-1073.993337$ \\
$6\,526$ & $-1075.437689$ & $-1074.047141$ \\
$12\,721$ & $-1075.443504$ & $-1074.089872$ \\
$24\,796$ & $-1075.448899$ & $-1074.123108$ \\
$53\,158$ & $-1075.454545$ & $-1074.160719$ \\
$103\,595$ & $-1075.459037$ & $-1074.200737$ \\
$201\,882$ & $-1075.463096$ & $-1074.239225$ \\
$393\,416$ & $-1075.466817$ & $-1074.273835$ \\
$843\,328$ & $-1075.470686$ & $-1074.314882$ \\
$2\,000\,000$ & $-1075.474562$ & $-1074.357745$ \\
$4\,000\,000$ & $-1075.477385$ & $-1074.392273$ \\
$8\,000\,000$ & $-1075.479988$ & $-1074.426653$ \\
$16\,000\,000$ & $-1075.482341$ & $-1074.460524$ \\
$32\,000\,000$ & $-1075.484561$ & $-1074.494091$ \\
$64\,000\,000$ & $-1075.486616$ & $-1074.527142$ \\
\end{tabular}
\end{ruledtabular}
\end{table}

\clearpage

\subsection{Fig.~4 data}
\label{sec:fig4_data}
Tables~\ref{tab:fig4_p_fe2s2}--\ref{tab:fig4_dmrg_d100} list the
raw data underlying Fig.~4 of the main text. The TrimCI~+~COO
top-1 probabilities are reported per system in
Tables~\ref{tab:fig4_p_fe2s2}--\ref{tab:fig4_p_pcluster}, sampled
at every doubling of $N_{\rm det}$ along the Phase~1$+$2
trajectory of each system. The $D\!=\!100$ DMRG anchor
(Table~\ref{tab:fig4_dmrg_d100}) is copied from
Ref.~\cite{Li2025}; the full $D$-dependence is plotted in Fig.~4
of the main text. The parameter axis in the main-text figure uses
$N_{\rm det}$ for TrimCI~+~COO and the nominal MPS count
$4\,L\,D^{2}$ for DMRG (see Sec.~\ref{sec:parameter_counting}).

\begin{table}[h]
\caption{Top-1 probability $p_{0} = |c_{\max}|^{2}$ along the
TrimCI~+~COO Phase~1$+$2 expansion of \textrm{[Fe$_{2}$S$_{2}$]}.
The first row ($N_{\rm det}\!=\!100$) is the Phase~0 output at
the 100-determinant starting core. Subsequent rows are sampled
at every doubling target $N_{\rm det}=100,\,200,\,400,\ldots$
from data.}
\label{tab:fig4_p_fe2s2}
\begin{ruledtabular}
\begin{tabular}{rc}
$N_{\rm det}$ & $p_{0}$ \\
\midrule
$100$ & $0.3564$ \\
$207$ & $0.3451$ \\
$408$ & $0.3224$ \\
$798$ & $0.3103$ \\
$1\,558$ & $0.3176$ \\
$3\,346$ & $0.3122$ \\
$6\,526$ & $0.3079$ \\
$12\,721$ & $0.3001$ \\
$24\,796$ & $0.2798$ \\
$53\,158$ & $0.2602$ \\
$103\,595$ & $0.2386$ \\
$201\,882$ & $0.2118$ \\
$393\,416$ & $0.1770$ \\
$843\,328$ & $0.0669$ \\
$2\,000\,000$ & $0.0569$ \\
$4\,000\,000$ & $0.0566$ \\
$8\,000\,000$ & $0.0518$ \\
$16\,000\,000$ & $0.0518$ \\
$32\,000\,000$ & $0.0518$ \\
$64\,000\,000$ & $0.0518$ \\
\end{tabular}
\end{ruledtabular}
\end{table}

\clearpage

\begin{table}[h]
\caption{Top-1 probability $p_{0} = |c_{\max}|^{2}$ along the
TrimCI~+~COO Phase~1$+$2 expansion of \textrm{[Fe$_{4}$S$_{4}$]}.
The first row ($N_{\rm det}\!=\!100$) is the Phase~0 output at
the 100-determinant starting core. Subsequent rows are sampled
at every doubling target $N_{\rm det}=100,\,200,\,400,\ldots$
from data.}
\label{tab:fig4_p_fe4s4}
\begin{ruledtabular}
\begin{tabular}{rc}
$N_{\rm det}$ & $p_{0}$ \\
\midrule
$100$ & $0.7637$ \\
$201$ & $0.7676$ \\
$397$ & $0.7299$ \\
$778$ & $0.6990$ \\
$1\,674$ & $0.6734$ \\
$3\,266$ & $0.6376$ \\
$6\,370$ & $0.6071$ \\
$12\,417$ & $0.5794$ \\
$26\,622$ & $0.5500$ \\
$51\,883$ & $0.4912$ \\
$101\,110$ & $0.3931$ \\
$197\,040$ & $0.3033$ \\
$422\,379$ & $0.2168$ \\
$823\,100$ & $0.1724$ \\
$2\,000\,000$ & $0.1600$ \\
$4\,000\,000$ & $0.1536$ \\
$8\,000\,000$ & $0.1478$ \\
$16\,000\,000$ & $0.1424$ \\
$32\,000\,000$ & $0.1371$ \\
\end{tabular}
\end{ruledtabular}
\end{table}

\clearpage

\begin{table}[h]
\caption{Top-1 probability $p_{0} = |c_{\max}|^{2}$ along the
TrimCI~+~COO Phase~1$+$2 expansion of the P-cluster
\textrm{[Fe$_{8}$S$_{7}$]}. The first row ($N_{\rm det}\!=\!100$)
is the Phase~0 output at the 100-determinant starting core.
Subsequent rows are sampled at every doubling target
$N_{\rm det}=100,\,200,\,400,\ldots$ from data.}
\label{tab:fig4_p_pcluster}
\begin{ruledtabular}
\begin{tabular}{rc}
$N_{\rm det}$ & $p_{0}$ \\
\midrule
$100$ & $0.8691$ \\
$207$ & $0.8527$ \\
$408$ & $0.8348$ \\
$798$ & $0.8213$ \\
$1\,558$ & $0.8072$ \\
$3\,346$ & $0.7890$ \\
$6\,526$ & $0.7705$ \\
$12\,721$ & $0.7521$ \\
$24\,796$ & $0.7346$ \\
$53\,158$ & $0.7126$ \\
$103\,595$ & $0.6926$ \\
$201\,882$ & $0.6711$ \\
$393\,416$ & $0.6484$ \\
$843\,328$ & $0.6200$ \\
$2\,000\,000$ & $0.5910$ \\
$4\,000\,000$ & $0.5665$ \\
$8\,000\,000$ & $0.5400$ \\
$16\,000\,000$ & $0.5128$ \\
$32\,000\,000$ & $0.4826$ \\
$64\,000\,000$ & $0.4510$ \\
\end{tabular}
\end{ruledtabular}
\end{table}

\clearpage

\begin{table}[h]
\caption{DMRG anchor at $D\!=\!100$ for the three iron-sulfur
systems, copied from Ref.~\cite{Li2025}. EMO = entanglement-minimized
orbitals; LMO = localized molecular orbitals. Nominal MPS
parameter count $N_{\rm params} = 4 L D^{2}$ at $D\!=\!100$ (see
Sec.~\ref{sec:parameter_counting}). The full $D$-dependence is
plotted in Fig.~4 of the main text.}
\label{tab:fig4_dmrg_d100}
\begin{ruledtabular}
\begin{tabular}{l c c c c c c}
System & $L$ & $N_{\rm params}^{D=100}$
       & $E_{\rm DMRG}^{\rm EMO}$ (Ha) & $p_{0}^{\rm EMO}$
       & $E_{\rm DMRG}^{\rm LMO}$ (Ha) & $p_{0}^{\rm LMO}$ \\
\midrule
\textrm{[Fe$_{2}$S$_{2}$]}      & $20$ & $8.0\!\times\!10^{5}$
   & $-116.6031$ & $0.0686$
   & $-116.6020$ & $5.78\!\times\!10^{-3}$ \\
\textrm{[Fe$_{4}$S$_{4}$]}      & $36$ & $1.44\!\times\!10^{6}$
   & $-327.2003$ & $0.0416$
   & $-327.1574$ & $1.75\!\times\!10^{-4}$ \\
P-cluster \textrm{[Fe$_{8}$S$_{7}$]} & $73$ & $2.92\!\times\!10^{6}$
   & $-1075.452$ & $0.0163$
   & $-1075.373$ & $1.28\!\times\!10^{-5}$ \\
\end{tabular}
\end{ruledtabular}
\end{table}

\textbf{Reproducing the $\boldsymbol{16}$--$\boldsymbol{78\times}$
parameter-count ratios (\textnormal{Table~I of the main text}).}
We anchor the comparison at the smallest reported DMRG bond dimension
$D = 100$, where the parameter count is smallest --- the regime
relevant for quantum-circuit state preparation.
For each system and each DMRG basis (EMO, LMO), the recipe is: read
$E_{\rm DMRG}^{D=100}$ from Table~\ref{tab:fig4_dmrg_d100}; locate
the smallest TrimCI~+~COO determinant count $N_{\rm det}^{\rm match}$
at which $E_{\rm var}^{\rm COO}(N) \leq E_{\rm DMRG}^{D=100}$ (linear
interpolation in $\log N$ when the target falls between two data
points; power-law extrapolation in $|E - E_{\rm FCI}|$ when COO
already beats the target at its smallest available $N$);
report the ratio
$N_{\rm params}^{D=100} / N_{\rm det}^{\rm match}$.
Table~\ref{tab:fig4_ratios_recap} reproduces the main-text
ratios via this recipe.

\begin{table}[h]
\caption{Reproduction of Table~I of the main text.
DMRG anchors are at $D = 100$ from
Table~\ref{tab:fig4_dmrg_d100}; $N_{\rm det}^{\rm COO}$ at matched
energy is from the trajectories of
Tables~\ref{tab:fig4_p_fe2s2}--\ref{tab:fig4_p_pcluster} (linear
interpolation in $\log N$, except $^{\dagger}$ which is power-law
extrapolation per the recipe above). EMO = entanglement-minimized
orbitals; LMO = localized molecular orbitals. The $^{\dagger}$ entry
of $23$ determinants for the P-cluster LMO row reflects the regime
where the COO 100-determinant starting core already lies below the
LMO $D\!=\!100$ DMRG energy, so the matched $N_{\rm det}^{\rm COO}$
is an extrapolation that quantifies the implied compression factor
rather than a directly attained det count. The right-most column
reproduces the $16\times$--$4700\times$ range of the main-text
Table~I within rounding.}
\label{tab:fig4_ratios_recap}
\begin{ruledtabular}
\begin{tabular}{l c c c c r}
System & basis & $E_{\rm DMRG}^{D=100}$ (Ha) & $N_{\rm params}^{D=100}$
       & $N_{\rm det}^{\rm COO\;match}$ & ratio \\
\midrule
\textrm{[Fe$_{2}$S$_{2}$]} (20o) & EMO & $-116.6031$ & $8.0\!\times\!10^{5}$
       & $5.1\!\times\!10^{4}$ & $\sim\!16\times$ \\
\textrm{[Fe$_{2}$S$_{2}$]} (20o) & LMO & $-116.6020$ & $8.0\!\times\!10^{5}$
       & $1.6\!\times\!10^{4}$ & $\sim\!49\times$ \\
\textrm{[Fe$_{4}$S$_{4}$]} (36o) & EMO & $-327.2003$ & $1.44\!\times\!10^{6}$
       & $2.0\!\times\!10^{4}$ & $\sim\!71\times$ \\
\textrm{[Fe$_{4}$S$_{4}$]} (36o) & LMO & $-327.1574$ & $1.44\!\times\!10^{6}$
       & $3.1\!\times\!10^{2}$ & $\sim\!4700\times$ \\
P-cluster (73o)                 & EMO & $-1075.452$ & $2.92\!\times\!10^{6}$
       & $3.8\!\times\!10^{4}$ & $\sim\!78\times$ \\
P-cluster (73o)                 & LMO & $-1075.373$ & $2.92\!\times\!10^{6}$
       & $2.3\!\times\!10^{1}\,^{\dagger}$ & $\sim\!1\!\times\!10^{5}\,^{\dagger}$ \\
\end{tabular}
\end{ruledtabular}
\end{table}

\section{Orbital mutual information analysis}
\label{sec:strict_mi}

\emph{Why mutual information.} A central question for the
parameter-efficiency story of TrimCI\,+\,COO is whether the
excitation weight in the converged wavefunction has any
1D-localizable structure --- i.e.\ whether some ordering of the
COO orbitals along a line would let DMRG compress the
wavefunction with a small bond dimension. The standard
quantitative tool in the DMRG-orbital-ordering literature
\cite{Legeza2003,Rissler2006} is the orbital mutual information
$I_{ij}$: a non-negative measure of the total correlation between
orbitals $i$ and $j$, vanishing exactly when their two-orbital
reduced density matrix factorizes, $\rho_{ij} = \rho_i \otimes
\rho_j$.
We compute $I_{ij}$ on the
\textrm{[Fe$_4$S$_4$]}\,(54e,\,36o) ground state and read it two
ways. \emph{Spatial pattern}: which orbital pairs carry the
largest $I_{ij}$, and how do they map back to atoms in the cluster
(intra-Fe, Fe--Fe, Fe--S)? \emph{1D-compressibility}: under the
best ordering of the orbitals along a line, can the matrix
$\{I_{ij}\}$ be collapsed into a narrow band along the diagonal
(MPS-friendly), or does it stay spread out (no MPS can compress
it)?

\emph{Definition of $I_{ij}$.} Following
Refs.~\cite{Legeza2003,Rissler2006},
\begin{equation}
I_{ij} \;=\; S(\rho_i) \;+\; S(\rho_j) \;-\; S(\rho_{ij}),
\label{eq:Iij}
\end{equation}
where $S(\rho) = -\mathrm{Tr}(\rho \log_2 \rho)$ is the von Neumann
entropy, and $\rho_i$, $\rho_{ij}$ are reduced density matrices of
the wavefunction:
\begin{itemize}
\item the \emph{one-orbital RDM} $\rho_i$ is the $4{\times}4$
  matrix in the local Fock basis
  $\{|0\rangle, |{\uparrow}\rangle, |{\downarrow}\rangle,
   |{\uparrow}{\downarrow}\rangle\}_i$ of orbital $i$, obtained by
  tracing $|\Psi\rangle\langle\Psi|$ over all other orbitals;
\item the \emph{two-orbital RDM} $\rho_{ij}$ is the $16{\times}16$
  matrix in the tensor-product Fock basis
  $\{0,\!{\uparrow},\!{\downarrow},\!{\uparrow}{\downarrow}\}_i
   \otimes
   \{0,\!{\uparrow},\!{\downarrow},\!{\uparrow}{\downarrow}\}_j$,
  obtained by tracing over the remaining $N_{\rm orb}-2$ orbitals.
\end{itemize}
If orbitals $i$ and $j$ were uncorrelated --- i.e.\
$\rho_{ij} = \rho_i \otimes \rho_j$ --- the entropies would add,
$S(\rho_{ij}) = S(\rho_i) + S(\rho_j)$, giving $I_{ij}=0$; any
positive $I_{ij}$ quantifies the correlation between the two
orbitals.

\emph{Computing $\rho_i$ and $\rho_{ij}$ from a CI expansion.}
Both RDMs are partial traces of $|\Psi\rangle\langle\Psi|$, which we
expand directly in the determinants. Writing $s_i^I$ for the
local configuration of orbital $i$ in $|I\rangle$ (one of
$|0\rangle,\,|{\uparrow}\rangle,\,|{\downarrow}\rangle,\,
|{\uparrow}{\downarrow}\rangle$):
\begin{equation}
\rho_i \;=\;
\mathrm{Tr}_{\,\text{not-}i}\,|\Psi\rangle\langle\Psi|
\;=\;
\sum_{I, J} c_I\, c_J^*\;
\mathrm{Tr}_{\,\text{not-}i}\,|I\rangle\langle J|.
\end{equation}
The orbital-trace
$\mathrm{Tr}_{\,\text{not-}i}\,|I\rangle\langle J|$ is non-zero only
when $|I\rangle$ and $|J\rangle$ agree on every orbital except $i$;
conservation of $N_\uparrow$ and $N_\downarrow$ in $|\Psi\rangle$
then forces them to agree on $i$ as well, i.e.\ $I = J$. So $\rho_i$
is diagonal,
\begin{equation}
\rho_i^{(s)}
\;=\;
\sum_{I\,:\,s_i^I = s} |c_I|^2,
\qquad
s \in \{|0\rangle,|{\uparrow}\rangle,|{\downarrow}\rangle,|{\uparrow}{\downarrow}\rangle\},
\label{eq:rho_i}
\end{equation}
and $S(\rho_i) = -\sum_s \rho_i^{(s)} \log_2 \rho_i^{(s)}$.

For $\rho_{ij}$ we expand the same way:
\begin{equation}
\rho_{ij} \;=\;
\mathrm{Tr}_{\,\text{not-}(i,j)}\,|\Psi\rangle\langle\Psi|
\;=\;
\sum_{I, J} c_I\, c_J^*\;
\mathrm{Tr}_{\,\text{not-}(i,j)}\,|I\rangle\langle J|,
\label{eq:rho_ij_expand}
\end{equation}
a $16\!\times\!16$ matrix. Let $s_{ij}^I$ denote the local Fock
configuration of orbital pair $(i,j)$ in $|I\rangle$ --- one of the
16 product states
$\{|0\rangle,|{\uparrow}\rangle,|{\downarrow}\rangle,|{\uparrow}{\downarrow}\rangle\}_i
\otimes
\{|0\rangle,|{\uparrow}\rangle,|{\downarrow}\rangle,|{\uparrow}{\downarrow}\rangle\}_j$
--- and let $\eta_I = \pm 1$ be the sign incurred when
anticommuting the $a^\dagger_{i,\sigma}$ and $a^\dagger_{j,\sigma}$
operators of $|I\rangle$ from their canonical positions to the front
of the determinant; equivalently, $|I\rangle = \eta_I\,
|s_{ij}^I\rangle \otimes |\text{rest}^I\rangle$ where
$|\text{rest}^I\rangle$ collects the remaining $N_{\rm orb}-2$
orbitals. Then
\begin{equation}
\mathrm{Tr}_{\,\text{not-}(i,j)}\,|I\rangle\langle J|
\;=\;
\eta_I\,\eta_J\;\langle\text{rest}^J|\text{rest}^I\rangle\;
|s_{ij}^I\rangle\langle s_{ij}^J|,
\label{eq:tr_pair}
\end{equation}
which is non-zero only when $|I\rangle$ and $|J\rangle$ agree on
every orbital outside $(i,j)$, and equals
$\eta_I\,\eta_J\,|s_{ij}^I\rangle\langle s_{ij}^J|$ in that case.

Substituting Eq.~(\ref{eq:tr_pair}) into
Eq.~(\ref{eq:rho_ij_expand}), the inner-product factor
$\langle\text{rest}^J|\text{rest}^I\rangle$ enforces that only pairs
$(I, J)$ sharing the same outside occupation survive. Group the
determinants accordingly --- $|I\rangle, |I'\rangle$ share a group
$g$ iff their bitstrings agree on every orbital outside $(i,j)$ ---
and split the double sum as
$\sum_{I, J} = \sum_g \sum_{I \in g}\sum_{J \in g}$, so that
$\rho_{ij}$ becomes a sum of one independent contribution per
group. Each group's contribution further factorises into a vector
times its Hermitian conjugate:
\begin{equation}
\rho_{ij}
\;=\;
\sum_g \Bigl(\sum_{I \in g} \eta_I\, c_I\, |s_{ij}^I\rangle\Bigr)
       \Bigl(\sum_{J \in g} \eta_J\, c_J^*\, \langle s_{ij}^J|\Bigr)
\;=\;
\sum_g v_g\, v_g^{\dagger},
\label{eq:rho_ij}
\end{equation}
where the 16-vector
\begin{equation}
v_g[s]
\;=\;
\sum_{I \in g\,:\,s_{ij}^I = s} \eta_I\, c_I
\label{eq:vg}
\end{equation}
collects the determinants in $g$ with local $(i,j)$ configuration
$s$. Diagonalising the $16\!\times\!16$ matrix in
Eq.~(\ref{eq:rho_ij}) gives $S(\rho_{ij})$; total cost is
$\mathcal{O}(N_{\rm det})$ per orbital pair.

\emph{Wavefunction snapshot.} The algorithm in
Eqs.~(\ref{eq:rho_i})--(\ref{eq:rho_ij}) is linear in $N_{\rm det}$
and quadratic in the number of orbital pairs ($\binom{36}{2}=630$
here). To keep the computation fast while preserving quality, we
use the Phase~2 round-0 checkpoint at
$N_{\rm det} = 2\!\times\!10^{6}$ (variational energy
$-327.231$~Ha).

\emph{Labeling COO orbitals by Fe atom.} The COO basis at this
checkpoint has been rotated from the canonical localized-MO (LMO)
basis by the cumulative orbital optimization of Phases~0+1; we use
the recorded rotation matrix to project each COO orbital back onto
the LMO basis, where the iron $d$-orbitals are unambiguously
localized. The Fe-LMO indices are taken from the [Fe$_4$S$_4$]
active-space construction of
Ref.~\cite{li2017fe4s4github}: Fe$_1$ owns LMOs $\{2,3,4,5,6\}$,
Fe$_2$ owns $\{7,8,9,10,11\}$, Fe$_3$ owns $\{24,25,26,27,28\}$,
Fe$_4$ owns $\{29,30,31,32,33\}$, and the remaining $16$ LMOs are
S 3p ($4$ S atoms $\times$ $4$ valence orbitals). For each COO
orbital, summing the squared LMO coefficients within each candidate
group ($\mathrm{Fe}_1,\dots,\mathrm{Fe}_4,\mathrm{S}$) gives that
group's projected weight; the orbital is labelled by the iron whose
weight is largest, exceeds $40\%$, and exceeds the total S weight,
and is labelled S otherwise. Of the 36 COO orbitals, $28$ are
$\geq\!90\%$ localized on a single label, $3$ more lie in
$[88\%,90\%)$, and the remaining $5$ (indices $5,\,12,\,13,\,22,\,23$)
carry $70$--$87\%$ on their dominant center. The heatmap of
Fig.~5(a) of the main text orders the 36 orbitals as
Fe$_1$\,$|$\,Fe$_2$\,$|$\,Fe$_3$\,$|$\,Fe$_4$\,$|$\,S
($5/5/5/5/16$) with Fe blocks separated by dashed lines.

\emph{Pair-magnitude statistics.} Indexing COO orbitals $0$--$35$,
the four largest pair-MI values are
\begin{equation*}
\begin{array}{lcl}
I_{0,5}   = 1.04 & \text{(Fe$_1$--Fe$_2$)}, \quad &
I_{25,33} = 1.01 \;\,\text{(Fe$_3$--Fe$_4$)},\\[1pt]
I_{10,19} = 0.50 & \text{(Fe$_2$--Fe$_4$)}, \quad &
I_{18,28} = 0.48 \;\,\text{(Fe$_1$--Fe$_3$)};
\end{array}
\end{equation*}
all four are direct Fe--Fe couplings. Below this lies a long
shoulder of weaker non-local correlations: out of the
$\binom{36}{2}=630$ pairs, $18$ exceed $I = 0.1$ and $120$ exceed
$I = 0.01$; of the $18$ strong pairs, $8$ are Fe--Fe and $10$ are
Fe--S, and \emph{zero} are intra-Fe.

To quantify how this MI graph stretches along a 1D ordering of
the orbitals, we use the off-diagonal MI sum
$M_{\rm tot} = \sum_{i \ne j} I_{ij}$ as the ``MI mass,'' and
define the \emph{$95\%$-mass bandwidth}
\begin{equation}
\label{eq:k95}
k_{95} = \min\Bigl\{\,k\;\Big|\;
        \tfrac{1}{M_{\rm tot}}\!\!\sum_{|i-j|\le k}\!\! I_{ij}
        \;\ge\; 0.95 \Bigr\}
\end{equation}
as the smallest half-bandwidth such that the band of $\pm k$
diagonals contains $\ge 95\%$ of the total MI mass. $k_{95}$ depends
on the ordering. We report it under Fiedler reordering of the MI
adjacency graph (sort by the second eigenvector of the graph
Laplacian), the standard spectral heuristic in DMRG-orbital-ordering
practice. Fiedler is not provably optimal but is empirically
near-optimal and serves as the canonical proxy for the bandwidth
that the best 1D ordering would achieve. For [Fe$_4$S$_4$], Fiedler
reordering still leaves a half-bandwidth of $k_{95} = 15$.

\begin{figure}[!t]
\includegraphics[width=\columnwidth]{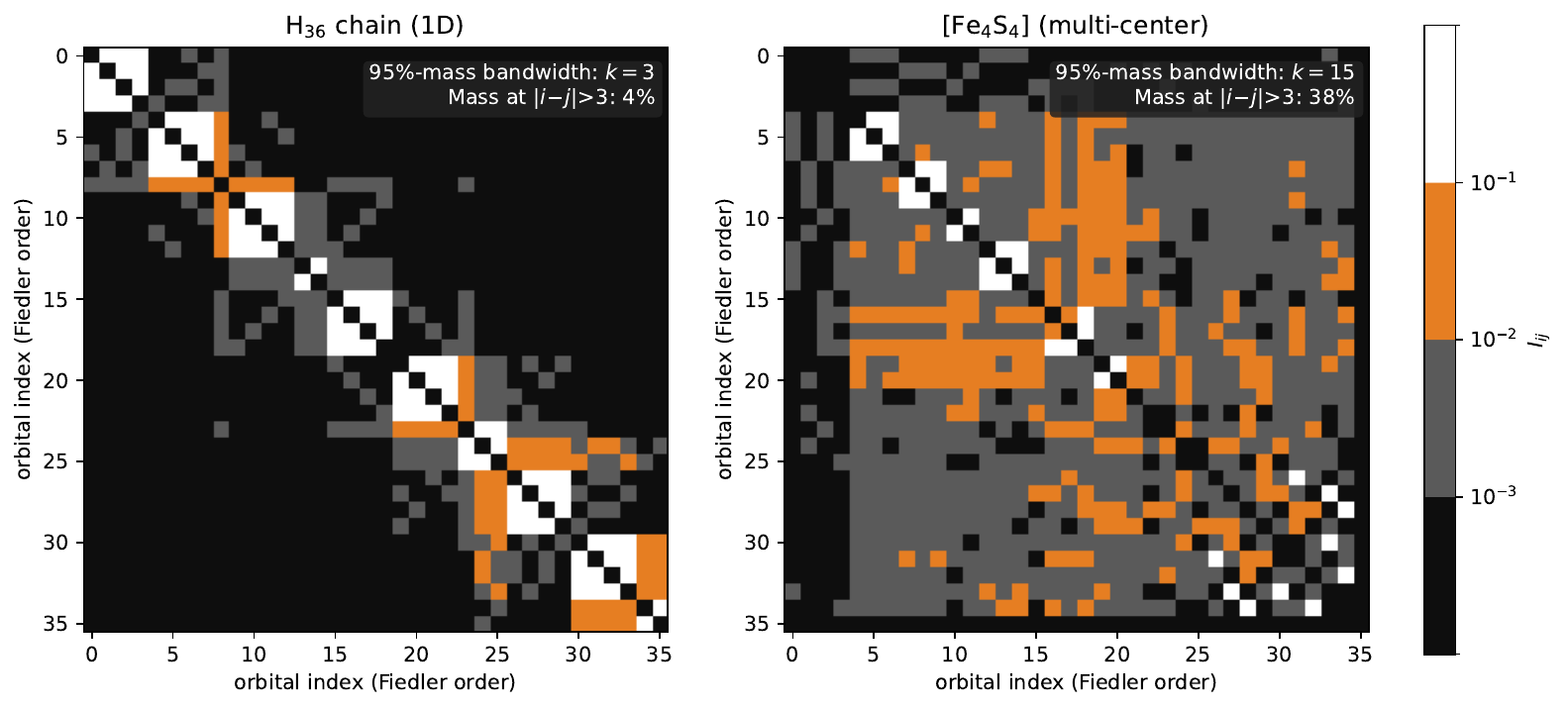}
\caption{Fiedler-reordered orbital mutual information for the
H$_{36}$ 1D chain (left) and \textrm{[Fe$_4$S$_4$]} multi-center
cluster (right), each at $N_{\rm det} = 2 \times 10^{6}$ in the
TrimCI~+~COO basis. Same color tiers and
boundaries as Fig.~5(a) of the main text. Annotations report the
95\%-mass bandwidth $k_{95}$ defined in the text and the fraction of
MI mass remaining beyond $|i-j|\!=\!3$.}
\label{fig:fiedler_compare}
\end{figure}

\textbf{Direct comparison with a 1D system: \textrm{[Fe$_4$S$_4$]}
has a $5\times$ wider Fiedler bandwidth than H$_{36}$
($k_{95} = 15$ vs.\ $3$).} To put this in context we ran the
identical TrimCI~+~COO pipeline (Phase~0 discovery on a
100-determinant core, Phase~1 refinement to $10^{6}$ dets with
orbital rotation enabled, Phase~2 frozen-orbital expansion to
$2 \times 10^{6}$ dets) on an H$_{36}$ 1D chain at 1.5~\AA{}
spacing in the STO-3G basis (36e, 36o), and computed the
von Neumann MI on the resulting Phase~2 round-0
wavefunction.

After Fiedler reordering, the two systems show qualitatively
different structure (Fig.~\ref{fig:fiedler_compare}):
\begin{itemize}
\item \textbf{H$_{36}$ (1D chain).} $k_{95} = 3$; only 4.1\% of
  the MI mass lies at $|i-j| > 3$. The matrix collapses into a
  chain of $\sim 5 \times 5$ near-block-diagonal cells, the
  canonical signature of a system an MPS can compress with small
  bond dimension.
\item \textbf{[Fe$_4$S$_4$] (multi-center cluster).} $k_{95} = 15$;
  $37.8\%$ of the MI mass remains at $|i-j| > 3$. No 1D ordering
  condenses the entanglement into a narrow band, so any MPS
  ansatz pays a substantially larger bond-dimension cost to
  represent the wavefunction.
\end{itemize}
The 5$\times$ bandwidth ratio quantifies the qualitative
difference between 1D and multi-center systems that motivates the
regime where TrimCI~+~COO outperforms MPS-based approaches, as
explored in the main text. Restricting the [Fe$_4$S$_4$]
wavefunction to its top $10^{4}$ or $10^{5}$ determinants by
$|c_I|^{2}$ leaves these conclusions unchanged: the number of
strong pairs ($I_{ij}>0.1$) is identical ($18$), the Fiedler-ordered
bandwidth shifts by one ($k_{95}=15$ at $2 \times 10^{6}$;
$k_{95}=16$ at $10^{5}$ and $10^{4}$), and the fraction of MI mass
at $|i-j|>3$ stays in the range $37$--$40\%$.

\section{Multi-center excitation analysis}
\label{sec:multicenter}

\begin{figure}[!t]
\centering
\includegraphics[width=\textwidth]{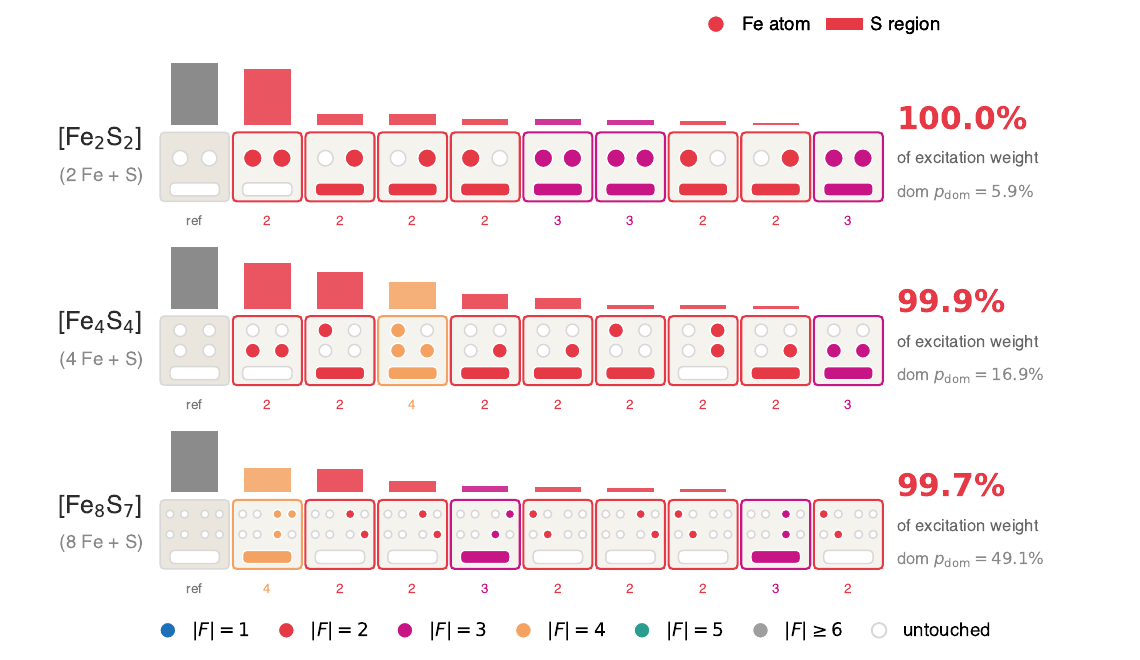}
\caption{Excitation pattern of the top high-weight determinants
for each iron-sulfur cluster. Each tile is a single determinant: Fe
atoms are drawn as circles and the combined S region as a pill below
them; an Fe atom is filled if its occupation in this determinant
differs from the dominant determinant (\textsc{ref}, the leftmost
tile in each row), and the S pill is filled if the excitation
touches \emph{any} non-Fe orbital. The colour encodes the number of
distinct centers touched,
$|\mathcal{F}(I)|=k$ with $\mathcal{C}=\{\mathrm{Fe}_1,\ldots,\mathrm{Fe}_{n_{\rm Fe}},\mathrm{S}\}$
(blue/red/magenta/orange/teal for $k=1,2,3,4,5$). The bar above
each tile is the weight $|c_I|^2$ on a logarithmic scale,
normalized so the dominant det fills the bar and all displayed
dets remain visible. Tiles are ordered by descending weight from
left to right. The bold percentage on the right is the
multi-center ($|F|\!\geq\!2$) fraction of the excitation weight;
the dominant-det weight $p_{\rm dom}$ is given just below.}
\label{fig:multicenter_distribution}
\end{figure}

\textbf{The excitation weight is overwhelmingly multi-center across
the iron-sulfur series ($99.7$--$100.0\%$).}
Sec.~\ref{sec:strict_mi} reached the multi-center conclusion
through the orbital mutual information matrix---a two-orbital
observable averaged over all determinants. Here we examine the
same question from a complementary angle, the excitation side,
asking of every high-weight determinant: \emph{how many distinct
atomic centers of the iron-sulfur cluster does this excitation
touch?} A determinant that flips electrons within a single Fe
atom's d-shell touches one center; a spin-exchange between two Fe
atoms touches two; an electron transfer between an Fe atom and
the S subsystem touches two; a single excitation that simultaneously
moves electrons across three Fe atoms and the S region touches
four; and so on.

\emph{Defining the centers.} For each iron-sulfur cluster we take
the set of centers to be the individual Fe atoms together with one
combined ``S region'' that covers the entire S subsystem,
\[
  \mathcal{C} = \{\mathrm{Fe}_1,\,\mathrm{Fe}_2,\,\ldots,\,\mathrm{Fe}_{n_{\rm Fe}},\,\mathrm{S}\},
  \qquad K \equiv |\mathcal{C}| = n_{\rm Fe}+1.
\]
We do not subdivide the S sublattice into individual sulfur atoms:
this is a coarse-graining that focuses on the Fe centers, where
the strong correlation lives, and ignores the internal structure
of the S region.

Let $\mathcal{F}(I) \subseteq \mathcal{C}$ denote the set of centers
\emph{touched} by determinant $|I\rangle$---those whose occupation
in $|I\rangle$ differs from that in the dominant determinant
$|\Psi_0\rangle = \arg\max_I |c_I|^2$. The multi-center count
$|\mathcal{F}(I)|$ takes values $0,\,1,\,\ldots,\,K$, with
$|\mathcal{F}(I)|=K$ meaning the excitation touches every Fe atom
\emph{and} the S region simultaneously.

\textbf{Methodology.} Given the wavefunction
$|\Psi_{\rm var}\rangle = \sum_I c_I |I\rangle$ in the COO basis,
the rotation $U_{\rm tot}$ from the LMO basis (where Fe-localization
is unambiguous) is recovered as in Sec.~\ref{sec:strict_mi}. The
classification proceeds in two steps.

\emph{Step 1 (orbital $\to$ center label, once per system).}
Each COO orbital expands as
$|p\rangle = \sum_i U_{{\rm tot},ip}|i\rangle$
in the LMO basis $\{|i\rangle\}$. We partition the LMO indices
into one set per center: every Fe atom $k$ owns a fixed set of
five $d$-LMOs $\mathrm{Fe}_k^{\rm LMO}$, and the remaining LMOs
form the S center's set $\mathrm{S}^{\rm LMO}$. The projected
weight of $p$ on a candidate center $c \in \mathcal{C}$ is
\[
w_c(p) \;=\; \sum_{i \in c^{\rm LMO}} |U_{{\rm tot},ip}|^2,
\]
and $p$ is labelled by the Fe of maximal $w_{\mathrm{Fe}_k}(p)$
when that weight exceeds both $0.4$ and $w_{\rm S}(p)$; otherwise
$p$ is labelled S. This is the same rule used in
Sec.~\ref{sec:strict_mi}.

\emph{Step 2 (det $\to$ touched-center set).} We restrict the
analysis to the top $10^4$ determinants by $|c_I|$, which capture
most of the wavefunction weight. Each top-$10^4$ determinant
$|I\rangle$ is an excitation relative to the dominant
$|\Psi_0\rangle$: comparing orbital occupations, we collect the
centers it touches into $\mathcal{F}(I)$, and count
$|\mathcal{F}(I)|$.

\emph{Fe-LMO indices per system.} All orbital indices in this
paper start from $0$. Note that the P-cluster reference
\cite{Li2019Pcluster} lists its active-space orbitals starting
from $1$, so we subtract $1$ when transcribing them. For
\textrm{[Fe$_4$S$_4$]} we use the partition
$\{[2{:}6], [7{:}11], [24{:}28], [29{:}33]\}$ already given in
Sec.~\ref{sec:strict_mi}. For the other two systems:
\begin{itemize}
\item \textrm{[Fe$_2$S$_2$]} (CAS 30e, 20o; LMO basis and Fe-d
  partition from Ref.~\cite{li2017fe4s4github}):
  \[
    \mathrm{Fe}_1^{\rm LMO} = \{2,3,4,5,6\}, \qquad
    \mathrm{Fe}_2^{\rm LMO} = \{13,14,15,16,17\}.
  \]
\item P-cluster \textrm{[Fe$_8$S$_7$]} (CAS 114e, 73o; LMO basis
  and Fe-d partition from Ref.~\cite{Li2019Pcluster}, converted
  from 1-based to 0-based as noted above):
  \begin{align*}
    \mathrm{Fe}_1^{\rm LMO} &= \{64,65,66,67,68\},     &
    \mathrm{Fe}_5^{\rm LMO} &= \{23,24,25,26,27\},     \\
    \mathrm{Fe}_2^{\rm LMO} &= \{45,46,47,48,49\},     &
    \mathrm{Fe}_6^{\rm LMO} &= \{28,29,30,31,32\},     \\
    \mathrm{Fe}_3^{\rm LMO} &= \{40,41,42,43,44\},     &
    \mathrm{Fe}_7^{\rm LMO} &= \{15,16,17,18,19\},     \\
    \mathrm{Fe}_4^{\rm LMO} &= \{59,60,61,62,63\},     &
    \mathrm{Fe}_8^{\rm LMO} &= \{3,4,5,6,7\}.
  \end{align*}
\end{itemize}
In all three systems, $\mathrm{S}^{\rm LMO}$ is the orthogonal
complement: every LMO not in any $\mathrm{Fe}_k^{\rm LMO}$. For
the P-cluster this comprises the $21$ S-3p plus $12$ peripheral
LMOs of Ref.~\cite{Li2019Pcluster}.

\textbf{Result for \textrm{[Fe$_{4}$S$_{4}$]}.} On the Phase~2 round-0
wavefunction ($N_{\rm det} = 2 \times 10^6$, $E = -327.231$~Ha; same
snapshot as the MI analysis of Sec.~\ref{sec:strict_mi}), the
distribution over $|\mathcal{F}(I)|$ is reported in
Table~\ref{tab:multicenter_fe4s4}: almost all the excitation
weight ($99.93\%$) lies on excitations that simultaneously
involve two or more atomic centers.
\begin{table}[h]
\centering
\begin{ruledtabular}
\begin{tabular}{c r r r}
\multirow{2}{*}{$|\mathcal{F}(I)|$ (centers touched)}
  & \multirow{2}{*}{\% of dets}
  & \multirow{2}{*}{\% of weight}
  & \% of excitation \\
  &  &  & weight \\
\midrule
0 (dominant only)        & $0.01\%$  & $16.94\%$ & ---       \\
1 (single-center)        & $0.04\%$  & $0.06\%$  & $0.07\%$  \\
2                        & $7.47\%$  & $33.85\%$ & $40.75\%$ \\
3                        & $27.84\%$ & $21.10\%$ & $25.40\%$ \\
4                        & $44.28\%$ & $20.84\%$ & $25.09\%$ \\
5 (every Fe + S)         & $20.36\%$ & $7.22\%$  & $8.69\%$  \\
\midrule
$\geq 2$ (multi-center)  & $\mathbf{99.95\%}$ & $\mathbf{83.01\%}$ & $\mathbf{99.93\%}$ \\
\end{tabular}
\end{ruledtabular}
\caption{Distribution of high-weight \textrm{[Fe$_{4}$S$_{4}$]}
determinants by the number of cluster centers
$|\mathcal{F}(I)|$ ($n_{\rm Fe}=4$ Fe atoms plus one combined S
region) involved in the excitation relative to the dominant
configuration. All three number columns sum to $100\%$ over their
filled rows: the first two normalize by the table total (count or
weight); the third normalizes by the total excitation weight, which
removes the dominant determinant from the denominator (hence the
dominant row carries no entry). Almost all the excitation weight
($99.93\%$) lies on excitations that simultaneously involve two
or more atomic centers.}
\label{tab:multicenter_fe4s4}
\end{table}

\textbf{Same conclusion across the iron-sulfur series.} We repeat
the classification for \textrm{[Fe$_2$S$_2$]} (30e, 20o, 2 Fe;
Phase~2 round-0, $N_{\rm det} = 2\!\times\!10^6$) and the P-cluster
\textrm{[Fe$_8$S$_7$]} (114e, 73o, 8 Fe; Phase~2 round-5,
$N_{\rm det} = 6.4\!\times\!10^7$, the same checkpoint as main-text
Fig.~3). Multi-center dominance is not an
\textrm{[Fe$_4$S$_4$]}-specific accident
(Fig.~\ref{fig:multicenter_distribution} and
Table~\ref{tab:multicenter_series}): of the excitation weight,
$\mathbf{100.0\%}$ (\textrm{[Fe$_2$S$_2$]}), $\mathbf{99.9\%}$
(\textrm{[Fe$_4$S$_4$]}), and $\mathbf{99.7\%}$ (P-cluster) lies on
multi-center excitations. Equivalently, the single-center weight
(an excitation localized either to a single Fe atom or to the S
region only) is at most $\sim\!1\%$ of the excitation weight in
every system, so any 1D ordering of the orbitals must visit
excitations whose support is spatially cluster-wide.

\clearpage
\begin{table*}[!t]
\centering
\begin{ruledtabular}
\begin{tabular}{l c c c c c c c c c c c}
$|\mathcal{F}(I)|$ & & 0 & 1 & 2 & 3 & 4 & 5 & 6 & 7 & 8 & 9 \\
\midrule
\multirow{2}{*}{\shortstack[l]{\textrm{[Fe$_{2}$S$_{2}$]}\\(2 Fe + S, $K=3$)}}
  & \%\,dets   & $0.01$  & $0.00$ & $9.81$  & $90.18$ & ---     & ---     & ---    & ---    & ---    & ---    \\
  & \%\,weight & $5.93$  & $0.00$ & $39.32$ & $54.75$ & ---     & ---     & ---    & ---    & ---    & ---    \\
\midrule
\multirow{2}{*}{\shortstack[l]{\textrm{[Fe$_{4}$S$_{4}$]}\\(4 Fe + S, $K=5$)}}
  & \%\,dets   & $0.01$  & $0.04$ & $7.47$  & $27.84$ & $44.28$ & $20.36$ & ---    & ---    & ---    & ---    \\
  & \%\,weight & $16.94$ & $0.06$ & $33.85$ & $21.10$ & $20.84$ & $7.22$  & ---    & ---    & ---    & ---    \\
\midrule
\multirow{2}{*}{\shortstack[l]{P-cluster\\(8 Fe + S, $K=9$)}}
  & \%\,dets   & $0.01$  & $0.13$ & $18.64$ & $32.99$ & $28.24$ & $15.91$ & $3.20$ & $0.88$ & $0.00$ & $0.00$ \\
  & \%\,weight & $49.10$ & $0.15$ & $25.85$ & $11.84$ & $9.45$  & $2.87$  & $0.65$ & $0.07$ & $0.00$ & $0.00$ \\
\midrule
\multicolumn{3}{l}{\textbf{$|F|\geq 2$, \% of total weight:}}
  & \multicolumn{3}{c}{Fe$_{2}$S$_{2}$: $94.07$}
  & \multicolumn{3}{c}{Fe$_{4}$S$_{4}$: $83.01$}
  & \multicolumn{3}{c}{P-cluster: $50.74$} \\
\multicolumn{3}{l}{\textbf{\hspace{1em}\% of excitation weight:}}
  & \multicolumn{3}{c}{$\mathbf{100.00}$}
  & \multicolumn{3}{c}{$\mathbf{99.93}$}
  & \multicolumn{3}{c}{$\mathbf{99.70}$} \\
\end{tabular}
\end{ruledtabular}
\caption{Distribution of high-weight determinants by the number
of cluster centers $|\mathcal{F}(I)|$ touched in the excitation,
for three iron-sulfur systems (same population as
Fig.~\ref{fig:multicenter_distribution}). Centers are the
$n_{\rm Fe}$ Fe atoms plus one combined S region, so
$|\mathcal{F}(I)|$ runs from $0$ (dominant determinant) to
$K = n_{\rm Fe} + 1$. Wavefunction snapshots:
\textrm{[Fe$_2$S$_2$]} and \textrm{[Fe$_4$S$_4$]} at
$N_{\rm det} = 2\!\times\!10^6$ (Phase~2 round-0); P-cluster at
$N_{\rm det} = 6.4\!\times\!10^7$ (Phase~2 round-5). The bottom
block reports the $|F|\geq 2$ weight as a fraction of the table
total (``of total'') and of the total excitation weight (``of
excitation weight''); the latter does not depend on the dominant
weight $p_{\rm dom}$ (which varies strongly across the series) and
sits at $99.70$--$100.00\%$ in all three systems.}
\label{tab:multicenter_series}
\end{table*}

\section{Hubbard-on-graph details}
\label{sec:hubbard_graph}

\textit{Motivation.} The iron-sulfur analyses of
Sec.~\ref{sec:strict_mi} and Sec.~\ref{sec:multicenter} establish
that the correlation in [Fe$_{n}$S$_{m}$] is
\emph{multi-center}: the strongly correlated entanglement is
spread across many atomic centers and cannot be localized along
a 1D path. To probe how the \emph{degree} of multi-center
character affects the parameter efficiency of selected CI versus
MPS, we introduce a controllable toy model in which connectivity
between centers is the only knob that varies.

We choose the Hubbard model at half-filling because each site
naturally plays the role of a center, the on-site interaction
$U$ supplies the strong-correlation physics, and its localized
site basis is the natural setting for multi-center analysis.
The multi-center degree is then tuned by adding
non-nearest-neighbour hopping with strength controlled by a
single parameter $\alpha\!\in\![0,1]$, on top of the bare
nearest-neighbour hopping $t$: at $\alpha\!=\!0$ each site only
couples to its 1D-chain neighbours; at $\alpha\!=\!1$ every site
couples to every other (the fully-connected, maximally
multi-center limit); the chain length $L$ (number of sites),
the ratio $U/t$, and filling are held fixed. The non-nn hopping
amplitudes $r_{ij}\!\sim\!\mathcal{U}[0.5,1.5]$ are independent
random variables that break the permutation symmetry of the
fully-connected graph, a realistic feature of molecular
clusters. With $L\!=\!8$ sites the FCI ground-state energy is
exactly reachable at every $\alpha$ (DMRG at $D\!=\!256$
converges to FCI), so every parameter-cost comparison below is
calibrated against the exact ground state.

\textit{Model definition.} The model is defined on $L\!=\!8$
sites at half-filling ($N_e\!=\!8$) with $U/t\!=\!4$:
\begin{equation}
\hat{H} = -t \sum_{\langle i,j \rangle, \sigma}
\hat{c}_{i\sigma}^{\dagger} \hat{c}_{j\sigma}
\;-\; \alpha\, t \sum_{\substack{i<j \\ \text{non-nn}}} \sum_{\sigma}
r_{ij}\left(\hat{c}_{i\sigma}^{\dagger} \hat{c}_{j\sigma} + \mathrm{h.c.}\right)
\;+\; U \sum_i \hat{n}_{i\uparrow} \hat{n}_{i\downarrow}\,.
\label{eq:hubbard_graph}
\end{equation}
The nearest-neighbour pairs $\langle i,j\rangle$ are those of an
open 1D chain. Non-nn hopping amplitudes $r_{ij}$ are drawn
independently from $\mathcal{U}[0.5, 1.5]$ and held fixed across
all~$\alpha$, leaving $\alpha$ as the single topology parameter.

\textit{Topology sensitivity.} \textbf{Sweeping $\alpha$ from
$0$ (1D chain) to $1$ (fully-connected graph) increases the
$N_{\rm DMRG}/N_{\rm COO}$ parameter ratio from
$\sim\!1\times$ to $\sim\!12\times$ at matched accuracy.}
Table~\ref{tab:topology_scan} reports the parameter count
required to reach $\Delta E < 0.1\,t$ from the FCI ground state
at four representative $\alpha$ values, the same threshold and
data underlying Fig.~5(f,g) of the main text. Both DMRG and
TrimCI are run in $S_z$ mode (block2 \texttt{SymmetryTypes.SZ}
for DMRG; TrimCI does not enforce SU(2)), so the comparison is
not biased by spin-symmetry adaptation.

\begin{table}[h]
\caption{Parameter count for matched accuracy
$\Delta E < 0.1\,t$ on the $L\!=\!8$ Hubbard-on-graph model,
sweeping the topology knob $\alpha$. ``DMRG'' is the smaller of
the natural-ordering and Fiedler-reordering bond-dimension scans
in block2 ($S_z$ mode). ``TrimCI (no-COO)'' is TrimCI in the
site basis; ``TrimCI (COO)'' is TrimCI with per-$\alpha$
orbital optimization. Each entry is the log--log interpolated
$N$ at which the method's convergence trajectory crosses
$\Delta E = 0.1\,t$.}
\label{tab:topology_scan}
\centering
\begin{tabular}{c r r r r r}
\toprule
$\alpha$ & DMRG & TrimCI (no-COO) & TrimCI (COO) &
  $\dfrac{N_{\rm DMRG}}{N_{\rm COO}}$ &
  $\dfrac{N_{\rm noCOO}}{N_{\rm COO}}$ \\
\midrule
$0.0$ (1D chain)        & $1{,}000$ & $1{,}246$ & $802$  & $1.25\times$  & $1.55\times$ \\
$0.4$                    & $2{,}198$ & $1{,}850$ & $1{,}050$ & $2.09\times$  & $1.76\times$ \\
$0.7$                    & $6{,}130$ & $1{,}982$ & $1{,}065$ & $5.76\times$  & $1.86\times$ \\
$1.0$ (fully-connected graph)   & $5{,}416$ & $1{,}546$ & $472$  & $11.46\times$ & $3.27\times$ \\
\bottomrule
\end{tabular}
\end{table}

The same picture in energy units: at fixed parameter count
(Table~\ref{tab:fixed_N_energy}), TrimCI (COO) sits closer to
FCI than either alternative across the topology scan, and the
gap widens at high $\alpha$. At $D\!=\!5$ ($\sim 800$ MPS
parameters), TrimCI (COO) is already $\sim\!7\times$ closer to
FCI than DMRG at the fully-connected end; at $D\!=\!10$
($\sim 3{,}200$ parameters) TrimCI (COO) is essentially
converged ($\Delta E \sim 10^{-3}\,t$) while DMRG still carries
$170\!\times\!10^{-3}\,t$ of error.

\begin{table}[h]
\caption{Energy gap to FCI, $\Delta E\!=\!E\!-\!E_{\rm FCI}$, in
units of $10^{-3}\,t$, at two fixed parameter counts. The DMRG
columns use bond dimensions $D\!=\!5$ and $D\!=\!10$
(corresponding to $4LD^{2}\!=\!800$ and $3{,}200$ parameters);
TrimCI columns are log-linearly interpolated to the same $N$
from the per-$\alpha$ convergence trajectory.}
\label{tab:fixed_N_energy}
\centering
\begin{tabular}{c|r r r|r r r}
\toprule
& \multicolumn{3}{c|}{$N\!=\!800$ ($D\!=\!5$ for DMRG)}
& \multicolumn{3}{c}{$N\!=\!3{,}200$ ($D\!=\!10$ for DMRG)} \\
$\alpha$ & TrimCI(COO) & no-COO & DMRG & TrimCI(COO) & no-COO & DMRG \\
\midrule
$0.0$  & $101$ & $228$ & $155$ & $0.1$ & $1.9$ & $10$  \\
$0.5$  & $143$ & $286$ & $285$ & $0.9$ & $20$  & $108$ \\
$1.0$  & $57$  & $254$ & $420$ & $1.6$ & $31$  & $170$ \\
\bottomrule
\end{tabular}
\end{table}

DMRG and TrimCI respond to topology in opposite ways. DMRG's
bond dimension must carry every correlation that crosses the 1D
path traversed by the MPS through the orbitals; in the
fully-connected limit every site couples to every other site,
so any 1D ordering necessarily cuts through many
strongly-correlated bonds, and the bond dimension required for
fixed accuracy grows accordingly. A TrimCI expansion has no
such 1D constraint --- it ranks determinants by importance,
independently of any spatial ordering --- and is therefore
nearly insensitive to whether the underlying graph is a chain
or fully connected: across the topology scan, the TrimCI (COO)
count stays in the range $\sim\!500$--$1{,}300$, while DMRG
grows from $\sim\!10^{3}$ to $\sim\!7\!\times\!10^{3}$. The \textrm{[Fe$_4$S$_4$]}
measured ratio ($N_{\rm DMRG}/N_{\rm COO} \approx 8$--$10\times$)
lies between the topology-scan values at $\alpha\!=\!0.7$ and
$\alpha\!=\!1.0$, consistent with the multi-center 4-Fe coupling
geometry sitting closer to the fully-connected end of the scan
than to the chain end.

The COO contribution itself
($N_{\rm noCOO}/N_{\rm COO}\approx 1.3$--$3.3\times$ across the
$\alpha$-scan) is small at low $\alpha$ because the site basis
is already a reasonable localized basis for the Hubbard model,
and grows toward $3.3\times$ at the fully-connected end where
orbital optimization extracts more compression. The orbital and
ansatz factors are comparable across the scan (cf.~Fig.~5(g) of
the main text); both contribute meaningfully to the total
DMRG/COO ratio. On molecular systems, where the starting HF or
LMO basis is far from optimal, COO is expected to contribute
more strongly.

\textit{Numerical details.}
\textbf{DMRG.} block2 in $S_z$ mode; bond dimensions
$D \in \{5, 10, 15, 20, 30, 50, 80, 120, 200, 256\}$, with both
natural (chain) and Fiedler orbital orderings; we report the
smaller parameter count of the two.
\textbf{TrimCI.} Phase~0 discovery on a 100-det core followed
by Phase~1 expansion to $N_{\rm det} = 4000$ with
growth factor $1.1$. The ``COO'' variant performs orbital
optimization (BFGS) in both Phase~0 and Phase~1; the ``no-COO''
control uses identical settings with orbital optimization
disabled in both phases. Phase~2 is skipped because the
FCI space is small enough ($\binom{8}{4}^{2}=4{,}900$) that
Phase~1 saturates it.
\textbf{Threshold.} The crossing $\Delta E\!=\!0.1\,t$ is
located by log--log interpolation of each method's $(N_k,
\Delta E_k)$ trajectory.
\textbf{FCI reference energies.} Table~\ref{tab:hubbard_fci}
lists $E_{\rm FCI}/t$ at every $\alpha$, the reference against
which all $\Delta E$ values in Fig.~5(c,d,e) of the main text are
measured.

\begin{table}[h]
\caption{Exact ground-state energies of the Hubbard-on-graph
model in units of $t$. The energy is non-monotonic at small
$\alpha$ due to disorder, then drops monotonically as more
hopping channels open at larger $\alpha$.}
\label{tab:hubbard_fci}
\centering
\begin{tabular}{cc}
\toprule
$\alpha$ & $E_{\rm FCI}\,/\,t$ \\
\midrule
0.0 & $-4.23581$ \\
0.1 & $-4.16583$ \\
0.2 & $-4.13544$ \\
0.3 & $-4.14472$ \\
0.4 & $-4.19532$ \\
0.5 & $-4.29052$ \\
0.6 & $-4.43660$ \\
0.7 & $-4.65420$ \\
0.8 & $-5.07920$ \\
0.9 & $-5.70051$ \\
1.0 & $-6.45590$ \\
\bottomrule
\end{tabular}
\end{table}

\section{Distributed Davidson via mini-task bundles: two-axis ($K\!\times\!Z$) scalability}
\label{sec:gpu_davidson_5120m}

\textit{Motivation.} At
$N_{\rm det}=5.12\!\times\!10^{9}$ --- the largest point of
main-text Fig.~2, $E=-327.2421555$~Ha, and the largest reported
variational selected-CI calculation --- the variational Davidson
loop is beyond the reach of any single computational node. The
dominant cost is solving the lowest eigenpair $(E_{\rm var}, v)$
of the CI Hamiltonian by Davidson iteration, and within each
Davidson iteration the matrix-vector product $\sigma = \hat{H}v$
dominates everything else by orders of magnitude.

The trial vector alone is $\sim\!41$\,GB
($N_{\rm det}\!\times\!8$~bytes per double-precision
coefficient). The Davidson subspace $\{V_k, HV_k\}_{k=1}^{m}$
at the typical depth $m\!=\!8$ adds another $\sim\!660$\,GB.
One matvec costs $\sim\!10^{16}$ multiply-add and auxiliary
bit operations.
This back-of-envelope follows from $\sim\!10^{10}$ determinants
$\times\!\sim\!10^{5}$ Slater--Condon connections each (singles +
doubles) $\times\!\sim\!10$ ops per connection (sign + integral
lookup + accumulate).
For context, a server CPU node delivers $\sim\!10^{12}$ FLOP/s
and a single data-center GPU card $\sim\!10^{13}$ FLOP/s in
double precision; the sparse, memory-bound access pattern of a
selected-CI matvec typically runs at $\lesssim\!10\%$ of peak,
so a single GPU card needs hours per matvec. Davidson at
$m\!=\!8$ takes $\sim\!8$ matvecs per outer loop, putting
single-GPU-card wall time at $\sim\!1$~day and a single CPU node
into the $\sim\!10$-day regime; and a single card cannot hold
the $\sim\!700$\,GB Davidson state in any case.
\emph{This calculation must be parallelised across multiple
nodes and multiple GPU cards.}

For the calculation to finish in reasonable time we need many
GPU cards working in parallel. Rather than treating this as a
demand on a dedicated HPC allocation --- which not every research
group has on hand --- we deliberately target \emph{opportunistic
backfill}: idle GPU capacity on shared academic clusters, which
is widely available, free at the point of use, and otherwise
wasted. Backfill is not a fallback; we treat it as a primary
compute source, because doing so opens billion-determinant
variational CI to any group with access to such platforms, not
only to those with dedicated GPU time. Looking forward, the
same design philosophy points to a longer-term ambition: a
single computation distributed across a planet-scale,
heterogeneous compute network --- pooling idle GPU capacity
across institutions and vendors --- as a path to computations
that exceed any single facility's reach. The scientific payoff
is concrete: ultra-large-scale variational calculations are
what we need to attack the strongly correlated electronic
structure problems on which classical methods still face
severe scaling and resource barriers ---
the FeMo-cofactor of nitrogenase and the catalytic cycle of
biological N$_2$ fixation, the manganese-oxygen complex of
photosystem~II and water oxidation, multi-iron cytochromes and
biological electron transport, and other multi-metal active
sites in catalysis, energy storage, and condensed-matter
materials.

Such an opportunistic, heterogeneous compute pool --- whether
hosted on one cluster or pooled across many --- comes with two
structural constraints that any viable architecture must
absorb. First, the pool is \emph{transient}: workers join,
leave, and fail on timescales of minutes to hours, with no
persistent inter-node interconnect, so algorithms that require
synchronous, fixed-size allocations simply cannot run. Second,
the pool is \emph{heterogeneous} --- T4, A100, H100, H200, and L40S
running simultaneously in our current deployment, with VRAM
$16$--$140$\,GB and per-card compute differing by
$10$--$100\times$, and even greater diversity expected as the
pool extends across institutions and vendors. A single program
optimized for one specific GPU model carries baked-in assumptions
about memory size, layout, and arithmetic throughput; on a
different model it either fails outright or runs at a small
fraction of peak, throwing away most of the available capacity.
Together with the multi-node multi-card parallelism required
for compute and memory, these constraints call for an
architecture that absorbs all of them at once and remains
scalable as $N_{\rm det}$ continues to grow.

We meet these requirements with a single design choice: factor
the matvec into many small, stateless \textbf{mini-task bundles}
that any GPU can execute independently, and let the system scale
along two orthogonal axes --- \emph{workers}, which consume
bundles, and \emph{factories}, which produce and aggregate them.
A worker is a stateless GPU process that pulls bundles, runs
them, and posts the results back; since each bundle carries its
own context, any GPU in the pool can run any bundle, and
workers can come or go without breaking correctness. A
\emph{factory} owns a contiguous determinant-index range, hence
the same row slice of $v$, $\sigma$, $V_k$, and $HV_k$; it hands
out bundles whose destination $\sigma$ rows lie in that range
and accumulates returned $\sigma$ contributions for that same
range. Adding workers ($Z$ per factory) raises compute throughput;
adding factories ($K$) shards the global state across more
nodes; the total worker pool is $K\!\times\!Z$. The $5.12$B run
sits at $(K\!=\!1, Z\!\approx\!20)$, and reaching $10^{12}$
amounts to deploying the same architecture at
$(K\!\approx\!200, Z\!\approx\!20)$, i.e.\ a
$K\!\times\!Z\!\approx\!4{,}000$ pool
(Sec.~\ref{sec:s8_path}).

Sec.~\ref{sec:s8_bundle} defines the bundle and its four
properties; Sec.~\ref{sec:s8_scaling} shows how those properties
yield the $K\!\times\!Z$ scaling; Sec.~\ref{sec:s8_matvec}
explains how the matvec inside a bundle turns determinant-level
connection finding into channel-specific searches over existing
$\alpha$ or $\beta$ groups;
Secs.~\ref{sec:s8_deployment}--\ref{sec:s8_checkpoint} report
the specific $(1, 20)$ deployment together with its out-of-core
Krylov and Ritz-checkpoint mechanisms; Sec.~\ref{sec:s8_path}
discusses how to apply the same architecture to reach
$10^{12}$ determinants.

\subsection{The mini-task bundle}
\label{sec:s8_bundle}

\textbf{The matvec decomposes into three nested units.
\emph{Channels} are the smallest physics-defined pieces of
work --- one per row per excitation type. \emph{Mini-tasks}
pack same-type channels into fixed-size task units.
\emph{Bundles} pack mini-tasks into hardware-tuned parcels for
GPU dispatch. Four bundle properties (small, stateless,
fixed-size, output-routable) drive the $K\!\times\!Z$ scaling.}

The Davidson inner loop is dominated by one matrix-vector
product (matvec) per iteration,
\begin{equation}
\sigma_i \;=\; \sum_{j} H_{ij}\, v_j,
\qquad i = 1, \ldots, N_{\rm det},
\label{eq:matvec_basic}
\end{equation}
where $H$ is the $N_{\rm det}\!\times\!N_{\rm det}$ sparse CI
Hamiltonian and $v$ the length-$N_{\rm det}$ trial vector. At
$N_{\rm det}\!=\!5.12\!\times\!10^{9}$ ($5.12$B), one matvec
costs $\sim\!10^{16}$ floating-point operations
(Sec.~\ref{sec:gpu_davidson_5120m} motivation), far beyond any
single node.

The key design choice is to decompose the Hamiltonian action
into determinant-row connections, not orbital blocks: each
contribution has a source determinant $j$, whose coefficient
$v_j$ is read, and a destination determinant $i$, whose
$\sigma_i$ is incremented. A worker therefore reads only the
$v$ entries referenced by its assigned mini-tasks and returns
sparse $(i,\Delta\sigma_i)$ updates. The decomposition exploits
the spin-bitstring structure of the determinant set. Each
$|D_i\rangle\!=\!|\alpha_i\rangle\!\otimes\!|\beta_i\rangle$,
and we sort the dets so that those sharing the same
$\alpha$-bitstring form a contiguous \emph{$\alpha$-group}
(likewise \emph{$\beta$-groups}). For each row $i$,
Slater-Condon rules limit $H_{ij}\!\neq\!0$ to at most a
two-electron transition between $|D_i\rangle$ and $|D_j\rangle$,
so the non-zero $j$'s fall into three \emph{types}:
\textbf{type (a)} --- $i$ and $j$ share an $\alpha$-group, so
only $\beta$ differs (single or double $\beta$-excitation);
\textbf{type (b)} --- $i$ and $j$ share a $\beta$-group, so
only $\alpha$ differs (single or double $\alpha$-excitation);
\textbf{type (m)} --- $\alpha_j$ differs from $\alpha_i$ by a
single excitation and $\beta_j$ differs from $\beta_i$ by a
single excitation (mixed double: one excitation in each spin).

Building on this classification, we decompose each row's work
into smallest units called \emph{channels}: a channel computes
$\sum_j H_{ij} v_j$ for $j$ in one source group, with $H_{ij}$
non-zero only for single/double excitations from $i$. For each
row $i$: type (a) contributes one channel
($j\!\in$~$\alpha$-group containing $i$, $\beta$-single or
$\beta$-double); type (b) contributes one channel
($j\!\in$~$\beta$-group containing $i$, $\alpha$-single or
$\alpha$-double); and type (m) contributes $\bar c$ channels
on average, one per $\alpha$-neighbor $g'$ of $i$'s
$\alpha$-group ($j\!\in\!g'$, $\alpha$-single combined with
$\beta$-single), where $\bar c$ is the average
$\alpha$-adjacency degree. Summed over all rows, the matvec
has $N_{\rm det}(2\!+\!\bar c)$ channels in total. For the
5.12B production run, $\bar c\!\approx\!220$
(Table~\ref{tab:bundle_counts}), so type (m) dominates by two
orders of magnitude ($\sim\!10^{12}$ type-(m) channels vs.\
$\sim\!5\!\times\!10^{9}$ each for types (a) and (b)).

We partition the channels into \textbf{mini-tasks} for
dispatch: each mini-task contains around $C\!=\!10^{5}$
same-type channels. Two design choices motivate this. First,
the \emph{fixed channel count per mini-task} makes every
mini-task carry approximately the same work --- the basis for
load balancing. Second, the \emph{single-type
restriction} lets each mini-task run on one specialized GPU
kernel: each type has its own kernel matched to its data
access pattern (details in Sec.~\ref{sec:s8_matvec}), so
mixing types in one mini-task would force kernel switches
inside a single GPU call with no compensating gain.
Concretely, the builder walks through the channels type by
type --- first all type-(a) channels, then all type-(b), then
all type-(m) --- grouping every $\sim\!10^{5}$ same-type
channels into one mini-task; when a type ends, the current
mini-task closes and a fresh one starts for the next type.
The total mini-task count is
$\approx\!N_{\rm det}(2\!+\!\bar c)/C$, dominated by type (m)
(Table~\ref{tab:bundle_counts}).

A mini-task is a task-level abstraction: it knows nothing
about hardware. A \textbf{mini-task bundle} adds the
hardware-aware layer --- a fixed-size group of $B$ mini-tasks,
sized to balance two physical constraints: the per-bundle GPU
compute time must \emph{exceed} the network dispatch and
state-transfer overhead by enough to keep the GPU busy, yet
stay short enough that work distributes evenly across
heterogeneous workers. Empirically $B\!=\!243$ works well
across our worker pool (T4, A100, H100, H200, L40S): a bundle
processes $\sim\!2.4\!\times\!10^{7}$ channels and takes, on
average, several seconds of GPU compute --- comfortably
exceeding the average network dispatch and state-transfer
overhead --- while the matvec's $\sim\!5\!\times\!10^{4}$
bundles give ample granularity for balanced distribution
across workers.

\begin{table}[h]
\centering
\small
\caption{Channel, mini-task, and bundle counts measured at two
production points. Per-row counts give type-(a) and type-(b)
totals of $N_{\rm det}$ each, and type-(m) total
$N_{\rm det}\bar c$. The implied average $\alpha$-adjacency
$\bar c$ stays near $220$ at both scales, so all counts scale
essentially linearly in $N_{\rm det}$. The $5.12$B figures are
the production run discussed in Sec.~\ref{sec:s8_deployment}.}
\label{tab:bundle_counts}
\begin{tabular}{lrrrrrrr}
\toprule
Run & $N_{\rm det}$ & type (a) & type (b) & type (m)
    & mini-tasks & bundles & $\bar c$ \\
\midrule
$640$M & $6.4\!\times\!10^{8}$
       & $6.4\!\times\!10^{8}$ & $6.4\!\times\!10^{8}$
       & $1.4\!\times\!10^{11}$
       & $1.41\!\times\!10^{6}$
       & $5{,}820$ & $\sim\!220$ \\
$5.12$B & $5.12\!\times\!10^{9}$
        & $5.12\!\times\!10^{9}$ & $5.12\!\times\!10^{9}$
        & $1.13\!\times\!10^{12}$
        & $1.17\!\times\!10^{7}$
        & $48{,}157$ & $\sim\!220$ \\
\bottomrule
\end{tabular}
\end{table}

A bundle's matvec needs three input classes. The first is the
trial vector $v$. Globally, at the 5.12B point, $v$ is a
roughly $41$\,GB object, but one bundle reads only the source
groups referenced by its fixed $B\!\times\!C$ channel list.
Thus the per-bundle $v$ working set stays at the fixed bundle
scale rather than scaling with global $N_{\rm det}$; workers
fetch or cache those group slices for the current Davidson
iteration and reuse cached slices across bundles. The second is
the $\alpha$- and
$\beta$-group metadata for the groups the bundle touches
and the corresponding group-index files; these structural data
are fixed during a Davidson iteration and are not a dense,
iteration-dependent vector like $v$. The third is the integral
payload, which is fixed by the orbital basis rather than
$N_{\rm det}$; for the \textrm{[Fe$_4$S$_4$]}
36-spatial-orbital case, even a dense spatial $g_{ijkl}$ array
is only $\sim\!13$\,MB. It is downloaded once per worker and
reused for the whole job. The worker runs the matvec on its
assigned mini-tasks via the bitstring-product implementation of
Sec.~\ref{sec:s8_matvec} and posts back the $\sigma$
contributions, each tagged by its destination row index for the
factory to aggregate.

Four properties of the bundle determine the architecture.
\textbf{(P1) Small}: the per-bundle working set is well below
$1$\,GB, comfortably within any production GPU's VRAM
($16$--$140$\,GB), so the same code runs unchanged on a T4,
A100, H100, H200, or L40S. \textbf{(P2) Stateless}: bundle
$k\!+\!1$ does not depend on bundle $k$'s output, so bundles can
be computed in any order, retried, redispatched, or duplicated
without affecting correctness. \textbf{(P3) Fixed-size}: a
bundle's compute and memory cost is set by the fixed bundle
shape ($B$ mini-tasks of $C$ channels each), not by the global
$N_{\rm det}$, so growing $N_{\rm det}$ multiplies the
\emph{number} of bundles, not their individual size.
\textbf{(P4) Output-routable}: each contribution carries its
destination row index, so multiple factories can aggregate
disjoint row ranges of $\sigma$ in parallel.

Each property carries one architectural consequence: P1 makes
the worker pool hardware-agnostic, P2 makes it dynamic, P3
makes throughput linear in worker count, and P4 lets the
factory pool itself shard. Together they yield the two-axis
scaling of the next subsection.

\subsection{Two-axis scaling: \texorpdfstring{$K$}{K} factories
\texorpdfstring{$\times$}{x} \texorpdfstring{$Z$}{Z} workers}
\label{sec:s8_scaling}

\textbf{The bundle abstraction makes the per-factory worker
count $Z$ and the factory count $K$ independent scaling axes,
with $K\!\times\!Z$ the total worker pool. The architecture is
invariant in $(K, Z)$; the production $5.12$B run is the single
$(K\!=\!1, Z\!\approx\!20)$ point of this design space.}

The matvec is orchestrated by two roles: \emph{workers} consume
bundles --- pull, run, post $\sigma$ contributions --- while
\emph{factories} dispatch them and aggregate the returned
contributions.

\emph{Worker dimension ($Z$).} Bundle properties (P1)--(P3)
make the worker pool hardware-agnostic, dynamic, and linearly
scalable: any GPU can run any bundle (we mix T4, A100, H100,
H200, and L40S in one calculation), workers can join or fail without
coordination (a dropped bundle is simply redispatched), and
per-factory throughput grows linearly in $Z$ until the
dispatcher itself saturates --- around $Z\!\sim\!100$ for the
HTTP-based design, well above our operating point of
$Z\!\approx\!20$.

\emph{Factory dimension ($K$).} The factory split is a
destination-row split. Let row $r$ label determinant $D_r$ and
the CI-vector component $\sigma_r$. Factory $a$ owns the
contiguous row interval $[r_a,r_{a+1})$ in the global
determinant ordering, and hence the corresponding local slices
of $v$, $\sigma$, $V_k$, and $HV_k$. Its work queue contains
the bundles for destination rows
$i\!\in\![r_a,r_{a+1})$. A worker assigned such a bundle
computes the relevant partial products $H_{ij}v_j$ and returns
sparse pairs $(i,\Delta\sigma_i)$ for those destination rows.
The source coefficients $v_j$ may come from groups outside the
factory's row interval, but these reads are fixed-scale
per-bundle inputs determined by the bundle's channel list; they
do not define the factory split. Property (P4) keeps every
returned pair explicitly addressed by destination row, so the
owning factory can accumulate its $\sigma$ slice without
ambiguity. Adding factories shrinks the heavy per-factory state
--- $v$ slice, $\sigma$ slice, Davidson scratch, and dispatcher
queue --- by $1/K$. Thus the memory-heavy vector state is
row-sharded. The main global synchronization is the set of
Davidson dot products used to form the Rayleigh--Ritz projected
Hamiltonian, orthogonalize new trial vectors, and compute
residual and normalization norms.

For the $5.12$B run we operate at $K\!=\!1$ because the
factory state ($\sim\!660$\,GB Davidson subspace on disk plus
auxiliary, Sec.~\ref{sec:s8_ooc}) fits a single high-memory
node with $\sim\!1$\,TB RAM and $\sim\!1$\,TB local scratch,
and a single dispatcher
comfortably feeds the $\sim\!20$ workers we run. The
architecture is symmetric in $K$ by property (P4);
Sec.~\ref{sec:s8_path} unrolls the $K\!>\!1$ deployment toward
$10^{12}$.

\subsection{Inside one bundle: connection finding as a search problem}
\label{sec:s8_matvec}

\textbf{In selected-CI algorithms, the key challenge in a matvec is
to find, for each destination determinant $D_i$, the connected source
determinants $D_j$ with $H_{ij}\ne 0$. At billion-determinant scale,
memory limits make it infeasible to store the determinant-level
connection graph. A typical selected-CI procedure would generate
all possible single/double excitations of $D_i$ and check whether
each candidate determinant is
present. When $N_{\rm det}$ is very large, this candidate-generation
and lookup route becomes inefficient because most such candidates
are absent from the wavefunction. Our main innovation is to use the
relevant $\alpha$ or $\beta$ group specified by the channel, scan
only determinants already present in that group, and use cheap bit
tests to identify the connected $j$'s.}

A bundle contains multiple mini-tasks. We write each mini-task as
$(c,g,g',r_s,r_e)$, where $c$ is the channel type, $g$ is the group
id, $g'$ is an adjacent-group id used only by the mixed type, and
$[r_s,r_e)$ is a local destination-row interval in that group. Choosing a
local row index $r\in[r_s,r_e)$ gives an atomic task, i.e., one
concrete channel instance $(c,g,g',r)$. This local index $r$ fixes
the target determinant $D_i$, and the task is to find, within the source group
specified by this channel, all source determinants $D_j$ for which
$H_{ij}\ne 0$, so that their coefficients $v_j$ contribute to
$\sigma_i$. The diagonal term $H_{ii}v_i$ is handled separately
after the off-diagonal contributions are summed.

For $c=1$ (same-$\alpha$), $g$ is the destination $\alpha$ group and
$D_i=(\alpha_g,\beta_i)$. Any source must have the same
$\alpha_g$, so the kernel scans the existing determinants
$D_j=(\alpha_g,\beta_j)$ in that group and keeps only
$\operatorname{popcount}(\beta_i\oplus\beta_j)=2$ or $4$.

For $c=3$ (same-$\beta$), $g$ is the destination $\beta$ group and
$D_i=(\alpha_i,\beta_g)$. The kernel scans
$D_j=(\alpha_j,\beta_g)$ in that group and keeps only
$\operatorname{popcount}(\alpha_i\oplus\alpha_j)=2$ or $4$.

For $c=2$ (mixed), $g'$ is the precomputed adjacent $\alpha$ group.
For $D_i=(\alpha_g,\beta_i)$, the kernel scans only the existing
sources $D_j=(\alpha_{g'},\beta_j)$ and keeps
$\operatorname{popcount}(\beta_i\oplus\beta_j)=2$.

In this way, a global determinant-level connection problem is
decomposed into many group-level searches. The channel metadata first
fixes the relevant existing $\alpha$ or $\beta$ group; the kernel then
scans only determinants in that group, applies the popcount test, and
evaluates Slater--Condon terms only for the surviving pairs. This is
the step that makes the matvec efficient without storing the full
connection graph.

\subsection{The
\texorpdfstring{$(K\!=\!1, Z\!\approx\!20)$}{K=1, Z=20}
deployment used for \texorpdfstring{$5.12$}{5.12}B}
\label{sec:s8_deployment}

\textbf{The $5.12$B calculation is the concrete production
instance of the $K\!\times\!Z$ architecture described above. It
realizes the $(K\!=\!1, Z\!\approx\!20)$ point: one persistent
factory on a high-memory node and $10$--$30$ stateless GPU
workers across two shared academic clusters, connected only by
HTTP. The rest of this section records the details of that run.}

The choice of $K\!=\!1$ follows from the factory-side state still
being large but localizable. In the successful deployment, the
prepared determinant data, checkpoints, and out-of-core Davidson
vectors all lived on the factory node's persistent local scratch.
The prepared data directory was already $\sim\!297$\,GB, dominated
by the $\alpha$- and $\beta$-group files and mini-task list; a Ritz
checkpoint was $\sim\!77$--$82$\,GB; and each additional $V/HV$
Krylov layer added $2N_{\rm det}\times8\ {\rm bytes}
\simeq81.9$\,GB. These numbers are too large for an in-memory
single-node Davidson iteration, but still small enough to keep
under one factory if vector storage is streamed to disk
(Table~\ref{tab:factory_state_budget}; Sec.~\ref{sec:s8_ooc}).

\begin{table}[h]
\caption{Factory-side state budget in the successful $5.12$B
deployment. The first two rows are measured deployment sizes on
persistent local scratch for the $K\!=\!1$ factory, the $V/HV$
row is the per-layer out-of-core vector increment, and the final
row gives an eight-layer storage example.}
\label{tab:factory_state_budget}
\centering
\begin{tabular}{lrl}
\toprule
Component & Size & Role \\
\midrule
Prepared data & $\sim\!297$ GB &
group files, mini-tasks, diagonal, permutations, and integrals \\
Ritz checkpoint & $\sim\!77$--$82$ GB &
restart state for the current Ritz vector and its image \\
One $V/HV$ layer & $\sim\!81.9$ GB &
one stored Davidson basis vector plus its matvec result in the
out-of-core store \\
\midrule
Example total with $8$ $V/HV$ layers & $\sim\!1.03$ TB &
prepared data plus Ritz checkpoint plus eight out-of-core layers \\
\bottomrule
\end{tabular}
\end{table}

The complementary single-node GPU route was not scalable. On the
earlier single-node $4\!\times\!\text{H200}$ path, the $2.56$B
matvec fit within one $4\!\times\!\text{H200}$ node, but a direct
linear projection to $5.12$B put the per-card footprint at about
$154$\,GB, above the nominal $140$\,GB available on an H200
(Table~\ref{tab:h200_stress_test}). This table is not the memory
model of the final HTTP workers, whose footprint is controlled by
the assigned bundle and local cache. Its role is instead to
record why the single-node GPU route was not the production path: even
if optimized to fit $5.12$B, it would have a hard scaling ceiling
set by the largest available single node, instead of scaling by
adding independent workers. However, the same experiment suggests
that a larger single node, for example an
$8\!\times\!\text{H200}$ server, could plausibly reach an
$10$B-determinant problem. That estimate is already near the
practical limit of currently available single-node GPU servers:
beyond it, further growth would require an even larger
single-node GPU server, which is not generally available, or the
distributed-GPU worker route.

\begin{table}[h]
\caption{Single-node $4\!\times\!\text{H200}$ memory stress test
for the earlier single-node route. The $2.56$B column is the measured
earlier single-node multi-GPU path; the $5.12$B column is a linear
projection in $N_{\rm det}$. Rounded component rows are shown to
explain the algorithmic scaling choice; the total rows report the
recorded measured/projection totals and do not describe the final
stateless HTTP workers.}
\label{tab:h200_stress_test}
\centering
\begin{tabular}{lrr}
\toprule
Component & 2.56B (measured) & 5.12B (projected) \\
\midrule
$\alpha$ CSR (replicated) & $\sim\!20$ GB & $\sim\!40$ GB \\
$\beta$ CSR (one of four shards) & $\sim\!5$ GB & $\sim\!10$ GB \\
permutation arrays & $\sim\!5$ GB & $\sim\!10$ GB \\
work buffers ($v$, $\sigma$, workspace) & $\sim\!47$ GB & $\sim\!94$ GB \\
\midrule
Total per GPU & $\sim\!79$ GB & $\sim\!154$ GB \\
H200 available & $140$ GB & $140$ GB \\
\bottomrule
\end{tabular}
\end{table}

In the production distributed-GPU route, by contrast, the persistent-scratch
factory is the stateful side of the calculation, while GPU
workers remain stateless. Its runtime role is coordination
through two HTTP services: a control endpoint assigns bundles and
receives partial $\sigma$ vectors, while a file-serving endpoint
on the same host streams the needed file chunks to workers. Across iterations the
factory drives the Davidson loop,
streams the on-disk $\{V_k, HV_k\}$ basis as needed, and
aggregates the workers' $\sigma$ contributions into the next
trial vector.

The workers themselves are stateless GPU processes: each pulls
a bundle, runs it, POSTs $\sigma$ back, and asks for the next.
We sustain $Z\!\approx\!20$ concurrent workers (raw count
varying $10$--$30$ with backfill availability) across
opportunistic GPU pools on two shared academic clusters, mixing
T4, A100, H100, H200, and L40S devices that come and go on
minute-to-hour timescales. By (P1)--(P3), a worker is not assigned
a fixed shard of the full $N_{\rm det}$ vector. It only downloads
the file chunks needed by its assigned bundle(s), so heterogeneous
devices can join the same queue and simply finish bundles at
different rates. Each matvec round splits into $48{,}157$ bundles;
workers request one or more bundles according to their capacity
and then return for more. This queue-based scheduling lets faster
devices consume a larger share of the work and limits a worker
failure to the bundle(s) it had already accepted. A fresh worker
reaches steady-state throughput after $\sim\!10$--$20$ bundles,
once repeated file chunks are already present on the worker node
and the CUDA runtime has initialized; already-running workers are
therefore kept busy when possible. In the final matvec, $26$ workers were
present at completion and processed the bundle stream at
$4.0$ bundles/s, i.e. $\sim\!1.4\times10^{4}$ bundles/h.

The wire protocol is plain HTTP \texttt{GET}/\texttt{POST}
end-to-end --- no MPI and no worker-side shared file system.
The operational advantage is that a temporary GPU node only needs
ordinary network access to the factory's control and file-serving
endpoints, so it can join from a backfill pool whose scheduler we do
not control.

Reaching $N_{\rm det}\!=\!5.12\!\times\!10^{9}$ also crosses the
32-bit indexing ceiling at $2^{31}\!\approx\!2.15\!\times\!10^{9}$:
determinant indices, sparse-matrix row pointers, and the dense
permutation tables used in the matvec all overflow \texttt{int32}.
This is a distinct engineering threshold from the algorithmic
scaling of selected-CI itself: implementation overhead grows
steeply at this boundary and requires a 64-bit data layout
throughout the pipeline.

\subsection{Out-of-core Krylov subspace}
\label{sec:s8_ooc}

\textbf{The bundle system moves the matvec work off the factory,
but the Davidson basis would still grow by one pair of full
vectors per iteration. Out-of-core (OOC) storage shifts this
growth from RAM to disk, capping factory memory.}

At $5.12$B determinants, one double-precision vector is
$N_{\rm det}\times8$ bytes, or $\sim\!41$\,GB. A Davidson basis
layer consists of both $V_k$ and $HV_k$, so each new layer adds
$2N_{\rm det}\times8$ bytes $\approx\!82$\,GB. A modest
$m\!=\!8$ subspace is therefore $\sim\!660$\,GB before counting
static data, the current $\sigma$, or temporary vectors. Keeping
that basis resident would make the factory memory grow by
$\sim\!82$\,GB every iteration.

The production factory instead stores the basis in a disk-backed
vector store on persistent local scratch. Each store is an append-only
binary file for the sequence of $V_k$ or $HV_k$ vectors. The
Rayleigh--Ritz matrix elements are formed by sequentially reading
$HV_j$ and the needed $V_i$ vectors from scratch and taking dot
products. The new trial vector is orthogonalized by the same
sequential scans over the stored $V_i$ vectors (Gram--Schmidt,
re-orthogonalized if the residual overlap check fails). Thus the
factory pays repeated full-vector reads, but it does not keep the
whole Krylov basis in RAM.

OOC moves the basis off RAM, but on-disk usage still grows by
$82$\,GB per Davidson iteration. The actual bound on basis
size comes from \emph{Davidson restart}: when the subspace
size $m$ reaches its cap, the factory discards the on-disk
basis and re-initializes with the surviving Ritz vector
($V_1$) plus its $HV_1$ from the next matvec --- so the
long-run on-disk footprint stays bounded at $m \times 82$\,GB
$\approx\!660$\,GB. The same
disk-backed structure scales to the multi-factory deployment
of Sec.~\ref{sec:s8_path}, where each factory's OOC store
covers only its assigned row range, so the per-factory disk
footprint scales as $1/K$.

\subsection{Checkpoint and restart}
\label{sec:s8_checkpoint}

\textbf{The OOC basis (Sec.~\ref{sec:s8_ooc}) lives on the
factory's local scratch, which is lost when the factory node
fails. To recover, we mirror state to a durable shared
filesystem --- but, instead of mirroring the full $660$\,GB
OOC basis, we save only the current Ritz pair.}

A single matvec round takes hours and the run lives on
machines that can disappear. Local scratch is fast but tied
to the factory node, so a node failure loses the OOC basis.
Recovery needs a copy on durable shared storage --- in our
deployment, the cluster's Network File System (NFS).
Mirroring the full OOC basis at every iteration would cost
$\sim\!660$\,GB of NFS write per checkpoint --- prohibitive
for frequent saves. Instead we save just the best current
Ritz pair: the Ritz vector $\hat v$ and its image
$\hat H\hat v$ ($\sim\!41$\,GB each) plus a small JSON of
energy, residual, and iteration counters. The two large
vectors are written by a background thread, so the NFS write
does not block the next Davidson matvec.

On resume the factory streams the saved Ritz pair from NFS
into a fresh local-scratch OOC store and restarts Davidson at
$k\!=\!1$. The logical matvec counter continues, but the
local subspace
dimension and Davidson iteration number reset. This is a
numerically meaningful continuation rather than a
bit-identical replay --- acceptable because the scientific
object is the converged Ritz energy, not the exact
intermediate Krylov path. The $5.12$B run used this mechanism
through several factory-disruption events.

\subsection{Path to \texorpdfstring{$10^{12}$}{10\^{}12}: keep the
per-factory configuration fixed}
\label{sec:s8_path}

\textbf{The $5.12$B run is an experimentally validated
per-factory configuration. Variational spaces beyond this
scale follow by holding that configuration fixed and adding
row-owning factories --- $K$ grows linearly with $N_{\rm det}$
while every factory stays near the memory and scratch
envelope already demonstrated, and $Z$ stays at $\sim\!20$
workers per factory.}

The convergence anchor is Table~\ref{tab:5120m_convergence}:
$5.12\!\times\!10^{9}$ determinants at $(K, Z)\!=\!(1,
\sim\!20)$, run on heterogeneous backfill GPUs through several
factory restarts (Sec.~\ref{sec:s8_checkpoint}). The
extrapolation below scales from this validated reference
point.

\begin{table}[h]
\caption{Convergence anchor for the largest run reported here:
the $5.12\!\times\!10^{9}$-determinant
\textrm{[Fe$_4$S$_4$]} BS-1 Davidson calculation at $K\!=\!1$
and $Z\!\approx\!20$. The run converged by the energy
criterion, $|dE|<10^{-5}$ Ha. Wall time
for matvec phases: $\sim\!24.2$~h; whole Davidson wall time
(matvec plus all factory-side phases --- Rayleigh--Ritz solve,
expansion, checkpoint): $\sim\!29.7$~h. Final-matvec throughput
was $4.0$ bundles/s, or $\sim\!1.4\times10^{4}$ bundles/h.}
\label{tab:5120m_convergence}
\centering
\begin{tabular}{rcrr}
\toprule
\texttt{matvec\_iter} & $E$ (Ha) & $|r|$ & $|dE|$ (Ha) \\
\midrule
0 & $-327.241615$ & $2.63\!\times\!10^{-2}$ & --- \\
1 & $-327.242051$ & $7.13\!\times\!10^{-3}$ & $4.36\!\times\!10^{-4}$ \\
2 & $-327.242096$ & $4.48\!\times\!10^{-3}$ & $4.52\!\times\!10^{-5}$ \\
3 & $-327.242109$ & $3.31\!\times\!10^{-3}$ & $1.30\!\times\!10^{-5}$ \\
4 & $-327.242125$ & $3.93\!\times\!10^{-3}$ & $1.56\!\times\!10^{-5}$ \\
5 & $-327.242141$ & $3.19\!\times\!10^{-3}$ & $1.64\!\times\!10^{-5}$ \\
6 & $-327.242152$ & $2.80\!\times\!10^{-3}$ & $1.09\!\times\!10^{-5}$ \\
7 & $\mathbf{-327.2421555}$ & $\mathbf{1.75\!\times\!10^{-3}}$ &
                              $\mathbf{3.42\!\times\!10^{-6}}$ \\
\bottomrule
\end{tabular}
\end{table}

What limits any one factory is its memory and OOC/scratch
envelope (Sec.~\ref{sec:s8_ooc}), not the worker pool size.
The natural scaling axis is therefore $K$, not single-factory
growth: hold each factory at the validated
$\sim\!5\!\times\!10^{9}$-determinant slice and shard
destination rows across more factories. Under this rule a
$10^{12}$ run is the $K\!\approx\!200$ point with
$Z\!\approx\!20$ workers per factory ---
$K\!\times\!Z\!\approx\!4{,}000$ total workers.
Table~\ref{tab:scaling_to_1T} lays out the full resource
extrapolation across $K\!=\!1\to 200$ (per-factory $v$,
Krylov scratch, and bundle throughput all stay near their
$5.12$B values; the aggregate $v$ grows linearly with
$N_{\rm det}$).

\begin{table}[h]
\caption{Resource extrapolation from the measured $5.12$B
factory-worker unit. $K$ scales with $N_{\rm det}$ to keep each
factory near the demonstrated per-factory vector and OOC-Krylov
footprint; $Z$ is kept at the observed $\sim\!20$ workers per
factory. This table is an architectural scaling envelope, not a
completed $10^{12}$ benchmark.}
\label{tab:scaling_to_1T}
\centering
\begin{tabular}{l r r r r}
\toprule
Resource & $5.12\!\times\!10^{9}$ (current) & $10^{10}$ & $10^{11}$ & $10^{12}$ \\
\midrule
$K$ (factories)                                      & 1     & 2     & 20    & 200 \\
$Z$ (workers per factory)                            & $\sim\!20$ & $\sim\!20$ & $\sim\!20$ & $\sim\!20$ \\
$K\!\times\!Z$ (total workers)                        & $\sim\!20$ & $\sim\!40$ & $\sim\!400$ & $\sim\!4{,}000$ \\
Per-factory $v$ slice                                & $41$\,GB & $40$\,GB & $40$\,GB & $40$\,GB \\
Per-factory Krylov on scratch ($m=8$)                & $660$\,GB & $640$\,GB & $640$\,GB & $640$\,GB \\
Per-factory bundle throughput (bundles/min, steady state) & $\sim\!2.4\times10^{2}$ & $\sim\!2.4\times10^{2}$ & $\sim\!2.4\times10^{2}$ & $\sim\!2.4\times10^{2}$ \\
Total $v$ across factories                           & 41\,GB & 80\,GB & 800\,GB & 8\,TB \\
\bottomrule
\end{tabular}
\end{table}

Two qualitative changes appear at $K\!>\!1$. First, each
factory serves its own row range of $v$, so when a worker's
bundle reads source dets owned by a different factory the
worker fetches those slices from that factory's file server.
Second, the global Davidson dot products require cross-factory
communication. Each operation of Sec.~\ref{sec:s8_scaling} --- forming the
Rayleigh--Ritz projected Hamiltonian
$\langle V_i, HV_j\rangle$, orthogonalizing new trial vectors
($\langle V_i, w\rangle$), computing the residual norm
$\|r\|\!=\!\sqrt{\langle r, r\rangle}$ for convergence, and
normalizing new basis vectors via $\|w\|\!=\!\sqrt{\langle w,
w\rangle}$ --- has the form $\langle a, b\rangle =
\sum_{n=1}^{N_{\rm det}} a_n b_n$. With the row index $n$
partitioned across $K$ factories ($R_1, \ldots, R_K$, each
factory $k$ holding the slices $a^{(k)}, b^{(k)}$ of $a, b$
over $R_k$), this becomes a per-factory partial sum plus a
sum of the $K$ resulting scalars across factories,
\begin{equation}
\langle a, b\rangle \;=\; \sum_{k=1}^{K} s_k,
\qquad s_k \;\equiv\; \sum_{n \in R_k} a^{(k)}_n\, b^{(k)}_n.
\label{eq:distrib_dot}
\end{equation}
Step 1 (computing each $s_k$) is embarrassingly parallel and
$O(N_{\rm det}/K)$ per factory; step 2 (summing the $K$
scalars and broadcasting the result back) is bandwidth-trivial
but forces every Davidson iteration through a global sync
point. The Rayleigh--Ritz $m^{2}\!=\!64$ partials are
independent and batch into a single $K$-way all-reduce of an
$m^{2}$-scalar vector; orthogonalization (classical
Gram--Schmidt + reorthogonalization) costs another one or two
calls; residual and normalization norms add one more. Each
Davidson iteration therefore needs only $\sim\!3$--$5$ global
sync points, not $\sim\!10^{2}$ separate ones.

All other Davidson per-factory work --- basis storage via OOC
and restart (Sec.~\ref{sec:s8_ooc},
\ref{sec:s8_checkpoint}), Ritz vector formation, residual
computation --- operates entirely on the per-factory slice
and costs the same per factory at $K\!=\!1$ as at $K\!>\!1$.

The extrapolation above is for the variational Davidson
matvec; semistochastic PT2 (Sec.~\ref{sec:fig2_data}) at this
scale requires a dedicated streaming/GPU implementation whose
memory scales with the per-factory slice rather than the full
external space, and will be reported in a separate paper.

The variational matvec architecture therefore carries over to
$10^{12}$ unchanged: a $10^{12}$ run is $K\!\approx\!200$ of
the validated factory--worker unit, with the matvec algorithm
itself unchanged. What remains is operational ---
sustaining $\sim\!4{,}000$ backfill workers reliably across
opportunistic compute pools --- but not algorithmic.

\section{Finding the P-cluster ground-state spin pattern from
random initial determinants}
\label{sec:pcluster_spin}

The P-cluster active space ((114e,\,73o), $M_s=0$) admits
$35$ distinct broken-symmetry spin patterns of its eight Fe(II)
centers~\cite{Li2019Pcluster}. The ground state, in the
enumeration of Ref.~\cite{Li2019Pcluster}, is configuration
No.~3,
$\uparrow\{\mathrm{Fe}_{1,2,3,8}\}\downarrow\{\mathrm{Fe}_{4,5,6,7}\}$.
The standard practice in the literature is to hand-pick initial
determinants that already encode the target spin pattern; this
commits the calculation to one basin from the start. In this
section we ask whether TrimCI~+~COO can locate the correct
ground-state spin pattern \emph{starting from random initial
determinants}, with no manual basin pre-selection. We make three
observations. First, with a small determinant budget
($100$ determinants), random-determinant CI alone in the LMO
basis can neither order basins nor reach a clean polarized
state. Second, COO collapses each randomly initialized
wavefunction onto a fully polarized four-up-four-down (4U+4D)
basin at that basin's intrinsic energy minimum within several
TrimCI~+~COO cycles. Third, the resulting ordering of plateau energies
puts No.~3 at the bottom --- so the random-start protocol
identifies the ground-state basin variationally, without
hand-picked spin-polarized guesses.

\begin{figure}[t]
\centering
\includegraphics[width=0.92\textwidth]{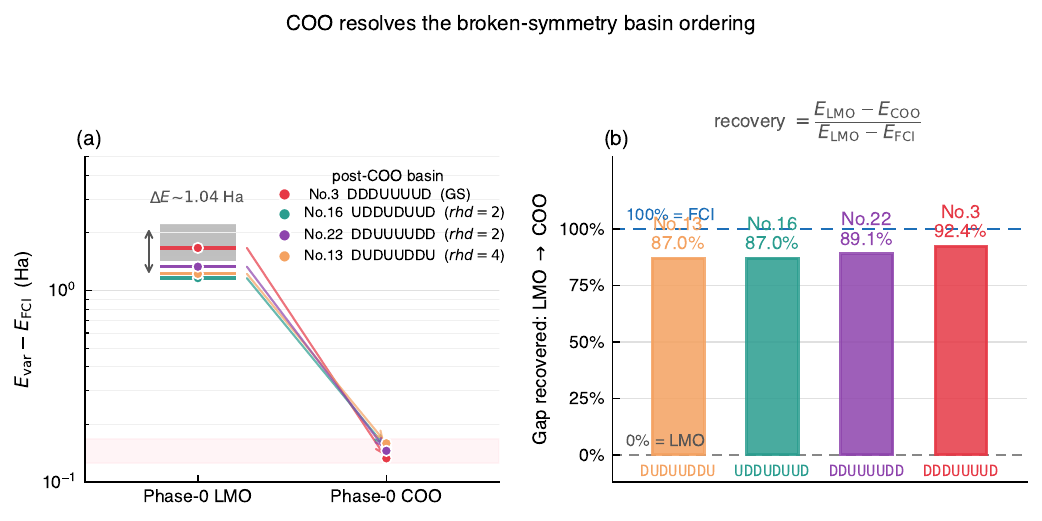}
\caption{COO resolves the broken-symmetry basin ordering on the
P-cluster (114e,\,73o).
(a) The $200$ Phase~0 runs in the LMO basis
(\texttt{orbital\_optimization=False}, $100$-det cap) span a
$\sim\!1$~Ha window without resolving basin structure
(grey ticks; vertical span annotated). Coloured ticks mark the
representative runs picked as COO seeds; arrows trace each
seed's descent to its post-COO plateau. Basins are tagged by their reduced Hamming distance
between spin-pattern bitstrings to the ground state No.~3
(\textit{rhd}).
(b) For each COO experiment, the bar shows the recovery
$(E_{\rm LMO} - E_{\rm COO})/(E_{\rm LMO} - E_{\rm FCI})$, i.e.\
the fraction of the LMO-seed-to-FCI energy gap closed by COO.}
\label{fig:basin_ordering}
\end{figure}

\textit{At the $100$-determinant scale, the LMO basis cannot
order basins or even resolve clean polarized states.}
We ran TrimCI Phase~0 with $\texttt{num\_runs}\!=\!200$
(independent random initializations) on the LMO
\cite{Li2019Pcluster} FCIDUMP with
\texttt{orbital\_optimization=False} and a $100$-determinant cap.
The final energies span a $\sim\!1.0$~Ha window from
$-17491.08$ to $-17490.03$~Ha (Fig.~\ref{fig:basin_ordering}a).
Only $4$ of the $200$ runs have a dominant determinant that
exactly matches one of the $35$ 4U+4D BS configurations of
Ref.~\cite{Li2019Pcluster} (No.~22, 8, 5, 17); \emph{none}
matches the ground state No.~3, and the lowest-energy run sits
$\sim\!1.0$~Ha above the COO basin minimum reported below. The
remaining $196$ runs either have at least one Fe with
$|S_z|\!<\!0.5$ (so the dominant determinant is partially
polarized) or lie outside the 4U+4D manifold (e.g.\ 5U+3D); the
random $100$-determinant search in the LMO basis does not
reliably reach the ground-state basin.
However, these random outputs are not wasted: they sample the
periphery of the broken-symmetry basins --- a continuous ridge
of partially-polarized Fe states surrounding the basin minima
--- and each provides a usable initial state for the COO step
that follows.

\textit{COO collapses each LMO wavefunction onto its nearest
4U+4D basin.}
We label each BS configuration by an $8$-character spin pattern
over Fe$_1$--Fe$_8$: U or D when the Fe center carries a clear
polarization ($|S_z|\!\geq\!0.5$ in the dominant determinant,
sign of $S_z$); 0 when $|S_z|\!<\!0.5$. The ground state No.~3
reads UUUDDDDU. Because $M_s\!=\!0$ admits global spin flip,
each BS configuration has an equivalent flipped form (No.~3
also reads DDDUUUUD), so we measure proximity between two
patterns by the \emph{reduced Hamming distance}
$\textit{rhd}$, the minimum count of differing Fe centers over
the two flip orientations.

Taking the $100$-determinant LMO Phase~0 wavefunction as the
initial state (seed) and running ten TrimCI~+~COO cycles
(\texttt{tracking\_dets=True} and
\texttt{loaded\_dets\_randomness=0.1}, see
Table~\ref{tab:bfgs_config}), the wavefunction lands in one of
the $35$ 4U+4D configurations of Ref.~\cite{Li2019Pcluster}
within the first several cycles.
The seed nearest to No.~3 ($\textit{rhd}\!=\!1$ to No.~3, with
one undecided Fe) converges to No.~3 itself at
$E=-17492.103$~Ha; a seed in basin No.~16
($\textit{rhd}\!=\!2$ to No.~3, single
Fe$_1\!\leftrightarrow\,$Fe$_5$ swap) plateaus at
$E=-17492.085$; a fully polarized seed in basin No.~22
($\textit{rhd}\!=\!2$ to No.~3 via
Fe$_3\!\leftrightarrow\,$Fe$_7$) plateaus at
$E\!\approx\!-17492.090$; a seed in basin No.~13
($\textit{rhd}\!=\!4$ to No.~3) plateaus at
$E\!\approx\!-17492.077$. The $\sim\!1.0$~Ha drop from LMO
Phase~0 seed to TrimCI~+~COO plateau, near-uniform across all
basins probed, is the energy gain from TrimCI~+~COO carrying
the wavefunction to the bottom of its basin.

\textit{The basin ordering after COO puts No.~3 lowest.}
Pre-COO, the seeds sit in a $\sim\!1$~Ha window with no clear
spin pattern (Fig.~\ref{fig:basin_ordering}a, grey ticks in the
LMO column). After ten TrimCI~+~COO cycles, each seed
plateaus inside its target basin, and the basin energies
spread out by $\sim\!90$~mHa
(Fig.~\ref{fig:basin_ordering}b). Quantifying with the gap
recovery toward the FCI reference (taken as the extrapolated
DMRG limit of Xiang \emph{et
al.}~\cite{xiangDistributedMultiGPUInitio2024},
$E_{\rm FCI}\!=\!-17492.236$~Ha),
$(E_{\rm LMO}-E_{\rm COO})/(E_{\rm LMO}-E_{\rm FCI})$, the
recovery is largest for No.~3 itself ($92.4\%$), and similar
for No.~16 ($87.0\%$), No.~22 ($89.0\%$), and No.~13
($87.0\%$). \textbf{In the LMO basis, the seeds are spread over a
$\sim\!1$~Ha window with no meaningful ordering; after COO,
they sort themselves into their physical basins, and the
ground state No.~3 emerges at the bottom of the ordering.}

Continuing on No.~3 to FCI: from the COO 100-det plateau at
$E=-17492.103$~Ha (the deepest in Fig.~\ref{fig:basin_ordering}b,
$92.4\%$ of the LMO-seed-to-FCI gap), Phase~1+2 expansion in
the COO basis to $N_{\rm det}=6.4\!\times\!10^{7}$ (main text
Fig.~3, panel f) closes another $\sim\!0.10$~Ha ($5.5\%$) to
$E_{\rm var}+\Delta E_{\rm PT2}=-17492.199$~Ha (with
semistochastic PT2 correction), leaving a residual
$\sim\!0.04$~Ha ($2.1\%$) to FCI. The bulk of the recovery
toward FCI thus comes from the orbital rotation in COO, with
determinant expansion serving as a smaller refinement on top.

\clearpage
\putbib[coo-ref]
\end{bibunit}

\end{document}